\documentclass{aa}
\usepackage{caption}

\usepackage[switch]{lineno}
\renewcommand\linenumberfont{\normalfont\tiny\sffamily}
\linenumbers

\usepackage{graphicx}

\usepackage{txfonts}

\usepackage[urlcolor=cyan, colorlinks=true, citecolor=blue, linkcolor=blue]{hyperref}
\usepackage{threeparttable}
\usepackage{comment}
\usepackage{cancel}
\usepackage[normalem]{ulem}
\usepackage{soul}
\usepackage{numprint} 
\usepackage{enumitem}
\usepackage{graphicx}
\usepackage{subcaption}

\newcommand{\EZ}{\texttt{pandora EZ }}
\newcommand{\zs}{$z_\mathrm{{spec}}$}
\newcommand{\cigale}{\texttt{CIGALE}}

\newcommand{\Mstar}{$\rm{M_\ast}$}
\newcommand{\lSFR}{$\log{\rm(SFR)}$}
\newcommand{\lMstar}{$\log{\rm(M_\ast)}$}
\newcommand{\SIPGI}{\textit{SIPGI}}

\begin{document}

   \title{Dissecting ultra-diffuse galaxies in the field:}
   \subtitle{Observational biases and spectroscopic determination of physical properties in low surface brightness systems}
   \titlerunning{Dissecting Ultra-Diffuse Galaxies in the Field.}

   \author{A.~Vanzanella\inst{1},
          K.~Malek\inst{1,2},
          Junais\inst{3,6},
          E.~Bernaud \inst{2},
          A.~Gargiulo\inst{4},
          H.~Thuruthipilly\inst{7,8},
          K.~Lisiecki \inst{1},
          A.~Pollo\inst{1},
          M.~Scodeggio \inst{4},
          P.~M. Weilbacher\inst{3},
          S.~Boissier \inst{2},
          P.~Franzetti \inst{4},
          G.~Lorenzon \inst{1},
          A.~Nanni \inst{1,9},
          C.~Adami  \inst{2},
          R.~Gavazzi \inst{2,5},
          O.~Ilbert \inst{2},
          S.~Pal \inst{1},
          N.~Principi~Cavaterra \inst{1}
          }
    
   \authorrunning{Vanzanella et al.}
   \institute{National Centre for Nuclear Research, Warsaw, Poland.
            \email{antonio.vanzanella@ncbj.gov.pl}
            \and Aix-Marseille Université, CNRS, CNES, LAM, Marseille, France.
            \and Leibniz-Institut für Astrophysik Potsdam (AIP), An der Sternwarte 16, 14482 Potsdam, Germany.
            \and INAF – Istituto di Astrofisica Spaziale e Fisica Cosmica Milano, Via A. Corti 12, I-20133 Milano, Italy.
            \and Institut d’Astrophysique de Paris, UMR 7095, CNRS, and Sorbonne Université, 98 bis boulevard Arago, 75014 Paris, France.
            \and Instituto de Astrofísica de Canarias, Vía Láctea S/N, 38205 La Laguna, Spain.
            \and Departamento de Física Teórica, Atómica y Óptica, Universidad de Valladolid, 47011 Valladolid, Spain.
            \and Laboratory for Disruptive Interdisciplinary Science (LaDIS), Universidad de Valladolid, 47011 Valladolid, Spain,
            \and INAF – Osservatorio Astronomico d’Abruzzo, Via Maggini SNC, 64100 Teramo, Italy.}
   \date{}

\abstract 
{Ultra-diffuse galaxies (UDGs) in the field, gravitationally isolated from galaxy clusters, are faint systems that remain poorly represented in the literature due to lack of spectroscopic confirmation and the difficulty of obtaining high signal-to-noise emission-line measurements.}
{We present a spectroscopic study of 17 blue UDG candidates in the field using optical emission lines to confirm their ultra-diffuse nature and characterise their physical properties.}
{Data were collected using the MODS instrument at the Large Binocular Telescope Observatory 
and the MISTRAL instrument at the Observatoire de Haute-Provence. 
Using the software \EZ, we derived spectroscopic redshifts (\zs) for our field UDG candidates. 
We then computed their effective radii ($r_\mathrm{{eff}}$) and central surface brightnesses ($\mu_\mathrm{{0,g}}$).
We measured the H$\alpha$ and H$\beta$ emission line fluxes in the 17 spectra and derived star formation rates (SFRs) from the line-luminosity relation.
We performed forced photometry on our sample to obtain photometric fluxes and applied an aperture correction on the H$\alpha$ integrated fluxes, propagating the correction to the derived SFRs.
Using colour relations and estimated dust attenuation, we then computed stellar masses (\Mstar) and corrected the SFRs accordingly.
Finally, we placed our sample on the star-forming main sequence, comparing its distribution with that of local star-forming galaxies. 
Two sources were further examined as potential giant low surface brightness galaxies (GLSBGs).}
{In our samples, we identify nine confirmed UDGs, eight other low surface-brightness galaxies (LSBGs), including one GLSBG.
The \zs~of our field UDGs span a range of $0.015 - 0.037$. Their $r_\mathrm{eff}$ ranges from $1.69 - 4.99~\mathrm{kpc}$, and their $\mu_{0,g}$ range from $24.05 - 24.98~\mathrm{mag\ arcsec^{-2}}$. Consistent with previous studies, the galaxies exhibit low to moderate dust content, with an average V-band attenuation of $0.29~mag$.
The spectroscopically confirmed UDGs presented in this work, after aperture correction, exhibit $-2.95\leq$ \lSFR $~\leq-1.60~\mathrm{M}_\odot\,\mathrm{yr}^{-1}$ as an upper limit and $7.20 \leq$ \lMstar~$\leq 8.08~\mathrm{M}_\odot$. These values place them along the star-forming main sequence.}
{Our results indicate that blue field UDGs are characterised by heterogeneous dust attenuation and occupy the same region of the star formation–stellar mass plane as dwarf LSBGs. This is consistent with the broader population of local star-forming galaxies and indicates no distinct global star-forming behaviour.}
   
   \keywords{Low surface brightness galaxies --
             Ultra-diffuse galaxies --
             Spectroscopy --
             Galaxy evolution --
             Galaxy properties.
               }

   \maketitle
\nolinenumbers 

\section{Introduction}
\label{introduction}
Observing the night sky, our eyes are mostly drawn to the brightest and shiniest bodies.
However, other types of celestial objects, equally abundant, remain hidden from sight.
Among these objects are low surface brightness galaxies (LSBGs).
Low surface brightness galaxies are defined as galaxies with an average optical surface brightness below the typical level of the night sky~\citep{Bothun1997}.
For instance, \cite{Martin2019} and \cite{Junais23} define LSBGs as galaxies with an average $r$-band surface brightness, $\overline{\mu}_r$, fainter than $23\ \mathrm{mag\ arcsec^{-2}}$.
Low surface brightness galaxies are estimated to contribute from 6\% to 11\% of the luminosity of the local observable universe. However, their contribution to the total number density of galaxies is estimated to be 40\% to 50\% \citep{Martin2019, ONeil2000}.
To date, studies on LSBGs have been limited by the depth thresholds of current instruments.
The completeness of surveys conducted over the past decades declines significantly below an effective surface brightness threshold of $\sim23\ \mathrm{mag\ arcsec^{-2}}$ in the \textit{r} band \citep[i.e. ][]{Driver2005, Blanton_2005, Zhong2008}.
With the advent of new generation wide-field instruments -- mainly the Vera C. Rubin Legacy Survey of Space and Time \citep[LSST,][]{LSST} and Euclid \citep{euclid2023} -- the previous effective surface brightness threshold can be overcome, ushering in a new era for the study of LSBGs.\\
Among LSBGs, a subsample of objects has been classified as ultra-diffuse galaxies (UDGs). These are LSBGs with a central surface brightness in the \textit{g} band ($\mu_\mathrm{0,g}$) fainter than $24\ \mathrm{mag\ arcsec^{-2}}$ and a half-light radius ($r_\mathrm{eff}$) larger than $1.5$~kpc \citep{Vandokkum2015}.
Ultra-diffuse galaxies occupy the faint end of the luminosity distribution of LSBGs, inhabiting field \citep{Prole_2019}, group \citep{Cohen_2018, Marleau_2021}, and cluster \citep{Koda_2015, Mihos_2015, Lim_2020, LaMarca} environments.
Previous studies \citep[e.g.] [] {Leisman_2017, Rong_2020} have established substantial differences in the properties of UDGs as a function of their environment.
The lack of atomic gas in UDGs found in high-density environments, such as clusters and groups, has led them to be referred to as red UDGs.
In contrast, gas-rich UDGs are referred to as blue UDGs, which mainly populate low-density environments~\citep{Prole_2019}.
Most UDGs identified in the literature to date have been found in galaxy clusters, largely due to observational biases. The UDGs in clusters predominantly lie on the red sequence~\citep{Koda_2015}, and their smooth stellar light distributions make them easier to detect from background noise in surface-brightness-limited surveys \citep{vanderburg2016udg}.
Moreover, galaxy clusters allow us to estimate the distance of UDGs by inferring their proximity to other objects in the cluster \citep[i.e. ][]{Koda_2015, vanderburg2016udg}. 
This method provides a reasonable estimate of the intrinsic surface brightness and size of the UDG candidates. 
It is more challenging to find UDG candidates in the field without broad photometric coverage. 
Additionally, without spectroscopic follow-up, it is impossible to estimate the proper redshift and, subsequently, the candidate's surface brightness and size.
To improve the study of UDGs, a larger statistical sample is needed, including both cluster and field UDGs.
Comparing properties across different UDG host environments would allow us to identify the most likely evolutionary paths of these galaxies, while factoring out the effect of episodic events in their surroundings.

Once a UDG candidate has been selected, spectroscopy is usually needed to measure the redshift and, hence, to confirm its physical size (e.g. verify whether it satisfies the \citealt{Vandokkum2015} criterion). However, the chances of obtaining a satisfactory signal-to-noise ratio (S/N) for the incoming flux are quite low (for further details, see Sect.~\ref{sub_section_MODS_data}).
This not only discourages further studies on UDGs in the field but also is the primary reason for the lack of a large sample of such galaxies in this environment.
The number of spectroscopically confirmed UDGs is $\sim80$ \citep{Jenny, Yiping2026}.
When considering UDGs in the field, this number is even smaller. 
In \cite{Leisman_2017} and \cite{Janowiecki2019}, comprehensive investigations of $HI$-bearing ultra-diffuse sources (HUDs) were conducted. These studies used $HI$ detections from the Arecibo Legacy Fast ALFA (ALFALFA) survey to determine distances and confirm the ultra-diffuse nature of the targets.
In contrast, \cite{Prole_2019} performed a photometric study of field UDGs using the publicly available Kilo-Degree Survey together with the Hyper Suprime-Cam Subaru Strategic Program to explore their abundance and intrinsic properties.
However, this analysis was performed without distance confirmation.
An automated approach is required to search for LSBG candidates in large new and archival surveys. This approach would enable a preselection of faint objects for further, more detailed analyses.
For example, \cite{Tanoglidis2018} used the support vector machine technique to obtain a catalogue of LSBG candidates with data from the Dark Energy Survey \citep[DES;][]{DES}. 
In a follow-up study, \cite{Thuruthipilly2024A} developed a machine learning (ML) pipeline based on the transformer model \citep{Attention} to improve the previously obtained catalogue. 
Among the 27\,000 LSBG candidates identified by \cite{Thuruthipilly2024A}, $\sim${4\,000} were new LSBG candidates never detected before. 
Furthermore, $\sim${317} UDG candidates were found in high-density environments. 
Accounting for the statistically significant number of UDGs in groups and clusters identified by \cite{Thuruthipilly2024A}, it is now possible to study the nature of UDGs in different environments.
In fact, in this work, we focus specifically on the field UDG candidates from \cite{Thuruthipilly2024A} without any known distance estimate.
In this paper, we first aim to
confirm the ultra-diffuse nature of UDG candidates through a spectroscopic follow-up. We then aim to extract the physical properties of field UDGs, such as star formation rates (SFRs) and stellar masses (\Mstar), and estimate their loci on the star-forming main sequence (SFMS) diagram.\\
This paper is structured as follows: 
In Sect.~\ref{2}, we describe the catalogue from which the UDG candidates were selected, outlining the motivation for our choices and the spectroscopic data collected. Here, we explain the observational strategy adopted and the reduction pipeline used. 
In Sect.~\ref{section_methodology}, we describe the measurements of the physical properties of the sources, including size, H${\alpha}$ line luminosity, the fitting procedure used, SFRs, and $\rm{M_{star}}$. 
In Sect.~\ref{4}, we present and discuss the SFMS plot and explain the corrections applied to obtain consistent results. 
Furthermore, we examine the two giant low surface brightness galaxies (GLSBGs) in our sample.
Finally, we draw our conclusions in Sect.~\ref{section_conclusion}.
Throughout this paper, we adopt a flat $\Lambda$CDM cosmology with $H_\mathrm{0} = 70\, \mathrm{km\,s^{-1}\,Mpc^{-1}}$, $\Omega_\mathrm{m} = 0.30$, and $\Omega_\mathrm{\Lambda} = 0.70$.
\section{Data}
\label{data}
In this section, we introduce the sample used in this work.
Our sample comprises data from two separate observation campaigns.
The first was performed at the Large Binocular Telescope Observatory (LBT; $2 \times 8.4$m) with the Multi-Object Double Spectrographs \cite[MODS, ][]{MODS} instruments and observed sources selected from the LSBGs catalogue by \cite{Thuruthipilly2024A}.\\
The second was performed at the T193 telescope of the Observatoire de Haute-Provence (OHP; $1.93$m) with the MISTRAL spectrograph \citep{MISTRAL} and observed UDG candidates from the SMUDGES catalogue \citep{Zaritsky_2023} and the COSMOS2020 catalogue \citep{cosmos2020}.
Both source ensembles were selected to identify the bluest field UDG candidates in their respective catalogues.
\label{2}    
\subsection{LBT/MODS sample selection}
\label{LBT/MODS sample selection}
For the LBT/MODS observations, we used UDG candidates from the \cite{Thuruthipilly2024A} catalogue. 
\cite{Thuruthipilly2024A} created the catalogue by deriving data from the DES and the work of
\cite{Tanoglidis2018}. 
In particular, \cite{Thuruthipilly2024A} obtained photometric data from DES data release~1 (DES DR1; \cite{DES}) and DES Y3 gold coadd object~\citep[DES Y3\_gold\_2\_2.1, ][]{des_object_catolog} to search for and select LSBG candidates, using transformer-based deep learning models.
 From these data, they performed several post-processing operations to populate the LSBGs and artefact dataset classes. 
To summarise, they extracted the sources from the DES Y3 gold coadd object catalogue (containing $\sim${319} million astronomical objects). 
First, they removed point-like objects and applied constraints on the \textit{g}-band half-light radius ($2.5'' < r_\mathrm{{eff}}<20''$) and on the average surface brightness within the effective radius values ($24.2 < \overline{\mu}_\mathrm{eff} < 28.8$). 
Then, they applied a colour cut on the sources:
\begin{equation}
\begin{aligned}
-0.1 &< g - i < 1.4,  \\
(g-r) &> 0.7 (g-i) - 0.4,\\
(g-r) &< 0.7 (g-i) + 0.4. 
\end{aligned}
\end{equation}
Their aim was to remove spurious detections due to optical artefacts detected in the \textit{g}, \textit{r}, \textit{z} DES bands and blends of high-redshift galaxies \citep{Greco2018}. 
They also imposed an axis ratio greater than 0.3 to remove artefacts, such as highly elliptical diffraction spikes. 
The training sample used to populate the dataset is the same as that used in \cite{Tanoglidis2018}.
The pipeline used on the DES DR1 data results in a catalogue of 27\,000 LSBG candidates, of which $4083$ had never been classified as LSBGs before.
The sample selection for the LBT observations was done starting from this subsample of $4083$ new galaxies.\\
The field UDG candidates were selected following the methodology of \cite{Zaritsky_2023}.
We first imposed a central surface brightness threshold, requiring $\mu_\mathrm{0,g} > 24$ mag arcsec$^{-2}$, and subsequently applied a constraint on the angular size, selecting only sources with effective radii larger than $5.3''$.
We then introduced two additional selection criteria:
\begin{itemize}
\item  a colour cut ($g - i < 0.4$ mag) to isolate the bluest galaxies in the catalogue;
\item  a declination constraint ($ > -10^\circ$) to ensure observability with the Large Binocular Telescope.
\end{itemize}
This selection preferentially targets galaxies with ongoing or recent star formation, which typically exhibit strong nebular emission lines (e.g. H${\alpha}$ and H$\beta$). The presence of such features facilitates the detection and measurement of reliable spectroscopic redshifts for these intrinsically faint systems.
Moreover, to select only UDGs in the field, the sources were cross-matched with the cluster and group catalogue in the same area \citep{Xu}, excluding those within the cluster and group virial radius.
The selection yielded 22 blue UDG candidates suitable for spectroscopic follow-up.
\subsection{OHP/MISTRAL sample selection}
\label{OHP/MISTRAL sample selection}
In this work, we also used another sample for the MISTRAL observations presented in a preliminary analysis~\citep{Mistral_proposal}. Sample selection was performed using the SMUDGES catalogue \citep{Zaritsky_2019}.
This catalogue contains 7\,070 UDG candidates in different density environments that satisfied the selection criteria based on an $r_\mathrm{eff}$  $> 5.3''$ and $\mu_\mathrm{0,g}$ larger than $24\ \mathrm{mag\ arcsec^{-2}}$.
To ensure the bluer sources from both catalogues, a colour cut of $(g-r) < 0.25$ and $r< 19~mag$ was applied to obtain the most star-forming candidates and increase the probability of H$\alpha$ detection.
Furthermore, sources with known spectroscopic redshifts listed in the \cite{Zaritsky_2019} catalogue were excluded, along with those lying outside the declination range of $-10^\circ < \mathrm{DEC} < 90^\circ$, imposed by observability constraints at OHP.
To select only field UDG candidates, all sources were cross-matched with the cluster and group catalogue in \cite{Xu}, and those located within the virial radius of any cluster or group were excluded.
An additional source in the sample was selected from the COSMOS2020 catalogue \citep{cosmos2020}.
This source was initially selected to check the feasibility of the observation of LSBGs with the MISTRAL instrument (this is also the reason for the different integration time with respect to the other OHP candidates stated in Table~\ref{tab:obs_details}).
The test, with this single source, showed the feasibility of the observation, so we decided to include the OHP-1 galaxy in our sample, although it does not satisfy our size and $\mu_\mathrm{0,g}$ selection cuts ($r_\mathrm{eff}=3.06''$ and $\mu_\mathrm{0,g}=21.43 ~\mathrm{mag\ arcsec^{-2}}$)\footnote{The OHP-1 values, $r_\mathrm{eff}$ and $\mu_\mathrm{0,g}$, were obtained by performing \textsc{GALFIT}~\citep{Peng_2002} Sérsic modelling on this target using the DEcalS~\citep{DECALs}~\textit{g}-band cutout.}.
The colour and magnitude values of this source are within our selection range, which we applied above for the SMUDGES catalogue.
This whole process results in a sample of  $20$ UDG candidates available for spectroscopic follow-up observation using the MISTRAL instrument.
\subsection{LBT/MODS data}
\label{sub_section_MODS_data}
We requested observational time for the selected 22 UDG candidates at LBT during the 2024B semester (proposal number AIP-2024B-008, with principal investigator P.~Weilbacher).  
We used the MODS instrument and its two channels -- blue (3200 -- 5800~\AA) and red (5800 -- 10000~\AA) -- to acquire the spectra of the sample (with an expected redshift $< 0.15$; \citealt{Greene2022}), detect the most prominent lines (H$\alpha$, H$\beta$), confirm the UDG nature of the candidates through spectroscopic redshift measurement, and compute the main physical parameters of those sources.
The observational run cycle yielded observations of 14 out of the 22 UDG candidates requested.
Due to the faintness of the UDGs and in accordance with previous work by \cite{Jenny} and \cite{Bellazzini2017}, we requested to observe the sources with the MODS instrument in binocular mode to increase the total integration time and, thereby, our chances of obtaining a sufficient S/N ratio and measuring the redshift of the targets. Unfortunately, 
both \textit{MODS1} and \textit{MODS2} instruments experienced intermittent problems during the 2024B cycle.
To see the observational details, see Appendix~\ref{Appendix_A}, Table~\ref{tab:obs_details}.
Since we aim to increase the total incoming flux, given the faintness of the UDG candidates, we chose to observe with a $2.4''$-wide long slit centred on the galaxies and aligned with their semi-major axis. 
All target observations were performed by applying a 10\textsc{"} dithering along the Y-direction of the slit during multiple exposures. 
For the targets observed with two exposures, a single dithering was performed in the positive Y direction of the slit. In contrast, for the targets observed with three exposures, dithering was performed along both the positive and negative Y directions of the slit ($+10$\textsc{"} and $-10$\textsc{"}).
For the spectrographs (\textit{MODS1} and \textit{MODS2}), observations were performed using G400L grating (400 lines~$\mathrm{mm}^{-1}$ blazed at $4.4^\circ$ centred on 4000~\AA\ )
in the blue channel (3200 -- 5800~\AA) and the G670L grating (250 lines~$\mathrm{mm}^{-1}$ blazed at $4.3^\circ$ centred on 7600~\AA)
in the red channel (5800 -- 10000~\AA), with a spectral resolution R$\sim400 -600$ depending on the grating. 

\subsection{Data reduction}
\label{section_data_reduction}
The spectroscopic data obtained from the AIP-2024B-008 run at the LBT resulted in 14 sources. 
We reduced these data using the Spectroscopic Interactive Pipeline and Graphical Interface \citep[\textit{SIPGI},][]{SIPGI}.
The spectra were corrected for bad pixels using flat-field files\footnote{We used the calibration files downloaded from the LBT archive (\url{http://archive.lbto.org/}) acquired the same night as the target observation. 
When unavailable, we revised our policy to use calibration files collected within a maximum of three nights from the observation night.} and for cosmic rays. 
Bias subtraction and flat-fielding were subsequently applied.
Wavelength calibration was performed using a standard \textsc{KrXe} arc lamp, achieving a median (mean) value of accuracy of $0.0555$\AA~($0.0497$\AA) for the blue channel of \textit{MODS1}, $0.0677$\AA~($0.065$\AA) for the red channel of \textit{MODS1}, $0.0533$\AA~($0.0522$\AA) for the blue channel of \textit{MODS2}, and $0.0785$\AA~($0.0878$\AA) for the red channel of \textit{MODS2}.
This accuracy represents the median root mean square distance between the analytical wavelength of the lines found by \textit{SIPGI} and the real wavelength of the lines in the slit used.
All the accuracy values are below $1/5$ of the pixel linear dispersion value ($0.1040$) and are therefore considered reliable.
We then sky-subtracted our targets using an \textit{ABBA} strategy~\citep{SIPGI}, leveraging the dithering applied across the different exposures.
The images obtained were also flux-calibrated during the final steps of the reduction using a sensitivity curve extracted from a standard star.
To create a valid sensitivity curve for our reduction, we selected several exposures ($90$ seconds) of a star observed specifically for this purpose, on the same night as the UDG candidates, in both the red and blue channels of the spectrographs.

{We applied the same reduction steps described above for the science data to the star data, using dedicated calibration files \footnote{To reduce the UDG candidates, we used \texttt{gd$153\_10$a.dat}, a spectro-photometric standard model. This is a file from the \SIPGI~database, which describes the spectrum of standard stars. In particular, \textsc{GD153} is a Hubble Space Telescope primary white dwarf star.}
\textit{SIPGI} determines the 2D spectrum profile and extracts the 1D spectra from all the sources it has automatically identified.
To accurately extract the 1D spectra from the regions containing our targets, we manually defined an extraction box around each source in the 2D spectra.
After reduction, all exposures corresponding to the same target and channel were stacked.
This results in a combination of 2D spectra for each slit and 1D spectra for each object.
In total, 13 spectra are obtained, hereafter referred to as LBT-UDG candidates (see Table~\ref{tab:obs_details}).
We discarded one source from the sample.
After reduction, this galaxy exhibited a spectrum that was too noisy for further analysis.
Cutouts of the sources with their respective reduced 1D spectrum are shown in Fig.~\ref{showing sample}.

\subsection{MISTRAL data and data reduction}
\label{MISTRAL data}
We used the data observed during the OHP run with the MISTRAL~\citep{MISTRAL} spectrograph (project id \textit{24B.PNCG.BOIS}, cycle 2024B, PI: S. Boissier), for which observational time for 20 UDG candidates was requested.
The MISTRAL instrument, mounted on the T193, a $1.93\mathrm{m}$ telescope at OHP, operates with a long slit of width $1.9''$.
We requested two 1800s exposures for all sources, except for the one selected in COSMOS, which was observed with a single 2700s exposure, using the blue configuration of MISTRAL \citep[4200-8000~\AA, spectral resolution R=700]{MISTRAL_reduction}.
Due to meteorological conditions, only 13 sources were observed: 12 from the SMUDGES and one from the COSMOS catalogue.
Data reduction was performed using the built-in pipeline provided by MISTRAL \citep{MISTRAL_reduction}.
For the 13 observed sources, we visually inspected the reduced spectra and discarded those without clearly identifiable emission lines. 
This selection reduced the sample to four sources that exhibit clear H$\alpha$ emission, hereafter classified as OHP-UDG candidates.
Observational details of both samples (LBT and OHP) are listed in Table~\ref{tab:obs_details}, and cutouts and spectra of all the galaxies in the samples are shown in Appendix~\ref{Appendix_B}.
 \subsection{Sample uniformity}
 \label{sample_consistency}
The $\mu_\mathrm{0,g}$ and $r_\mathrm{eff}$ values for the LBT and OHP samples were taken from two different catalogues (\citealt{Thuruthipilly2024A} and \citealt{Zaritsky_2023}, respectively) except for OHP-1, which was selected from the \cite{cosmos2020} catalogue and its $\mu_\mathrm{0,g}$ and $r_\mathrm{eff}$ derived through an independent \textsc{GALFIT} measurement, as mentioned in Sect.~\ref{OHP/MISTRAL sample selection}. 
In both studies, the structural parameters were derived by fitting a single Sérsic profile using \textsc{GALFIT}. 
In the case of \cite{Zaritsky_2023}, the fitting procedure consists of two stages: an initial fit with the Sérsic index fixed to $n=1$ to provide robust initial estimates, followed by a second fit in which $n$ is left free to vary. 
The final catalogue values correspond to this latter fit.
Similarly, \cite{Thuruthipilly2024A} adopted a single Sérsic model with the Sérsic index left free during the final fit.
The main difference between the two analyses resides in the cutouts used for the fitting.
\cite{Thuruthipilly2024A} derived the structural parameters from the \textit{g}-band cutout, whereas \cite{Zaritsky_2023} measured them from stacked \textit{g} and \textit{r} cutouts.
To assess the impact of these different approaches, we ran \textsc{GALFIT} on the OHP galaxies following the same procedure adopted by \cite{Thuruthipilly2024A}, using only the \textit{g}-band images.
We find differences in the effective radius of $1.23''$, $0.11''$, and $0.12''$ and in the central surface brightness of 0.08, 0.16, and 0.44 $\mathrm{mag\,arcsec^{-2}}$ for OHP-2, OHP-3, and OHP-4, respectively.
These comparisons indicate that the use of stacked \textit{g} + \textit{r} images has only a minor impact on the derived values. 
The largest difference is found for OHP-4, whose central surface brightness would fall below the conventional UDG threshold of $\mu_\mathrm{0,g}=24~\mathrm{mag\,arcsec^{-2}}$ when measured from the \textit{g}-band image alone.
However, it is worth mentioning that OHP-4 was already on the border of the UDG definition in the \cite{Zaritsky_2023} catalogue, with a brightness of $24.07$, slightly fainter than the threshold.
Throughout this work, we retained the $\mu_\mathrm{0,g}$ and $r_\mathrm{eff}$ reported in the original catalogues to maintain consistency with the published analyses and other quantities derived from catalogues.

Furthermore, the central surface brightness values used throughout this work are not deprojected to face-on values.
We explored the effect of applying this correction following the approach adopted by \cite{Junais2020}, based on Eq.~2 of \cite{Trujillo2020}\footnote{Following \cite{Junais2020}, we assumed a ratio of scale height to scale length of $z_0/h = 0.12$, as listed in Table~1 of \cite{Trujillo2020}.}. As expected, the de-projected central surface brightnesses become systematically fainter, with an average (median) increase of $0.58\mbox{ }(0.53)~\mathrm{mag\,arcsec^{-2}}$.
Nevertheless, we chose not to adopt these corrected values in the following analysis.
The structural parameters used here were taken directly from the original catalogues and are therefore consistent with the measurements reported in previous studies. 
Moreover, the comparison samples considered throughout this paper likewise do not apply inclination corrections. 
Using de-projected values only for our sample would therefore introduce a systematic offset and impede a fair comparison between datasets.

\subsection{Final sample of this work}
\label{Final sample of this work}
The series of selection criteria described in the previous sections yields a final sample of 17 sources. 
Of these, 13 were observed with the LBT/MODS instrument and exhibit clear H$\alpha$ emission line detections. These are hereafter referred to as LBT-UDG candidates. 
The remaining four sources were observed with the MISTRAL instrument and likewise show H$\alpha$ emission line detections. These are hereafter referred to as OHP-UDG candidates.
The differences in the selection cuts adopted for the two samples, as described throughout Sect.~\ref{data}, are primarily driven by the lack of \textit{i}-band photometry for the OHP-UDG candidates in the SMUDGES catalogue, as well as by the differing spectral resolutions of the two instruments.

\section{Methodology}
\label{section_methodology}
\subsection{Redshift measurement}
\label{Redshift Measurement}
We measured spectroscopic redshifts (\zs) of the observed UDG candidates using \EZ software \citep{Garilli2010_EZ}. 
The \EZ was originally developed for the VVDS survey \citep{LeFevre2005_VVDS} and then used for other large ESO spectroscopic surveys, such as zCosmos \citep{Lilly2007_zcosmos} or VIPERS \citep{Guzzo2014_VIPERS}.
Primarily, the tool was designed to work in fully automatic mode; however, it can also be used manually for a single spectrum.
The software is optimised for the redshift measurements of galaxies (star-forming and quiescent), quasars, and active galactic nuclei (AGNs).
The main functionality of \EZ is to cross-correlate spectral templates with observational spectra and to compare spectroscopic features, such as absorption and emission lines, with the continuum. 
Moreover, in manual mode, the user can define the redshift range, select a set of templates, and indicate the most prominent lines to the software, among other options. 
A detailed description of the code can be found in \cite{Garilli2010_EZ}.\\
The \EZ has already been successfully used for LBT observations \citep[i.e. ][]{Rabitz2017, Polletta2023}, but primarily for high-redshift, star-forming galaxies. 
We used \EZ to select a redshift range from 0 to 0.5 with a step of 0.0001, along with galaxy templates.  
We performed \zs~measurements using both red and blue channels. 
Unfortunately, as reported in \cite{Jenny}, the \textit{MODS} red channel suffers from higher sky backgrounds.
To reduce potential contamination, we chose to use the redshift estimated from the blue channel in our subsequent analysis.
We find that all LBT-UDG candidates are very local ($0.003 \leq$ \zs $\leq 0.065$), with a mean \zs~value of $0.023$ $\pm0.016$.
For the OHP-UDG candidates, we find a similar redshift range to that of the LBT-UDG candidates ($0.005 \leq$ \zs $ \leq 0.076$), characterised by a mean \zs~value of $0.031$ $\pm0.027$.
After determining \zs, for each target we converted angular sizes (arc second) to physical sizes (kiloparsec) using the estimated redshift to confirm their ultra-diffuse nature.
Then, we also corrected for the observed central surface brightness in the g band (\(\mu_\mathrm{0,g}\)) for the cosmological dimming by using an equation:

\begin{equation}
\label{eq:cosmological_dimming}
\mu_\mathrm{0,g}^{\mathrm{corr}} = \mu_\mathrm{0,g} - 10\times \log(1+z).
\end{equation}
Although Eq.~\ref{eq:cosmological_dimming} does not explicitly include a $K$-correction, this effect is expected to be marginal given the low redshifts of our galaxies.
To verify this assumption, we computed the $g$-band $K$-corrections for the LBT sample using the best-fit spectral energy distribution (SED) obtained with the \cigale~tool \citep[see Sect.~\ref{Stellar_mass_estimation}]{CIGALE_Boquien2019}.
We find a mean (median) $K$-correction of 0.067 (0.068) mag, respectively, confirming the minor impact of this correction on the derived surface brightnesses.

After correcting the $\mu_\mathrm{0,g}$ values of the LBT-UDG candidates for cosmological dimming, we find a mean (median) difference with the observed values, equal to $0.098$ ($0.100$). For the OHP-UDG candidates, the mean (median) difference is equal to $0.139$ ($0.079$).
We applied the selection criterion presented by \cite{Vandokkum2015} to identify which of our targets can be spectroscopically confirmed UDGs (see Fig.~\ref{Mu0VSreff}).
This selection criterion is purely observational and does not include the cosmological dimming correction. 
However, two sources (LBT-9 and OHP-4) that previously satisfied the UDG definition became too bright to be classified as such after the correction.
Given the significant impact of the cosmological dimming on the classification of galaxies as UDGs, we reassessed the \cite{Vandokkum2015} criterion using the corrected values. 
In the following part of this paper, $\mu_\mathrm{0,g}$ refers to the values corrected for cosmological dimming.}
Of the 13 LBT-UDG candidates studied, eight (66.7\%) are identified as UDGs, and the rest as LSBGs, which includes an interesting case (LBT-9) of a GLSBG candidate. 
Studying the four galaxies of the OHP-UDG sample, we find a single confirmed UDG (OHP-3) and three LSBGs, while OHP-4, with an effective radius of $8.9$~kpc, is considered another GLSBG candidate.
We discuss the properties of LBT-9 and OHP-4 in Sect.~\ref{Giant-LSBG candidate}.
It should be noted that LBT-5 is on the edge of the UDG definition and could be considered a UDG if the selection criterion for the required size is relaxed, as has been done previously in the literature \cite[i.e.][]{Di_cintio2017, Cardona-Barrero}.
Finally, although the face-on deprojection correction was not applied to the values shown in Fig.~\ref{Mu0VSreff}, the average correction reported in Sect.~\ref{sample_consistency} indicates that all galaxies in our sample, except OHP-1, would satisfy the surface brightness threshold adopted by \cite{Vandokkum2015}.
\begin{figure}[h]
\hspace{2em}
\includegraphics[width=0.40\textwidth, height=0.35\textwidth]{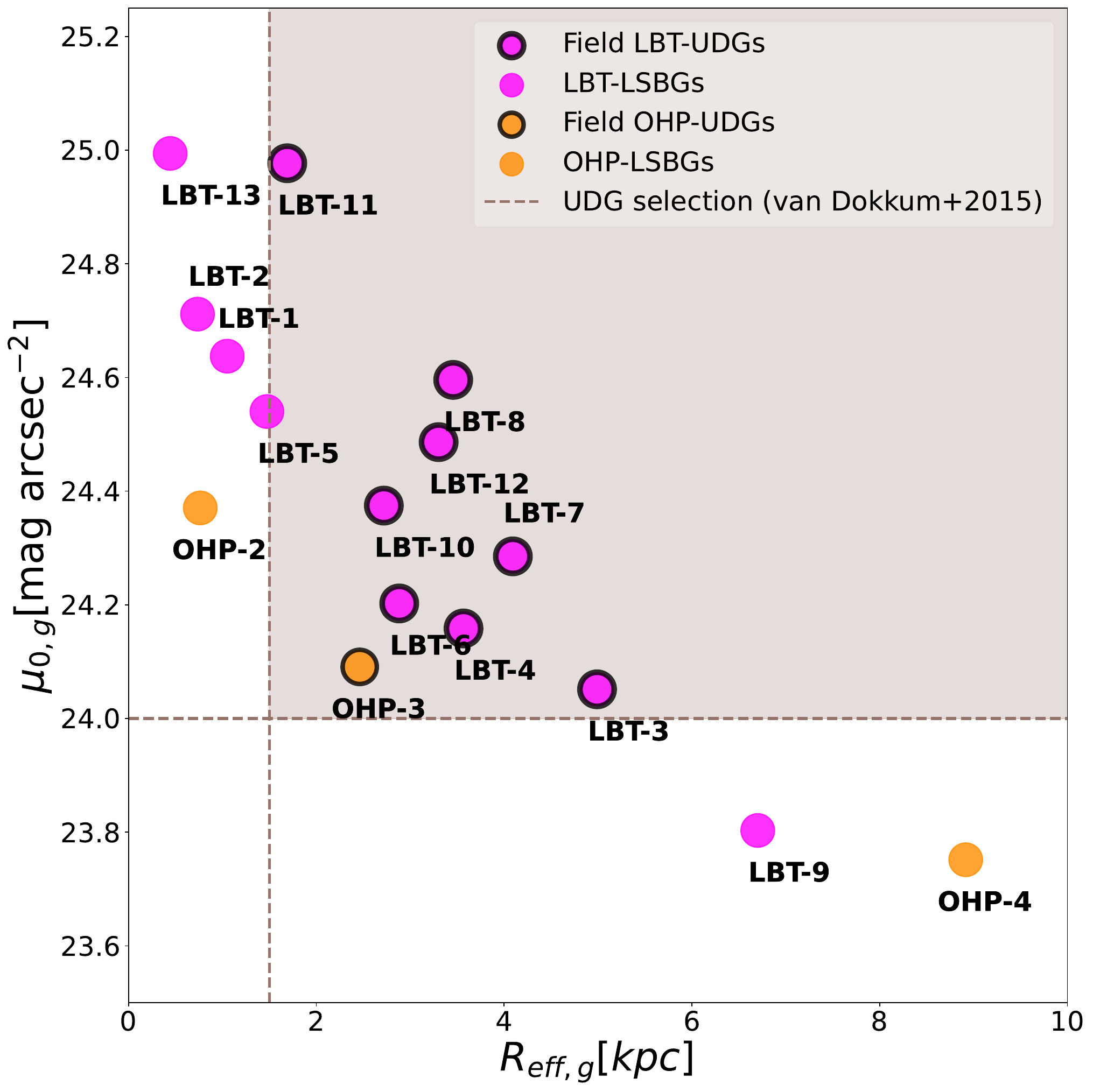}
  \caption{Targets of this study in the \textit{g}-band central surface brightness vs effective radius plot. The magenta dots represent the targets from the LBT observational campaign; and the orange dots represent the targets from the OHP observational campaign. 
  The dashed lines represent the \cite{Vandokkum2015} selection criterion for the UDGs and delimit the opaque area where the UDGs reside. The targets that satisfy the UDG selection criteria after dimming correction are outlined in black. Object names are listed in Table~\ref {tab:Halpha_flxs}. The OHP-1 source was removed from the graph due to its low $\mu_\mathrm{0,g}$, which renders it insignificant, even though it is an LSBG, and to emphasise the fundamental region of the graph.}
     \label{Mu0VSreff}
\end{figure}

\subsection{Effects of local peculiar velocity}
\label{peculiar velocities}
Our sample consists of nearby galaxies for which peculiar motions might affect the inferred distances. Since any error in the distance estimate can propagate into the determination of the effective radius -- which enters the criterion for classifying a galaxy as a UDG -- peculiar velocities can potentially affect the classification of our sample.
At the same time, our specific sample is composed of field UDG candidates that cannot be associated with any groups/clusters with known redshift, as mentioned in Sects.~\ref{LBT/MODS sample selection} and~\ref{OHP/MISTRAL sample selection}, which would enable a direct measurement of their peculiar velocities.
As an alternative, to assess the impact of peculiar velocities on our analysis, we used the reconstructed velocity field created by \cite{Carrick2015}. 
Except for OHP-4, which lies outside the reconstructed volume ($200 h^{-1} \mathrm{Mpc}$), we find 13 galaxies (LBT-1, LBT-3, LBT-4, LBT-5, LBT-6, LBT-7, LBT-8, LBT-9, LBT-10, LBT-11, LBT-12, OHP-1, and OHP-3) with reconstructed peculiar velocities corresponding to only $1-7\%$ of their observed recessional velocities. 
Only three galaxies (LBT-2, LBT-13, and OHP-2) show larger corrections, within $20 -35\%$ of their recessional velocities. However, given that the uncertainties of estimated peculiar velocities for individual objects in low-density environments are large and that the estimated effect in most cases is small, we decided not to apply the corresponding correction in our analysis.
A more detailed discussion can be found in Appendix ~\ref{Appendix_D}.
These results indicate that peculiar velocities may noticeably affect the inferred distances of only a few nearby systems, while for the majority of the sample their impact is expected to be modest.

\subsection{Flux estimation}
\label{flux estimation}
LBT-UDG candidate fluxes were extracted from the 1D spectra produced by the \textit{SIPGI} pipeline, using the corresponding 2D spectra and several built-in analysis tools. 
OHP-UDG candidate fluxes were instead extracted using the MISTRAL built-in pipeline \citep{MISTRAL_reduction}. 
Using the expected wavelength of the lines based on the measured redshift, we fitted and estimated fluxes of the H$\alpha$ and H$\beta$ emission lines in the \textit{Python} environment.
We obtained the rest-frame spectra using the redshift values measured in Sect.~\ref{Redshift Measurement}. 
As shown in Fig.~\ref{showing sample} in Appendix~\ref{Appendix_B}, the extracted spectra show the presence of all standard nebular emission lines (O$\rm[II]_{3727}$, H$\beta$, O$\rm[III]_{5007}$, H$\alpha$). In this work, we focused only on the H$\alpha$ and H$\beta$ measurements, because these are needed to estimate the SFR and the Balmer decrement.
For the LBT-UDG candidates, we re-binned the spectra using the \textsc{FluxConservingResampler} method from the \textsc{Spectrum1D} package~\citep{Earl2024} with a re-binning factor of three.
We performed this re-binning to increase the S/N of the spectra, as preliminary tests showed that, in some cases, the fitting procedure for the H$\beta$ line resulted in non-detections.
Finally, we fitted the processed spectra.
We avoided re-binning the spectra of the OHP-UDG candidates, given the quality of the data.
We performed a fit of the emission lines using Python routines that implement a Markov chain Monte Carlo (MCMC) method on a Gaussian line profile to obtain the peak wavelength, flux, and associated error bars of each emission line \citep[see Appendix~A in][]{Junais_lines}.
We iterated this process to obtain the H$\alpha$ fluxes and errors for all targets.
To ensure an accurate estimate of the integrated flux of H$\alpha$, we performed an additional fit using a three-component Gaussian model. 
In this approach, the H$\alpha$ emission line was simultaneously fitted with two Gaussian components representing the adjacent [NII] lines. 
This method minimises the risk of overestimating the flux of H$\alpha$ due to contamination from the emission of [NII].
Although the [NII] lines are resolved in most of our spectra, the multi-component fitting indicates that their contribution to the H$\alpha$ integrated flux is negligible (less than 1\% in all cases). 
Therefore, we adopted the single-component Gaussian fit for all targets.
The fluxes and errors of H$\beta$ were successfully extracted for ten of the 13 LBT-UDG candidates that exhibit an emission of H$\beta$ strong enough to be measured by our algorithm.
For the OHP-UDG candidates, the H$\beta$ emission lines were too embedded in the noise to be precisely measured.
After estimating the H$\alpha$ and H$\beta$ fluxes, we noticed that the Balmer decrement is clearly affected by underlying stellar absorption; therefore, we adopted the approach of \cite{Junais_lines}, applying standard equivalent width (EW) corrections. Specifically, we used EW (H$\alpha_\mathrm{abs}$) = $1.3$~\AA ~from \cite{Gavazzi2011} and EW (H$\beta_\mathrm{abs}$) = $2.5$~\AA ~from \cite{Moustakas2010}.
For the LBT-UDG candidates, we verified that the red and blue channels used in the spectral reduction were consistently calibrated to an absolute flux scale.
We computed magnitudes by integrating the extracted spectra over the SDSS \textit{g} and \textit{r} band filters, and compared these magnitudes with the corresponding values from the photometric values computed in this work (see Sect.~\ref{forced_photometry}).
In cases of significant discrepancies between the two magnitudes, we rescaled the differing spectral fluxes to align them with the photometric values.
Finally, for the OHP-4 source, we obtained a low S/N, only slightly above the one-sigma level.
Despite the faintness of the measured emission of H$\alpha$, we consider this a tentative detection, motivated by its extended $r_\mathrm{eff}$ and the possibility that it may be confirmed as a GLSBG.
The fluxes with associated errors are listed in Table~\ref{tab:Halpha_flxs}.
Finally, the H$\alpha$ and H$\beta$ fluxes were corrected for the Milky Way foreground Galactic extinction
\citep{Schlegel1998} using the standard \cite{Cardelli89} dust extinction law.
\npdecimalsign{.}
\nprounddigits{6}
\begin{table*}[ht]
\centering
\begin{threeparttable}
\caption{Spectroscopic and physical properties of the sample presented in this work.}
\small
\setlength{\tabcolsep}{3pt}
\begin{tabular}{lcrrrrrrrrrrr}
\hline
ID &\textit{z}& \textit{r$_\mathrm{eff}$} & \textit{$\mu_\mathrm{0,g}$} & A$_\mathrm{{g}}$ & \textit{$\mu^\mathrm{Dust}_{0,g}$} & F$_\mathrm{\text{H}\alpha}$[$\times10^{-16}$]  & F$_\mathrm{\text{H}\beta}$[$\times10^{-17}$] & \lSFR & $\rm\log(SFR_\mathrm{Ap.})$ & A$_\mathrm{H_\alpha}$ & $\rm\log(SFR_\mathrm{Dust})$  & \lMstar \\
\hline
\textit{(1)}&\textit{(2)}&\textit{(3)}&\textit{(4)}&\textit{(5)}&\textit{(6)}&\textit{(7)}&\textit{(8)}&\textit{(9)}&\textit{(10)}&\textit{(11)}&\textit{(12)}&\textit{(13)}\\
\hline
\\
LBT-1     & $0.0076$ & $1.05$ & $24.64$ & & & $2.96$ $\pm0.05$  &   $10.29$ $\pm1.07$    & $-3.71$ $\pm0.01$ &  $-2.71$ $\pm0.01$ & $ $&  & $6.87$ $\pm0.01$\\ 
LBT-2     & $0.0035$ & $0.73$ & $24.71$&  &  & $1.47$ $\pm0.06$  &   $5.51$ $\pm0.92$    & $-4.69$ $\pm0.02$ &  $-3.67$ $\pm0.02$ &  & &$6.58$ $\pm0.01$\\
\textbf{LBT-3}   & $0.0373$ & $4.99$ & $24.05$  & $0.81$& $23.25$& $1.56$ $\pm0.12$  &   $3.37$ $\pm0.59$ & $-2.59$ $\pm0.04$ & $-1.83$ $\pm0.04$ & $1.29$  & $-1.31$ $\pm0.01$& $7.99$ $\pm0.03$\\
\textbf{LBT-4}     & $0.0264$ & $3.57$ & $24.16$  & $0.01$ & $24.16$ & $1.13$ $\pm0.04$  &   $3.89$ $\pm0.34$    & $-3.04$ $\pm0.02$ & $-2.46$ $\pm0.02$ &$0.02$&$-2.45$ $\pm0.01$& $7.90$ $\pm0.02$\\
LBT-5               & $0.0126$ & $1.47$ & $24.54$  &  &  & $1.96$ $\pm0.09$   &   $10.40$ $\pm1.17$    & $-3.45$ $\pm0.02$ & $-2.73$ $\pm0.03$ &   &   & $7.17$ $\pm0.05$\\
\textbf{LBT-6}     & $0.0233$ & $2.88$ & $24.20$  & $0.05$  & $24.15$& $4.76$ $\pm0.07$ & $15.68$ $\pm0.06$ & $-2.52$ $\pm0.01$ & $-2.00$ $\pm0.01$     & $0.09$ & $-1.96$ $\pm0.01$ & $7.83$ $\pm0.02$ \\
\textbf{LBT-7}     & $0.0207$ & $4.09$ & $24.29$  &  &  & $1.18$ $\pm0.11$   &  & $-3.24$ $\pm0.04$ & $-2.18$ $\pm0.04$ &   &  & $8.08$ $\pm0.03$\\
\textbf{LBT-8}     & $0.0294$ & $3.46$ & $24.60$ & $0.06$ & $24.54$ & $2.10$ $\pm0.07$   & $6.83$ $\pm0.88$ & $-2.67$ $\pm0.02$ & $-2.41$ $\pm0.02$ & $0.09$ & $-2.37$ $\pm0.01$ & $7.90$ $\pm0.02$\\
LBT-9     & $0.0651$ & $6.70$ & $23.80$  &  &  & $1.18$ $\pm0.08$   &       $4.77$ $\pm1.02$ & $-2.21$ $\pm0.03$ & $-1.77$ $\pm0.03$ &   &  & $8.52$ $\pm0.05$\\
\textbf{LBT-10}    & $0.0249$ & $2.72$ & $24.37$ &  & & $2.13$ $\pm0.07$   & $9.15$ $\pm1.17$   & $-2.81$ $\pm0.01$ &$	-2.17$ $\pm0.02$ & $ $ &  & $7.51$ $\pm0.04$\\
\textbf{LBT-11}    & $0.0151$ & $1.69$ & $24.98$  &  &   & $0.57$ $\pm0.08$ &     & $-3.83$ $\pm0.06$ & $-2.87$ $\pm0.06$ &  &   & $7.20$ $\pm0.04$\\
\textbf{LBT-12}    & $0.0289$ & $3.30$ & $24.49$  & $0.78$ & $23.71$ & $0.36$ $\pm0.01$  &   $0.76$ $\pm0.08$    & $-3.46$ $\pm0.02$ & $-2.95$ $\pm0.02$ & $1.24$ &$-2.46$ $\pm0.01$&$7.83$ $\pm0.03$\\
LBT-13    & $0.0027$ & $0.44$ & $24.99$  &   &   & $1.84$ $\pm0.09$  &  & $-4.82$ $\pm0.03$ & $-3.78$ $\pm0.02$   &   &   & $5.78$ $\pm0.03$\\
OHP-1    & $0.0251$ & $1.55$ & $21.32$ &   &   &  $14.90$ $\pm1.14$   &          & $-1.96$ $\pm0.03$  & $-1.33$ $\pm0.03$   &   &  &$7.75$ $\pm0.01$ \\
OHP-2   & $0.0045$ & $0.76$ & $24.37$ &   &    & $33.10$ $\pm2.30$  &   & $-3.12$ $\pm0.03$ &$-1.98$ $\pm0.03$ &  &   & $6.85$ $\pm0.01$ \\
\textbf{OHP-3}    & $0.0184$  & $2.46$ & $24.09$ &   &      & $11.29$ $\pm1.15$     &  & $-2.36$ $\pm0.05$        & $-1.60$ $\pm0.05$&  &     & $7.75$ $\pm0.01$ \\
OHP-4   & $0.0761$ & $8.92$ & $23.75$  &   &     & $10.87$ $\pm9.63$ &   & $-1.10$ $\pm0.40$ & $0.15$ $\pm0.38$  &   &   & $9.40$ $\pm0.01$ \\
\\
\hline
\end{tabular}
\begin{tablenotes}
\footnotesize
\item \textbf{Notes.} Galaxies confirmed as UDGs are shown in bold. (1) IDs of the galaxy. (2) Estimated redshifts (see Sect.~\ref{Redshift Measurement} for details). (3) Effective radii in kpc. (4) Central surface brightness in the \textit{g} band after dimming correction in $\mathrm{mag\ arcsec^{-2}}$. (5) Dust attenuation in the g band expressed in unit of mag. The dust attenuation was measured using Balmer ratio (Sect.~\ref{Dust Correction}). (6) Central surface brightness in the g band after dimming and dust attenuation correction in $\mathrm{mag\ arcsec^{-2}}$. (7) and (8) Observed H$\alpha$ and H$\beta$ fluxes and their associated errors, uncorrected for Milky Way Galactic extinction, in $\mathrm{erg}\,\mathrm{cm}^{-2}\,\mathrm{s}^{-1}$. (9) Log of star formation and star formation errors expressed in $\mathrm{M}_\odot\,\mathrm{yr}^{-1}$. (10) Log of star formation and star formation errors after aperture correction expressed in $\mathrm{M}_\odot\,\mathrm{yr}^{-1}$ (Sect.~\ref{Aperture correction for SFRs}).
(11) Dust attenuation for H$\alpha$ expressed in units of mag.
(12) Log of star formation and star formation errors after dust attenuation correction expressed in $\mathrm{M}_\odot\,\mathrm{yr}^{-1}$ (Sect.~\ref{Dust}).
(13) Log of stellar masses and errors in $\mathrm{M}_\odot$ (Sect.~\ref{Stellar_mass_estimation} ).
\end{tablenotes}
\label{tab:Halpha_flxs}
\end{threeparttable}
\end{table*}
\npnoround
\subsection{H$\alpha$ line luminosity and star formation rate}
\label{H_alpha line luminosity}
After obtaining the H$\alpha$ emission line fluxes, we computed the line luminosity ($L_{\mathrm{H}\alpha}$) in $\mathrm{erg}\,\mathrm{s}^{-1}$ using
\begin{equation}
L_{\mathrm{H}\alpha} = 4\pi d_L^2 F_{\mathrm{H}\alpha},
\label{eq: line luminosity}
\end{equation}
where $F_{\mathrm{H}\alpha}$ are the respective flux of H$\alpha$ in $\mathrm{erg}\,\mathrm{cm}^{-2}\,\mathrm{s}^{-1}$ and $d_L^2$ is the luminosity distance in centimetres  (used to match the H$\alpha$ flux unit measure). 
The luminosity distance was computed starting from the spectroscopic redshift of each target measured with the \EZ~tool, using the \textsc{\(luminosity\_distance\)} method of the 
\texttt{\(astropy.cosmology.FlatLambdaCDM\)} python package~\citep{Astropy}, assuming the same cosmology as stated in Sect.~\ref{introduction}.
Using H$\alpha$ line luminosities, we estimated the SFRs using the relation from \cite{Samuel}:
\begin{equation}
\mathrm{SFR({M}_\odot\,\mathrm{yr}^{-1})} = 5.1 \times 10^{-42} \times L_{\mathrm{H}\alpha}({erg~s^{-1}}).
\label{eq: SFR Boissier}
\end{equation}
This relation assumes the \cite{Kroupa} initial mass function (IMF).
The SFR values obtained are listed in  Table~\ref{tab:Halpha_flxs} (\lSFR, column~9).
\subsection{Photometry}
\label{forced_photometry}

\begin{figure*}[t]
\centering

\begin{subfigure}{0.17\textwidth}
    \includegraphics[width=\linewidth]{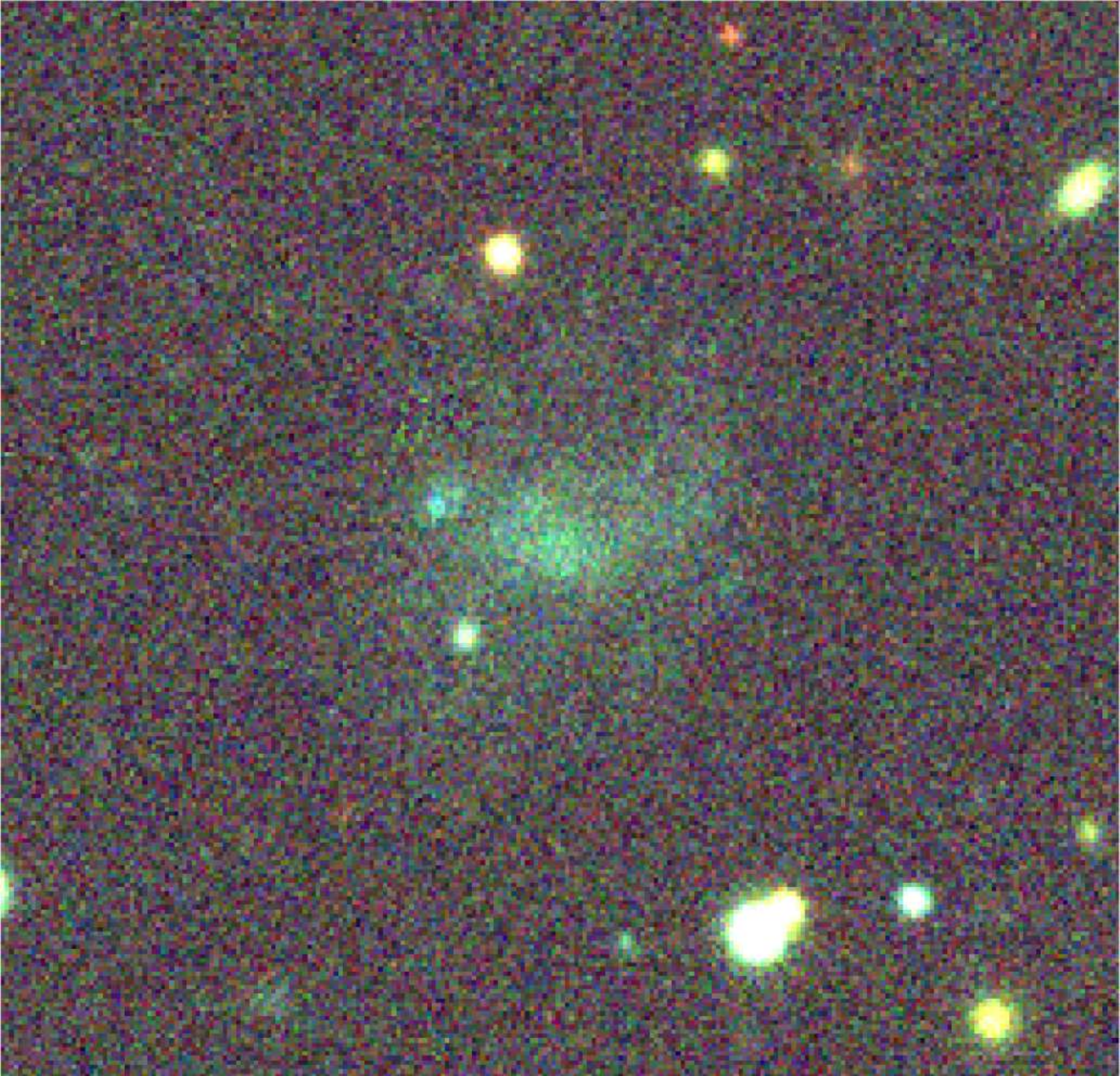}
    \subcaption{}
\end{subfigure}
\hspace*{0.01\textwidth}
\raisebox{-2.9em}{
\begin{subfigure}{0.12\textwidth}
  \includegraphics[width=\linewidth]{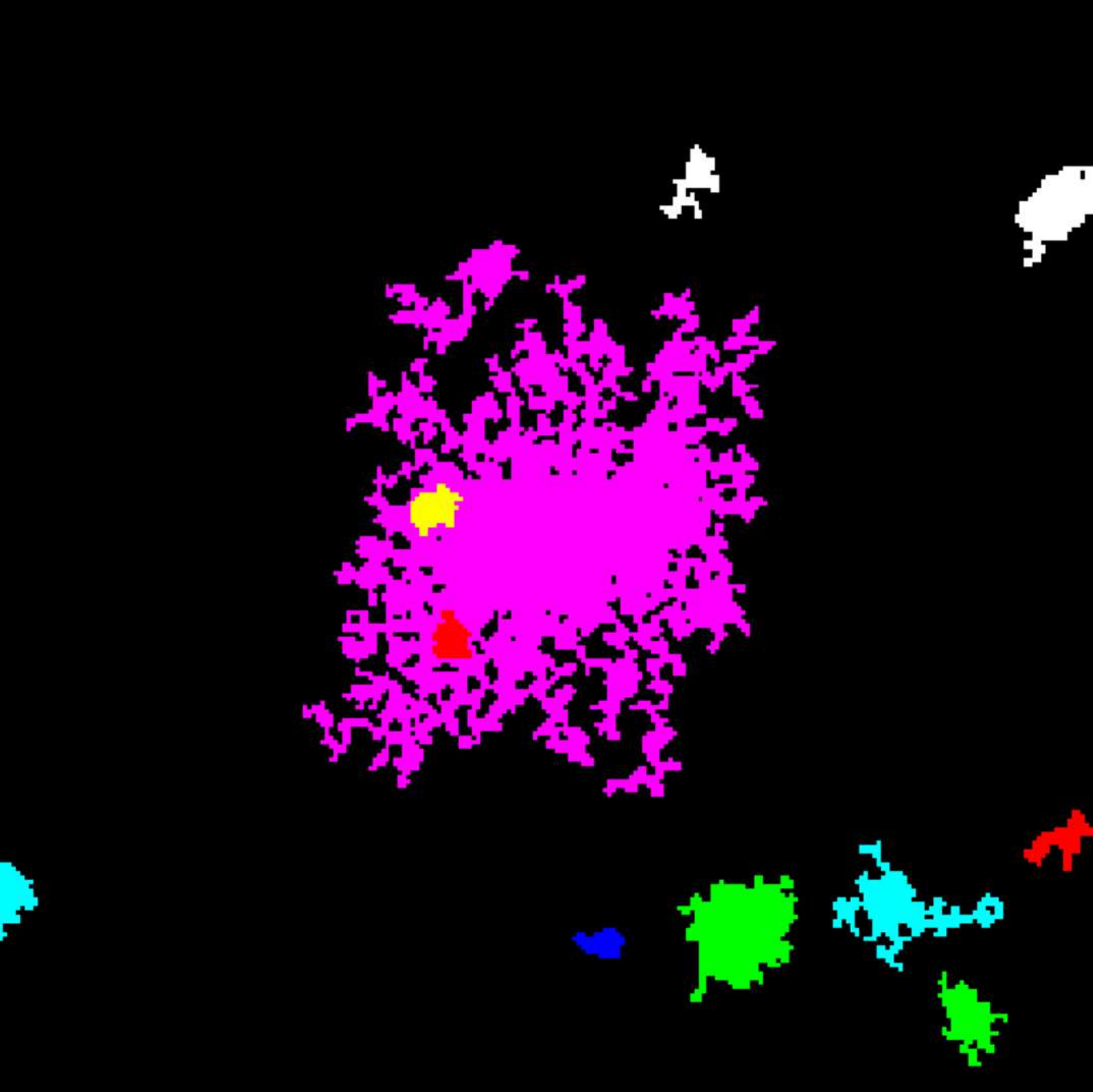}
  \includegraphics[width=\linewidth]{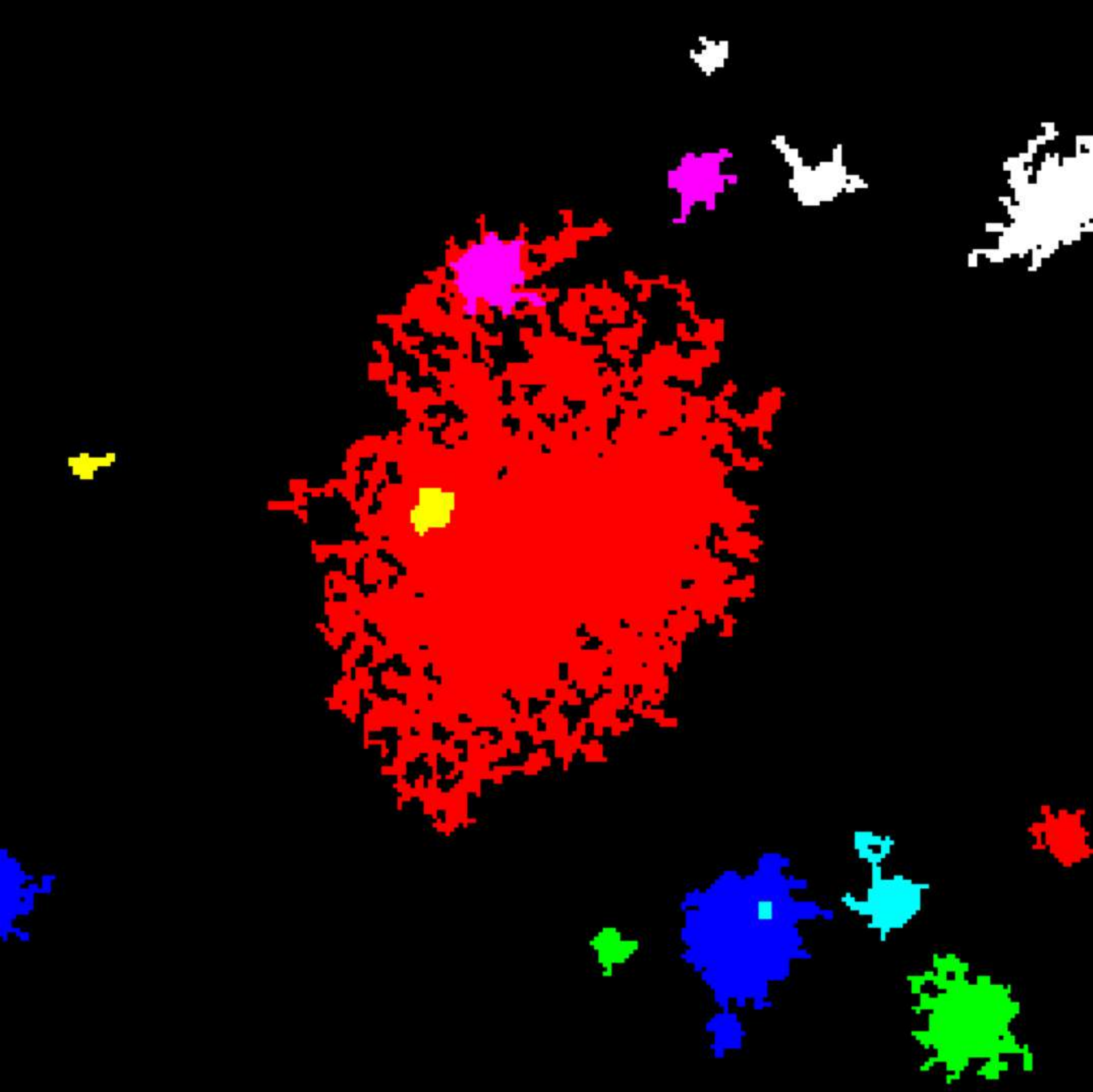}
  \subcaption{}
\end{subfigure}
}
\raisebox{-2.9em}{
\begin{subfigure}{0.12\textwidth}
  \includegraphics[width=\linewidth]{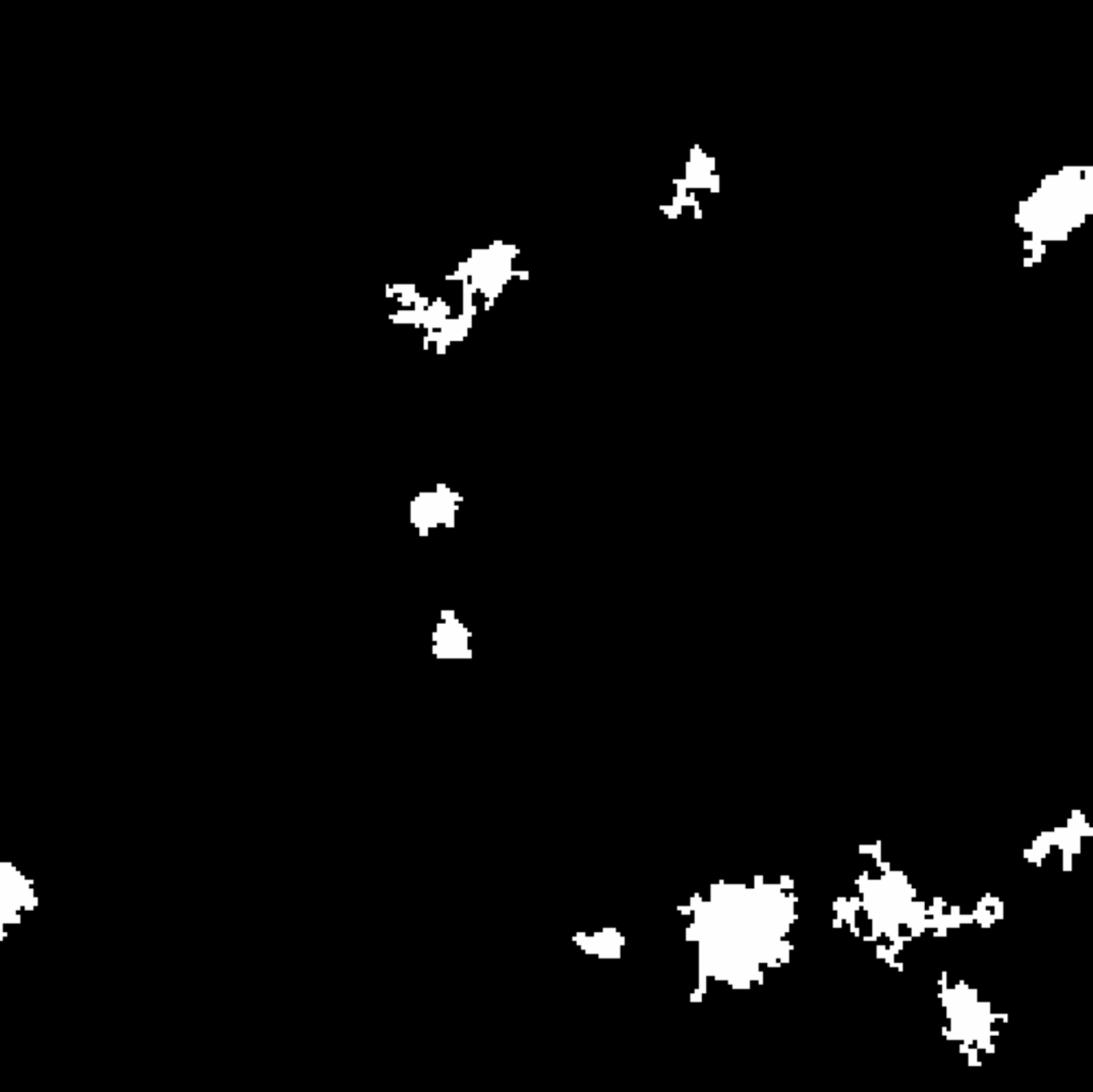}
  \includegraphics[width=\linewidth]{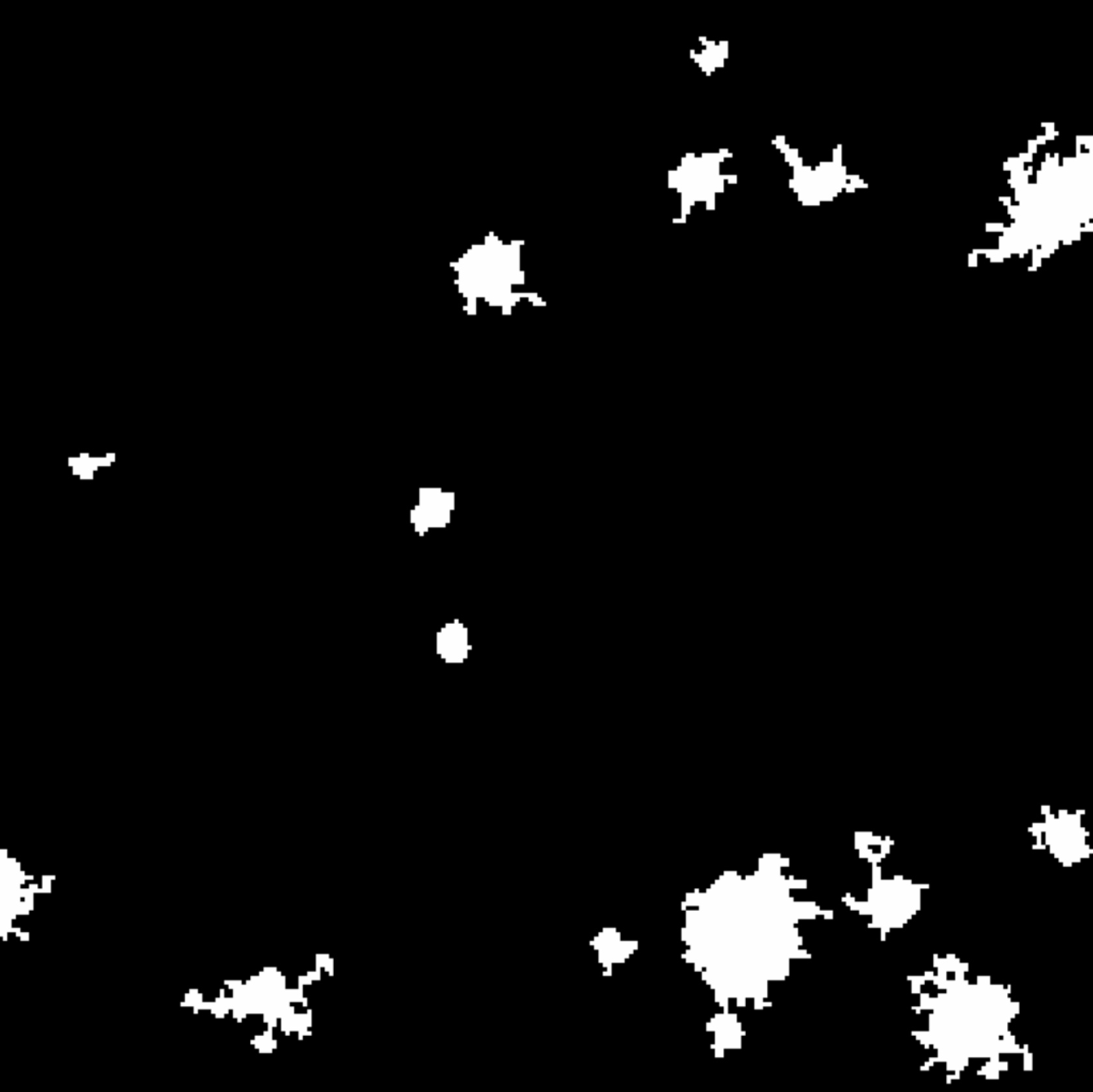}
  \subcaption{}
\end{subfigure}
}
\hspace*{0.01\textwidth}
\begin{subfigure}{0.17\textwidth}
    \includegraphics[width=\linewidth]{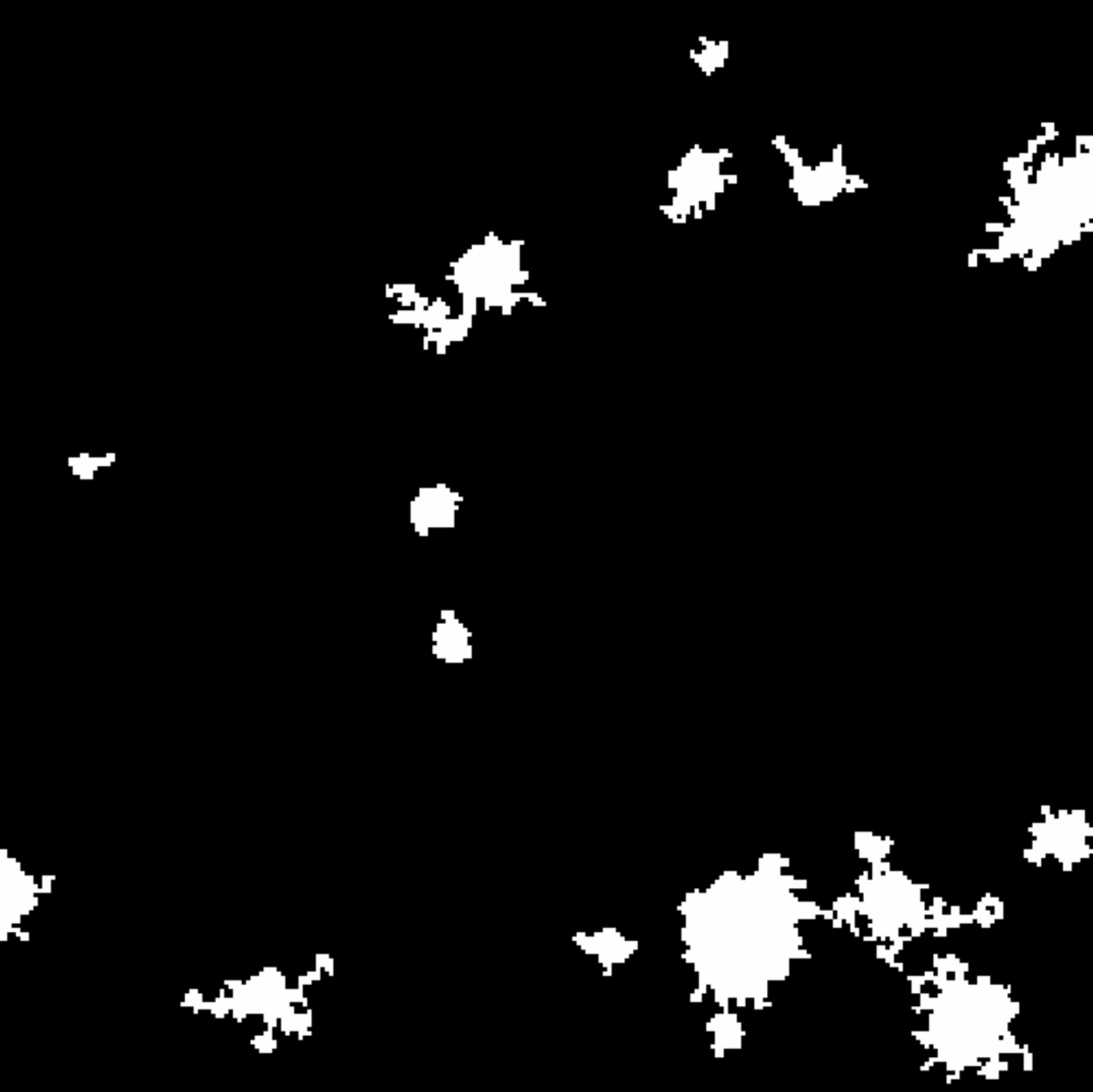}
    \subcaption{}
\end{subfigure}
\hspace*{0.01\textwidth}
\begin{subfigure}{0.1774\textwidth}
    \includegraphics[width=\linewidth]{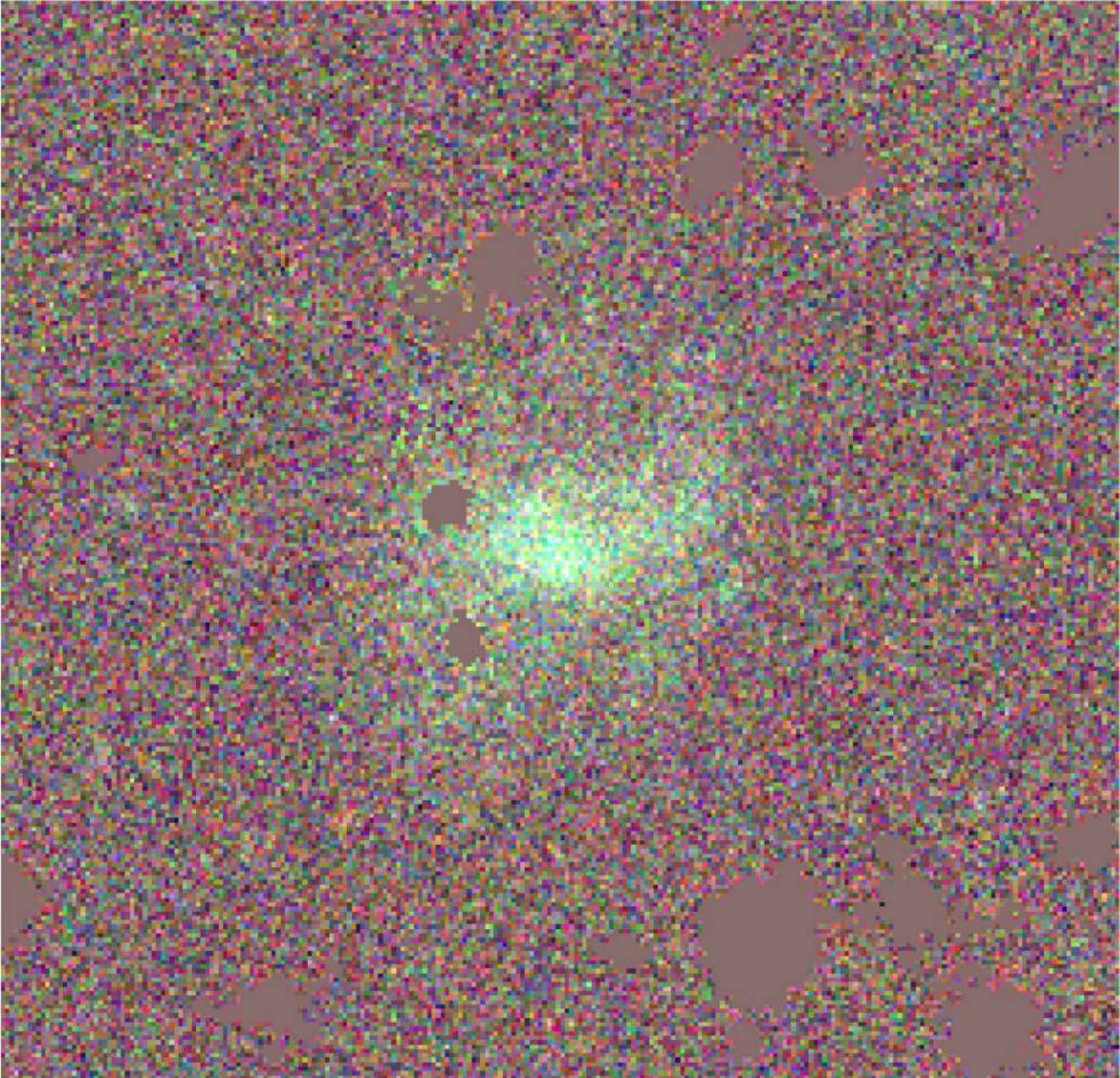}
    \subcaption{}
\end{subfigure}

\caption{Images showing the results of the different photometry steps applied for the \textbf{LBT-7} source. (a) \textbf{LBT-7} RGB $60''$ cutout made using the \textit{g}, \textit{r}, and \textit{z} bands. (b, c) Segmentation maps and corresponding masks for the \textit{r} (top) and \textit{g} (bottom) bands. Each colour identifies a unique object in the cutout. (d) \textit{g}- and \textit{r}-band combined masks, along with the final mask. (e) RGB cutout after removal of other objects.}
\label{fig:segmentation}

\end{figure*}

Originally, the fluxes in the \textit{g} and \textit{r} bands for the LBT-UDG candidates were automatically computed by \cite{Thuruthipilly2024A} using the \textsc{GALFIT} tool. 
For the OHP-UDG candidates, instead, fluxes were taken directly from the SMUDGES catalogue~\citep{Zaritsky_2019}.
To avoid systematic differences and obtain uniform photometry for both samples in this work, we performed forced photometry for all candidates.
The photometric fluxes for all the UDG candidates are computed following the same approach. 
A cutout centred on the target was taken from DES DR1 (for the LBT-UDG candidates) in the \textit{g}, \textit{r}, \textit{i}, \textit{z}, and \textit{Y} bands and from DEcalS (for the OHP candidates) in the \textit{g}, \textit{r}, and \textit{z} bands. 
The cutout was selected with a radius of $33''$, which corresponded, after conversion to the common pixel scale of 0.262, to an image of 252 by 252 pixels.
Using the \textit{MTObjects} package \citep{MTObjects} with a moving factor equal to $0.3$, the cutouts were segmented into several objects and subsequently masked, after excluding the central UDG candidate from the mask.
Due to the faintness of our sources, using a single band for object masking in the cutout can result in incorrect masking. 
To avoid potential errors, we combined the \textit{g}- and \textit{r}-band masks by summing their pixel values to create the final mask for each galaxy.
Through \textsc{photutils}~\citep{larry_bradley_2025_14889440}, we extracted fluxes from the masked cutout in each band, applying an elliptical aperture using the \textsc{\textit{SkyEllipticalAperture}} function.
We used the position angle (PA), the angle used during the observation to place the slit on the target, and further assumed a semi-major axis equal to twice the half-light radius of the UDG candidate.
The PA of each source was determined based on \textsc{GALFIT} measurements derived from the analysis presented in \cite{Thuruthipilly2024A}, although the PAs themselves are not explicitly reported in that work.
For the LBT-2, LBT-4, and LBT-12 targets, these values were subsequently adjusted to less than $10^\circ$ at the time of observation to ensure that the long slit also intersected a suitable guide star (see Table~\ref{tab:obs_details}).
All the steps of the described procedure are shown in Fig.~\ref{fig:segmentation}. 
We manually inspected the obtained mask to ensure the correctness of the process.
In more detail, when we found poorly masked sources, we edited the masks by comparing the respective segmentation maps with the final masks.
As the final step, we corrected the measured fluxes for Galactic extinction by applying a foreground Galactic extinction correction based on the dust maps of~\cite{Schlegel1998} and adopting the \cite{Cardelli89} Milky Way dust extinction law.
\subsection{Stellar mass estimation}
\label{Stellar_mass_estimation}
We derived the stellar masses of our UDG candidates using the mass-to-light ratio–colour relation (MLCR) found by \cite{Du_2020}.
In their work, \cite{Du_2020} studied UV and optical photometry for~1\,100 LSBGs using SED fitting. 
From the SED fitting, they estimated stellar masses and mass-to-light ratios in each band for the entire LSBG sample.
From the measured colour and the estimated physical properties, they extracted the MLCR relation.
\cite{Du_2020} concluded that LSBGs follow the same relation as normal galaxies.
Here, we used the equation in \cite{Du_2020} (Eq.~\ref{eq: m_l_r}) to obtain the mass-to-light ratio in the \textit{r} band for our UDG candidates.
We used the $g-r$ colour of each galaxy and the \textit{r}-band coefficients derived from the bi-weight fit presented in \cite{Du_2020}, adopting $a_r = -0.7$ and $b_r = 1.252$ (see their Table~3).
\begin{equation}
\log(\gamma_\mathrm{r}) = a_\mathrm{r} + \mathrm{b_r} \times \text{colour}.
\label{eq: m_l_r}
\end{equation}
Here, $\gamma_\mathrm{r}$ is the stellar mass-to-light ratio in the \textit{r} band.
We then computed the luminosity in the \textit{r} band:
\begin{equation}
\label{Lum_r-b}
\log(L_\mathrm{r}) = -0.4\, (M_\mathrm{r} - M_\mathrm{r,\odot}).    
\end{equation}
where $M_\mathrm{r}$ is the absolute magnitude of the galaxy in the \textit{r} band and $M_\mathrm{r,\odot}$ is the absolute magnitude of the Sun in the \textit{r} band for the DES instrument. It is equal to $4.42$ mag (AB system).
Once these quantities were obtained, we computed the stellar masses for our targets.
As already stated in Sect.~\ref{H_alpha line luminosity},  we assumed a \cite{Kroupa} IMF, while in \cite{Du_2020}~a \cite{Chabrier2003} IMF was assumed.
To ensure consistency between our results and theirs, we applied an appropriate conversion factor.
To compare our results with colour relations for standard galaxies, we also computed the UDG candidate stellar masses using the relation by \cite{Ebrov_2025} and \cite{Taylor2011}, in a similar way we did for \cite{Du_2020} values, finding a negligible average difference with $\rm\log{(M_\mathrm{\ast \ Ebrova})}$ ($-0.10\pm0.07$ for LBT and $-0.19\pm0.07$ for OHP) and with $\rm\log{(M_\mathrm{\ast \ Taylor})}$ ($-0.07\pm0.07$ for LBT). 
For the OHP sources, we were unable to estimate stellar masses using the \cite{Taylor2011} relation because \textit{i}-band photometry is unavailable for these objects.
Furthermore, we performed an SED fitting with the Code Investigating GALaxy Emission \citep[\cigale,][]{CIGALE_Burgarella2005, CIGALE_Noll2009, CIGALE_Boquien2019} using photometric fluxes for the DES DR1 bands of the LBT-UDG candidates.
We find that \cigale performs a satisfactory fit to the photometric fluxes and yields consistent stellar mass estimates (average difference between $\rm\log{(M_\mathrm{\ast \ Du})}$ and $\rm\log{(M_\mathrm{\ast \ CIGALE})}$ equal to $0.15\pm0.08$).
A detailed comparison between the SED fitting estimates (\Mstar, SFR) is given in Appendix~\ref {Appendix_C}.
For the final analysis, we used the relation from \cite[see Eq.~\ref{eq: m_l_r}]{Du_2020}, as it is explicitly derived from the LSBGs and can be homogeneously calculated for the entire sample.

\subsection{Aperture correction}
\label{Aperture correction for SFRs}
In Sect.~\ref{H_alpha line luminosity} we obtained SFR measurements from the H$\alpha$ line luminosity relation; 
Although reliable and widely used in the literature, these derived fluxes do not account for the galaxy's specific morphology or the percentage of the surface area covered by the long slit used for spectroscopic observations.
Given UDGs' intrinsic faintness, during observations, slits were positioned with a PA that maximised the number of observed star-forming clumps at the centre of the galaxy and the incoming flux from the source.
This was done to ensure an S/N that is high enough to enable redshift measurements from the emission lines.
To improve SFR measurements, accounting for the entire galaxy and its less bright regions, we applied an aperture correction to our fluxes.
We recomputed the photometry for each target.
This time, we used the \textsc{\textit{SkyRectangularAperture}}, another \textsc{photutils} function, to measure the incoming flux of the sources, mimicking the slit area used for the spectroscopic observations.
We applied a rectangular aperture centred on each UDG candidate. 
Its height was set equal to the size of the custom boxes ($3.72-17.4$ arcsecs) used during data reduction to extract the 1D spectra (see Sect.~\ref{section_data_reduction}), while its width matched the slit width of each spectrograph ($2.4''$ for MODS and $1.9''$ for MISTRAL).
We performed photometry correction only for the \textit{g}-band UDG candidate cutouts, which are the most sensitive among all available bands to recent star formation events.
Then, for each UDG candidate, we divided the flux calculated with the elliptical aperture for the \textit{g} band (discussed in Sect.~\ref{forced_photometry}) by the flux calculated with the rectangular aperture, obtaining a correction factor, $C_\mathrm{ap}$.
The $C_\mathrm{ap}$ factors were then simply multiplied by the H$\alpha$ emission line fluxes, as shown in Eq.~\ref{correction_factor_applied}:

\begin{equation}
\label{correction_factor_applied}
F^{\mathrm{corrected}}_{\text{H}\alpha}
= F_{\text{H}\alpha} \times C_{\mathrm{ap}}, \qquad \text{where} \quad
 C_{\mathrm{ap}} = \frac{F^{\mathrm{Ellip}}_{g}}{F^{\mathrm{Rect}}_{g}}.
\end{equation}

$F^\mathrm{Ellip}_g$ and $F^\mathrm{Rect}_g$ stand for the photometric fluxes computed with elliptical and rectangular apertures, respectively.
In the final step, updated fluxes were used to compute corrected H$\alpha$ line luminosities and SFRs from Eq.~\ref{eq: SFR Boissier}.
Corrected fluxes and corrected SFRs (SFR$_\mathrm{Ap}$) are reported in Table~\ref{tab:Halpha_flxs}.
Finally, we stress that the aperture correction, implemented by comparing the flux extracted from the central region of the galaxy with that from a larger elliptical aperture, assumes a uniform luminosity distribution across the galaxy and should therefore be regarded, along with the SFR$_\mathrm{Ap}$, as an upper limit.

\subsection{Intrinsic dust correction}
\label{Dust Correction}
As mentioned in Sect.~\ref{flux estimation}, we measured H$\beta$ integrated fluxes for LBT-UDG candidates. 
Reliable H$\beta$ measurements were not possible for the OHP-UDG sample. 
The H$\alpha$ and H$\beta$ integrated fluxes were corrected for the Milky Way foreground Galactic extinction \citep{Schlegel1998} using the standard \cite{Cardelli89} dust extinction law.
After measuring the S/N for H$\beta$, we find that for only ten LBT-UDG candidates the S/N is greater than three (see Table~\ref{tab:Halpha_flxs}, column 8).
Furthermore, we find that the Balmer decrement was equal to or greater than the theoretical value for case B recombination \citep[$\text{H}\alpha/\text{H}\beta = 2.86$,][]{Osterbrock}, in only five sources.
Following the same approach as \cite{Junais}, we computed the intrinsic dust attenuation values in these UDG candidates using the equation derived by \cite{Yuan_2018}:

\begin{equation}
\label{att_value}
A_\lambda = -2.5\, \frac{k(\lambda)}{k(\mathrm{H}\alpha) - k(\mathrm{H}\beta)}
\log\left( \frac{(\mathrm{H}\alpha/\mathrm{H}\beta)_{\mathrm{obs}}}{2.86} \right).
\end{equation}
where (H$\alpha/\text{H}\beta)_\mathrm{obs}$ is the observed Balmer decrement, $k_\lambda$ is the value of the attenuation curve at a wavelength $\lambda$.
Equation~\ref{att_value} allows one to calculate the attenuation value, $A_\lambda$, at a given wavelength, $\lambda$, by comparing the theoretical value of the Balmer ratio (H$\alpha$ $/$H$\beta$ =$2.86$) with the observed one.
Assuming \cite{Calzetti2000} attenuation law with $R_\mathrm{V} = 4.05$, we computed the $k(\text{H}\alpha)$ and $k(\text{H}\beta)$ values, respectively equal to $3.33$ and $4.60$.
For all sources with measured H$\mathrm{\alpha}$ and H$\beta$ fluxes, we computed the attenuation at the H$\mathrm{\alpha} $ wavelength (hereafter, $A_\mathrm{H\alpha}$).
The $A_\mathrm{H_\alpha}$ values, which range between $0.02$ and $1.29$ mag, are reported in Table~\ref{tab:Halpha_flxs}.
In the next step, we corrected for previously measured fluxes of H$\alpha$ by $A_\mathrm{H_\alpha}$ using Eq.~\ref{att_coor}. 
We used corrected fluxes,  $F^\mathrm{dust_{corr}}_{H_\alpha}$,  to recompute the H$\alpha$ luminosities, and to obtain the attenuation-corrected SFRs (see Sect.~\ref{H_alpha line luminosity}): 

\begin{equation}
\label{att_coor}
F^\mathrm{dust_{corr}}_\mathrm{H_\alpha}=F_\mathrm{H_\alpha}\times10^{(0.4*\mathrm{A_{\text{H}\alpha}})}.
\end{equation}

We also corrected for the $\mu_\mathrm{0,g}$ values using the measured dust attenuation.
We used, once again, Eq.~\ref{att_value} to obtain $A_\lambda$ for the \textit{g} band.
In this case, we computed $k(g)=4.73$ by adopting the central wavelength of the \textit{g}-band filter (4720~\AA) from DES DR1 and applying a correction factor of $0.44$ to convert the line-flux attenuation to the continuum attenuation in the \textit{g} band.
This band was used to obtain the cutout from which the original $\mu_\mathrm{0,g}$ values were computed by \cite{Thuruthipilly2024A} through \textsc{GALFIT} software.
We corrected each LBT-UDG central surface brightness in the \textit{g} band, having a S/N satisfactory for the dust attenuation value and discovered that two (LBT-3, LBT-12) of the five confirmed UDGs that previously fulfilled the \cite{Vandokkum2015} criterion are now too bright to fall under the threshold of $24\ \mathrm{mag\ arcsec^{-2}}$. 
Unfortunately, the spectral resolution of the MISTRAL instrument does not allow us to resolve H$\beta$, which was found embedded in the noise of the OHP-UDG spectra.
It is worth mentioning that the dust attenuation correction performed does not account for the $[NII]$ contamination in the H$\alpha$ flux, since we find a negligible total change in the final flux.
From here on, all the $\mu_{0,g}$ values used in this work are corrected for cosmological dimming.

\section{Physical properties}
\label{4}
As summarised in Table~\ref{tab:Halpha_flxs}, for the LBT-UDG candidates we find an average $\mu_\mathrm{0,g}$ of $24.45$ $\mathrm{mag\ arcsec^{-2}}$ and an average effective radius of $2.85$ kpc. 
Among the 13 targets studied, we conclude that eight are confirmed UDGs that meet the definition of \cite{Vandokkum2015} after applying the dimming correction.
For the OHP-UDG candidates, we instead observe an average $\mu_\mathrm{0,g}$ of $23.38$ $\mathrm{mag\ arcsec^{-2}}$ and an effective radius of $3.42$ kpc.
Among the four targets studied, we confirm only one UDG.
We have two interesting cases of GLSB candidates, LBT-9 and OHP-4.
Their effective radii are, respectively, $6.70$ and $8.92$ kpc. A detailed discussion is presented in Sect.~\ref{Giant-LSBG candidate}.
\subsection{Star-forming main sequence}
\label{SFMS}
\begin{figure*}[ht]
\centering
\begin{minipage}{0.9\textwidth}
\centering
\includegraphics[width=\linewidth]{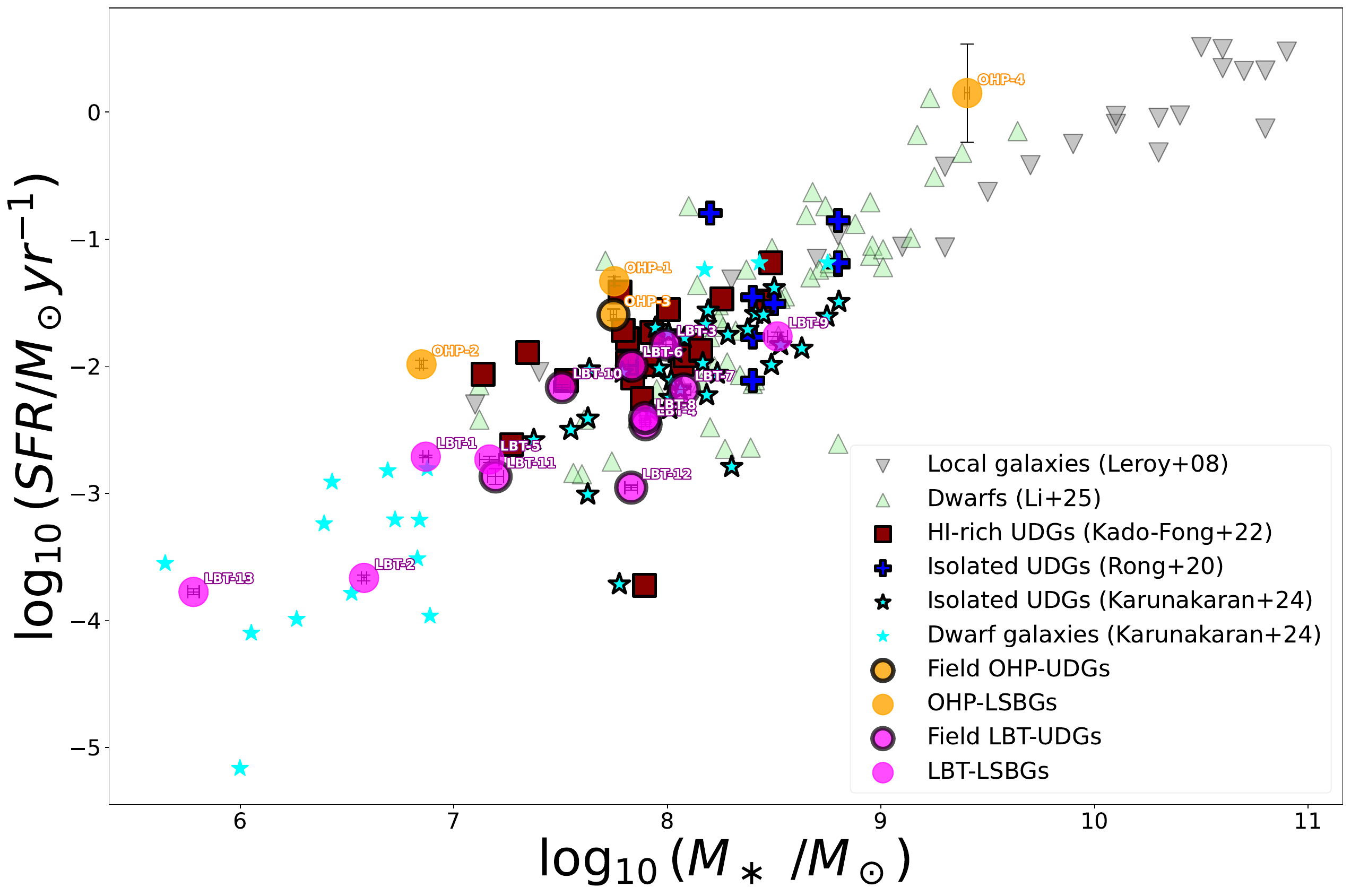}
\end{minipage}
\caption{SFMS of the local and normal star-forming galaxies. 
The grey triangles represent the local galaxy sample from \cite{Leroy}; the light-green triangles denote the dwarf galaxy sample from \cite{Li_2025}; and the cyan stars represent dwarf galaxies from \cite{Karunakaran}.
The LBT and OHP samples of this work are shown as magenta and orange dots, respectively. 
Spectroscopically confirmed UDGs in the LBT and OHP samples are represented by the outlined magenta and orange dots. 
For comparison, other isolated UDGs are shown: 
the outlined red squares are $HI$-rich UDGs from \cite{Kado-Fong};
the outlined blue crosses are isolated UDGs from \cite{Rong_2020}; and
the outlined cyan stars are isolated UDGs from \cite{Karunakaran}.
LSBG numbering follows that in Table~\ref{tab:obs_details}.
}
\label{MS}
\end{figure*}
Using the SFRs computed in Sect.~\ref{H_alpha line luminosity} and the \Mstar~computed in Sect.~\ref{Stellar_mass_estimation}, we analysed the stellar-mass-to-star-formation relation, the so-called SFMS, presented in Fig.~\ref{MS}, where we plot all objects from our sample against a background of local and normal star-forming galaxies from \cite{Leroy}.
All field LSBGs from our analysis, including UDGs, follow the normal local galaxy distribution.
Our results also align with previous findings in the literature.
\cite{Kado-Fong} studied UDGs rich in $HI$, where the same tendency is observed.
In addition, we compared our results with \cite{Karunakaran}. 
In their work, \cite{Karunakaran} studied dwarf galaxies and UDGs in isolated environments. 
We also compared our findings with eight field UDGs from \cite{Rong_2020}.
A similar correction to the aperture correction we applied to all the UDG candidates (Sect.~\ref{Aperture correction for SFRs}) was used for the SFRs of the \cite{Rong_2020} UDGs.\\
Figure~\ref{MS} clearly shows that the field UDGs tend to follow the relation of local and standard galaxies.
They also populate a limited stellar mass range, $7\leq \rm\log\left({M_*}/{M_\odot}\right) \leq9$.
All the other LSBGs, which are not found to be UDGs, populate either the bottom-left corner of the plot or are distributed in a high stellar masses-SFRs space (in the top-right of the plot). 
Our results support the general idea in the literature \citep[e.g.][]{Lim_2020} that UDGs belong to the same distribution as dwarf LSBGs.\\ 
Comparing the two samples used in this work, we can clearly notice that the OHP-UDG candidates are, in general, more efficient in star-forming processes with respect to the LBT-UDGs.
The measurements and conversions we used were the same for all candidates, so the main reason for this behaviour lies in the colour cut of our candidates. 
OHP-UDG candidates were selected to be less than 19 mag in the \textit{r} band due to the instrumental limitations of the MISTRAL spectrograph.
Analysing the \textit{r}-band distribution of the sample, we find that the LBT-UDG candidates are fainter than the OHP-UDG sample on average by $1.52$ mag.
The H$\alpha$ line luminosity-to-SFR conversion we used is based on H$\alpha$ fluxes, which are a good tracer of recent star formation.
The \textit{r} band is usually less sensitive to these properties than other bands, which is why we also compare the distribution in the \textit{g} band, the most sensitive band, to the recent star formation process among the available bands. 
The \textit{g}-band distribution of the two samples shows the same offset as the \textit{r}-band distribution (LBT-UDG candidates are fainter on average by $1.62$), proving that the higher SFR values result from the initial selection bias.
These results indicate that the different behaviour observed in the MS relation for the two samples is primarily a consequence of selection effects. 
The OHP-UDG candidates are not only those with clear detections in the MISTRAL observations, but also correspond to the bluest sources in the parent sample.
The smaller aperture of the telescope, compared to the LBT, effectively biases the sample towards galaxies with higher star formation activity, likely introducing a threshold in colour for sources detectable with a 2-metre-class instrument.
\begin{figure}[h]
\includegraphics[width=0.4\textwidth]{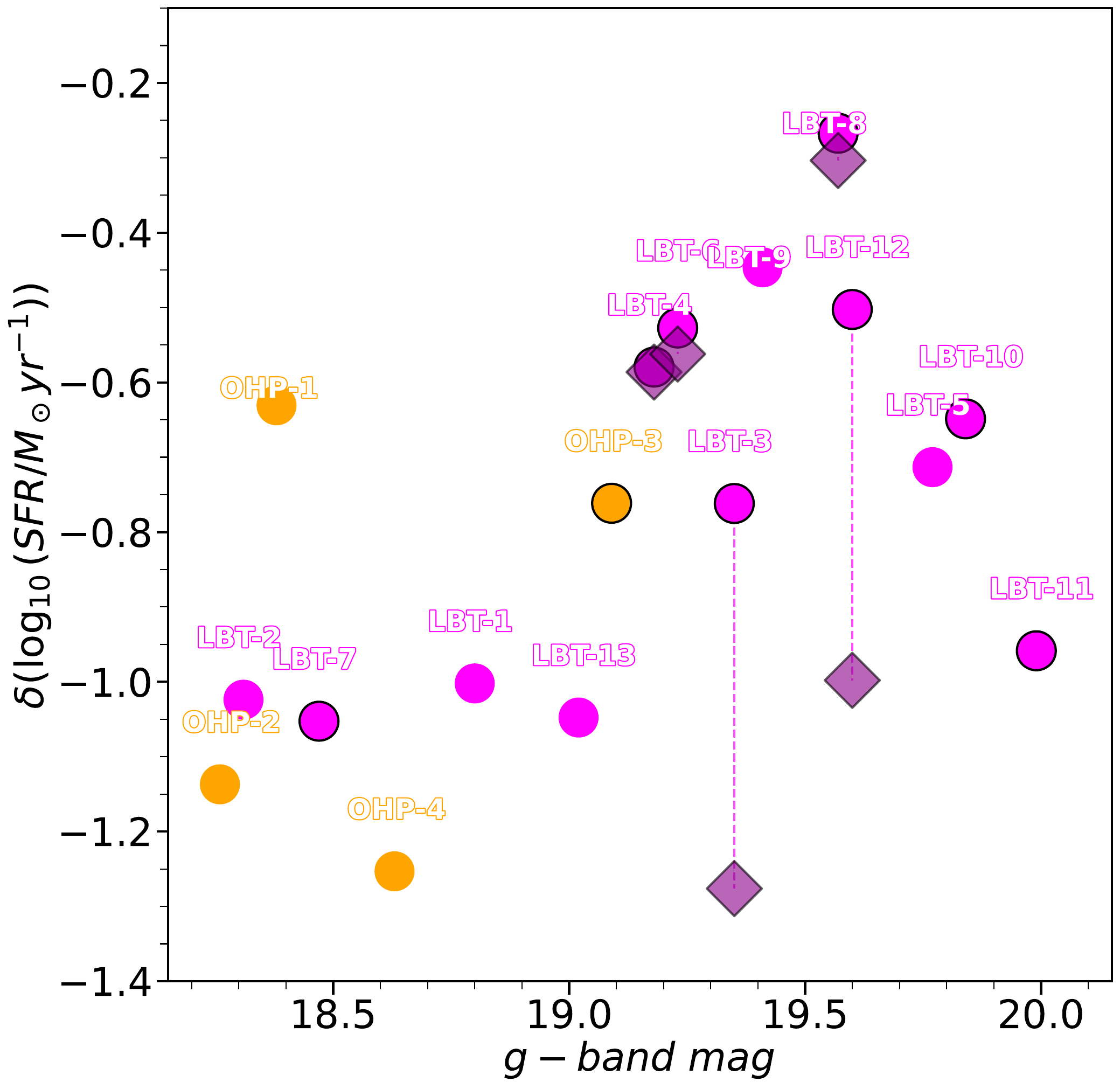}
  \caption{Magnitude in the \textit{g} band vs the differences in the SFR values before and after aperture correction for our sample, represented as magenta (LBT) and orange (OHP) circles.
  The purple diamonds indicate the differences between the SFR values before and after aperture and dust correction for the sources with clear H$\alpha$ and H$\beta$ detections.
  All the outlined dots in the plot represent confirmed UDGs.}
\label{dust and aperture correction}
\end{figure}
In Fig.~\ref{MS}, our MS is compared to a dwarf galaxy sample from \cite{Li_2025}. 
This sample spans a range of stellar masses that also overlaps our UDGs, reinforcing the previously stated hypothesis that UDGs inhabit the same population distribution as dwarf galaxies.
We performed a two-sample, two-sided, and 1D Kolmogorov-Smirnov (KS) test~\citep{Press2007} to compare the distributions of the specific SFR of our LBT-UDGs (we do not take into account the OHP-UDG due to the scarcity of the sample) and the \cite{Li_2025} sample (55 galaxies).
The KS test quantifies the maximum absolute difference between the cumulative distributions of the two samples and computes the associated parameter, $p$. 
If the value of $p$ is below a theoretical threshold of 0.05, the null hypothesis that the two samples are drawn from the same distribution is rejected.
In our case, we find a $p$-value of 0.61.
Therefore, we cannot reject the null hypothesis, so we suggest that the two samples may be statistically consistent with being drawn from the same parent distribution.
We performed an additional KS test (analogous to the one already performed) between our LBT-UDGs and the sample of local and normal star-forming galaxies of \cite{Leroy} (23 galaxies).
Although smaller in size, this sample extends to higher stellar masses (\Mstar ~$ >10^{9.5}M_\odot$) compared to typical dwarf galaxies.
In this case, we find a higher value of $p$ = $0.94$.
This, again, means that we cannot reject the null hypothesis.
This finding is consistent with the previous assumption that the two samples are drawn from the same underlying distribution.

\subsection{Aperture correction}
\label{Ap.Corr}
\begin{figure}[h]
\includegraphics[width=0.45\textwidth]{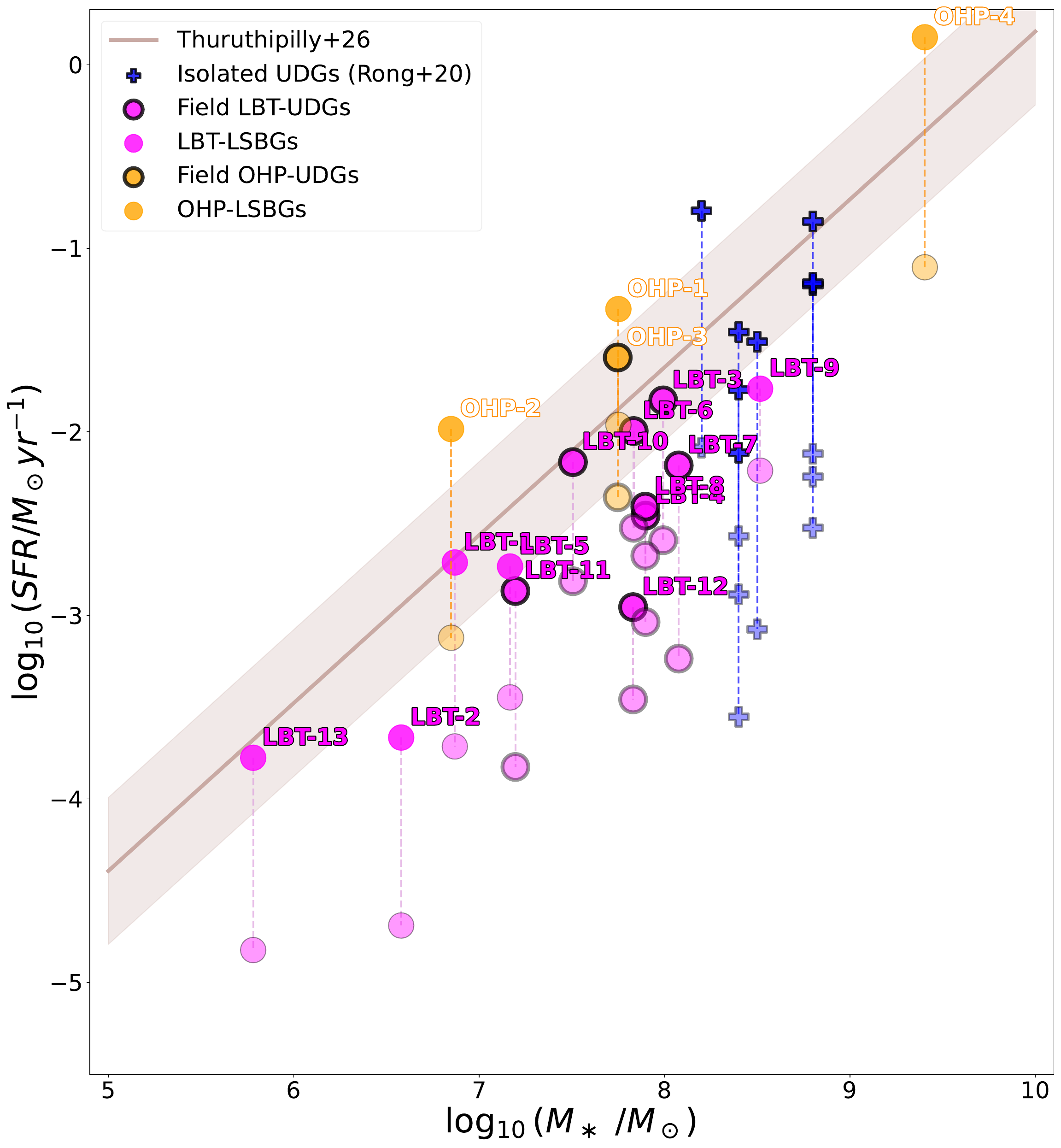}
  \caption{\Mstar~log vs the SFR log for our sample before and after aperture correction (Sect.~\ref{Aperture correction for SFRs}).
  The LBT sample is shown as magenta and light-magenta dots for the SFR values after and before aperture correction, respectively. The OHP sample is shown as orange and light-orange dots, using the same colour map. The blue crosses show the UDGs from \cite{Rong_2020}; the light-blue crosses represent the SFR values obtained using a fiber; and the blue crosses represent the estimated total SFR.
  The dotted lines connect each pair of the same sample.
  All the UDGs in the plot are shown as black-outlined dots.
  For comparison, we show the star-forming LSBG relation from \cite{Hareesh+26} as a brown line within the uncertainty interval ($\pm0.4$ dex) depicted as a grey-shaded region.}
     \label{aperture correction}
\end{figure}
In Fig.~\ref{dust and aperture correction}, the difference in the SFRs before and after aperture correction is shown for all targets.
The aperture correction has been described in Sect.~\ref{Aperture correction for SFRs}.
The plot clearly shows the necessity of correction to obtain reliable SFR values. 
Our targets were observed with a long slit passing through the centre of the galaxy, sampling the brightest, and thus the most star-forming regions.
Nevertheless, simply plotting the SFR values derived from these fluxes would have led to a significant underestimation, leaving a population of low star-forming LSBGs below the standard MS.
For all UDG candidates in this work, prior to aperture correction, we find $\mathrm{-4.82 \leq \log{(SFR)} \leq -1.10}$ and a $\rm\log(SFR_\mathrm{mean})$ equal to $-3.03\pm0.94$. After aperture correction, we find $\mathrm{-3.78 \leq \log(SFR) \leq 0.15}$ and a $\rm\log(SFR_\mathrm{mean})$ equal to $-2.25\pm0.89$ in $\mathrm{M}_\odot\,\mathrm{yr}^{-1}$, respectively.
With the correction applied, we account for each galaxy's shape and its brightness distribution across its surface.
The result is not negligible: all the targets lie fully on the MS after aperture correction, demonstrating that UDGs in the field are as active as standard local galaxies and dwarf galaxies, rather than being 'low efficiency galaxies' in the star formation regime~\citep{Vandokkum2015}.
For comparison, we plot our sources alongside the \cite{Rong_2020} sample (blue crosses).
In their article, the authors stacked isolated UDG spectra obtained from fiber observations using the seventh data release of SDSS data \citep{Abazajian_2009}. 
Using a similar line luminosity relation~\citep{Kennicutt94} to that we used to compute the SFR of our sources in Sect.~\ref{H_alpha line luminosity}, they obtain only partial values of the SFR.
Then, \cite{Rong_2020} find the total SFRs of the UDGs by comparing the ratio of the luminosity of the entire galaxies and the region covered by the fiber.
In the paper, the authors highlight the total SFRs as an upper limit.
Figure~\ref{aperture correction} shows the SFR values of LBT- and OHP-UDGs, before and after correction, together with the SFRs from \cite{Rong_2020} sample with fiber and total SFR values.
In Fig.~\ref{aperture correction}, we plotted the star-forming LSBG relation from \cite{Hareesh+26}. After applying aperture corrections, we find that our sources are in good agreement with the relation.
\cite{Hareesh+26} derived this relation using a sample of galaxies in the range $0.0 < z\leq0.03$, which overlaps with the average redshift of our samples (LBT: $0.023\pm0.016$, OHP: $0.031\pm0.027$).
The SFRs obtained by \cite{Hareesh+26} were derived through SED fitting and therefore already account for the total flux of the galaxies without requiring additional corrections.
The results indicate that the correction magnitude order is reasonable across both samples and that both targets move from below the usual SFMS into the MS loci.
We stress here again that the values computed, discussed, and also shown in Fig.~\ref{Ap.Corr}, with the aperture correction represent an upper limit for the SFRs.
\subsection{Internal extinction correction}
\label{Dust}
In Sect.~\ref{Dust Correction}, we used the Balmer ratio to derive the dust attenuation values of the LSB-UDG sample.
Using Eq.~\ref{att_value}, we computed dust corrections for the H$\alpha$ and DES \textit{g}-band wavelength ($A_{\text{H}\alpha}$ and $A_\mathrm{g}$, respectively).
The retrieved values ($1.29$, $0.02$, $0.09$, $0.09$, and $1.24$ mag for \textbf{LBT-3}, \textbf{LBT-4}, \textbf{LBT-6}, \textbf{LBT-8}, and \textbf{LBT-12}, respectively) indicate a small amount of dust in LBT-UDG candidates, with an average $A_{\text{H}\alpha}$ of $0.55$ mag, consistent with previous work \citep[i.e.][]{Hinz, Rahman} that expect low dust content in LSBGs.
\cite{Junais23} studied a sample of $\sim$1000 local LSBGs ($z < 0.1$) and report minimal dust attenuation ($< 0.1$) for the majority of galaxies in the sample, with a small percentage (4\%) having significant attenuation (a $0.8$ mean attenuation value in the \textit{V} band in mag).
Other recent studies at intermediate redshift have drawn similar conclusions.
~\cite{Shields}, for example, studied a large ($\sim1\,100$) LSBGs in a redshift range between 0.4 and 0.8, finding that 92\% of high-z LSBGs have $A_V$ < 1.0 mag and 73\% of low-z LSBGs have $A_V$ < 1.0 mag.
For our five sources, for which dust attenuation could be measured, we find in their star-forming regions values of $A_V$ equal to $0.69$, $0.01$, $0.05$, $0.05$, and $0.66$ mag for \textbf{LBT-3}, \textbf{LBT-4}, \textbf{LBT-6}, \textbf{LBT-8}, and \textbf{LBT-12}, respectively, with an average $A_V$ of $0.29\pm0.31$ mag.
We computed $A_\mathrm{V}$ in the same way as $A_\mathrm{H_\alpha}$ and $A_\mathrm{g}$ in Sect.~\ref{Dust Correction}.
Comparing them with the $A_V$ values from \cite{Shields}, we find that generally, our UDGs are aligned with those studied in the literature.
According to that, we corrected (in Sect.~\ref{Dust Correction}) the $\mu_\mathrm{0,g}$ values by subtracting the dust attenuation for the \textit{g} band.
After this correction, two out of the eight LBT-UDGs satisfying the \cite{Vandokkum2015} criteria exhibit $\mu_\mathrm{0,g}$ values less than $24\ \mathrm{mag\ arcsec^{-2}}$ and can no longer be classified as UDGs.
The definition of UDGs does not account for the dust component in these galaxies.
This inconsistency is primarily due to selection bias or to a lack of spectroscopic information required to measure the attenuation directly from the lines.
However, although these two sources do not reach the UDGs faintness threshold, they still exhibit the diffuse, structureless signature typical of UDGs.
For this reason, despite being slightly brighter than the canonical threshold, we still consider all eight galaxies among the LBT sources to be UDGs.
This interpretation is motivated by the fact that the various definitions of UDGs in the literature (e.g.,~\citealt{Vandokkum2015, Gannon2024, Chan18, Marleau_2021}) are based on observed central surface brightness values ($\mu_{0}$) rather than intrinsic or dust-corrected quantities.

Finally, as shown in Appendix~\ref{Appendix_C}, SFRs measured based on the H$\alpha$ line luminosity after dust and aperture correction were compared with the estimated SFRs based on optical broadband photometry, finding good agreement (the obtained scatter is $<$ 0.10 dex).
However, we would like to stress that the estimated SFRs are sensitive to the SFH used for SED fitting. 
The appropriate selection of the SFH module was only possible thanks to the ground-truth SFRs derived from the spectra.

\subsection{Giant low surface brightness galaxy candidates}
\label{Giant-LSBG candidate}
\begin{figure}[h]
\centering
\hspace{-2em}
\includegraphics[width=0.45\textwidth]{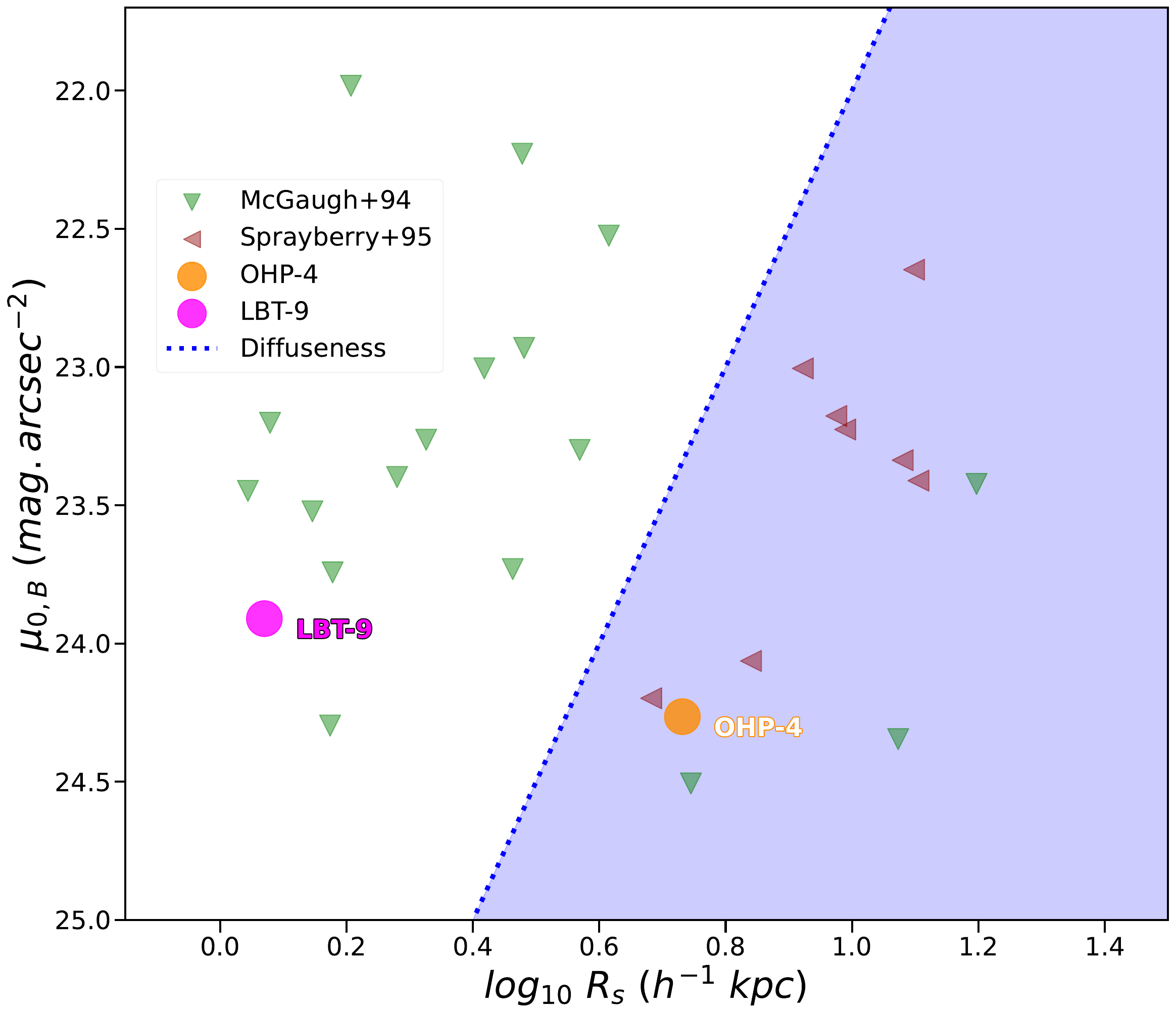}
  \caption{Classification of OHP-4 and LBT-9 from \citet{sprayberry1995}, using the scale length ($R_s$) and central surface brightness ($\mu_{0,g}$) of the disk.
  The diffuseness (dotted blue line) separates the LSB galaxies from the GLSBGs (blue area). In the background, sources from \cite[green inverted triangles]{McGaugh} and \cite[dark red sideways triangles]{sprayberry1995} are plotted.}
     \label{GLSBG}
\end{figure}

The GLSB galaxies are a subclass of LSB galaxies, first characterised by an unusually extended optical disk \citep{sprayberry1995, Beijersbergen1999A}. 
In an attempt to distinguish regular spirals, GLSBGs (as massive but more extended than regular spirals) and LSB galaxies (less massive), \citet{sprayberry1995}, proposed a photometric classification using the diffuseness ($d_\mathrm{i}$) parameter, which depends on the disk scale length $(R_\mathrm{s})$ and disk central surface brightness ($\mu_\mathrm{0, B}$), as in the Eq.~\ref{Giant}:

\begin{equation}
\label{Giant}
d_\mathrm{i} = \mu_\mathrm{0,B} + 5* \log{(R_\mathrm{s}}).
\end{equation}

Galaxies with a diffuseness index greater than 27 are GLSBGs, while lower values are found for regular spiral galaxies \citep{sprayberry1995}. However, additional criteria have been proposed since then, including a high $HI$ mass, above $10^{10}$ M$_{\odot}$ \citep{Saburova2021} and a $HI$ radius (R$_\mathrm{HI}$) > 50 kpc \citep{2023MNRAS.523.3991Z}.
As mentioned in Sect.~\ref{Redshift Measurement}, LBT-9 and OHP-4 present effective radius in kpc of 6.70 and 8.92, and their stellar masses are 8.52 and 9.40 in $\log$ \Mstar $/$ M$_\odot$, respectively. 
Given their sizes -- much larger than UDGs and LSBs -- and their masses, the aforementioned sources can be considered as GLSBG candidates. 
To verify the GLSB nature of these galaxies, using the \cite{sprayberry1995} definition, we proceeded in the same way as in \citet{2025BERNAUD} to compute their surface brightness profiles and estimate the central surface brightness and scale length.
The results are shown in Fig. \ref{GLSBG}. 
OHP-4 lies within the region of GLSBGs with a massive stellar disk passing the criteria of \citet{sprayberry1995}. LBT-9 is below the diffuseness threshold of 27.
It is not considered a GLSBG according to the original definition. Despite its relatively large size, it is not large enough for its central surface density to pass the criteria. 
\section{Conclusions}
\label{section_conclusion}
In this work, we studied two samples of field UDG candidates (LBT-UDGs and OHP-UDGs).
The sample of LBT-UDG candidates was observed with the LBT/MODS spectrograph and comprises 13 targets.
The OHP-UDG candidate sample was observed with the OHP/MISTRAL instrument and comprises four sources.
Our main results are as follows:
\begin{itemize}
  \item  We performed data reduction, extracted 1D spectra, and measured the redshifts of all sources with prominent H$\alpha$ emission.
  \item We find eight spectroscopically confirmed UDGs among the 13 in the LBT-UDG sample, according to the definition of \cite{Vandokkum2015} after applying cosmological dimming correction. All other sources have been confirmed as LSBGs, including one LSBG close to the GLSBG definition but not extended enough to be considered as such (see Sect.~\ref{Giant-LSBG candidate}).
  Among the OHP-UDG candidates, we confirm a single UDG. The other candidates are identified as LSBGs, including a confirmed GLSBG (see Sect.~\ref{Giant-LSBG candidate}). 
  \item For all sources, we computed forced photometry using \textsc{photutils} with an elliptical aperture in all available bands. 
  \item We estimated the SFRs using the H$\alpha$ line luminosity and corrected the flux integrated with a long slit, accounting for the morphology of the corresponding source.
  \item We also estimated the stellar masses of UDG candidates using the mass-to-light ratio colour relation.
  \item We analysed their position (after aperture correction) compared to the main-sequence relation and find that field UDGs follow the same relation as local and normal star-forming galaxies, assuming the SFRs are computed as an upper limit. 
  \item We measured the H$\beta$ emission line fluxes for ten LBT sources. Five of these have sufficient S/Ns to allow the use of the Balmer ratio to estimate dust attenuation at different wavelengths (see Sect.~\ref{Dust Correction}).
  The attenuation of the H$\alpha$ emission line was used to correct the SFRs.
  We also computed dust attenuation at the DES central \textit{g}-band wavelength for our sources, used to correct the central surface brightness in the \textit{g} band. After correction, two sources fall below the faintness threshold imposed by the UDG definition. This highlights that dust attenuation can be significant in a subset of field UDGs and should therefore be accounted for when deriving their intrinsic photometric properties.

\end{itemize}
Finally, we emphasise that the corrections discussed in this work, particularly those for dust attenuation and face-on de-projection, demonstrate that a classification based solely on observed photometric quantities may overlook galaxies that would otherwise be classified as UDGs once these effects are accounted for.
\begin{acknowledgements}
We thank Wilfried Mercier and Cedric Dubois for carrying out the observations with the MISTRAL instrument at the Observatoire de Haute-Provence used in this work. We are also grateful to Veronique Buat and Denis Burgarella for helpful discussions and insightful comments, and to Priscilla J. Pessi for her assistance with the computation of the galaxy peculiar velocities. This research was supported by the Polish National Agency for Academic Exchange under the CLEVER, Strategic Partnership programme BPI/PST/2024/1/00019, the Polish National Science Centre grant no 2023/50/A/ST9/00579 and UMO-2024/53/B/ST9/00230, the PRELUDIUM grant UMO-2023/49/N/ST9/00746 and the SONATA BIS grant UMO-2020/38/E/ST9/00077. Additional support was provided by the European Union through the MSCA EDUCADO project (GA 101119830) and WIDERA ExGal-Twin (GA 101158446). This research used the facilities of the Italian Centre for Astronomical Archive (IA2) operated by INAF at the Astronomical Observatory of Trieste. Part of this work was supported by the German \emph{Deut\-sche For\-schungs\-ge\-mein\-schaft, DFG\/} project number Ts~17/2--1.
\end{acknowledgements}
\bibliographystyle{aa}
\bibliography{bibliography}

@ARTICLE{ciesla2017,
       author = {{Ciesla}, L. and {Elbaz}, D. and {Fensch}, J.},
        title = "{The SFR-M$_{{\ensuremath{*}}}$ main sequence archetypal star-formation history and analytical models}",
      journal = {A\&A},
         year = 2017,
        month = dec,
       volume = {608},
          eid = {A41},
        pages = {A41},
          doi = {10.1051/0004-6361/201731036},
archivePrefix = {arXiv},
       eprint = {1706.08531},
 primaryClass = {astro-ph.GA},
       adsurl = {https://ui.adsabs.harvard.edu/abs/2017A\&A...608A..41C}
}

@ARTICLE{Bothun1997,
       author = {{Bothun}, G. and {Impey}, C. and {McGaugh}, S.},
        title = "{Low-Surface-Brightness Galaxies: Hidden Galaxies Revealed}",
      journal = {PASP},
         year = 1997,
        month = jul,
       volume = {109},
        pages = {745-758},
          doi = {10.1086/133941},
       adsurl = {https://ui.adsabs.harvard.edu/abs/1997PASP..109..745B}
}

@INPROCEEDINGS{ONeil2000,
       author = {{O'Neil}, K.},
        title = "{Gas, Stars and Baryons in Low Surface Brightness Galaxies}",
    booktitle = {Stars, Gas and Dust in Galaxies: Exploring the Links},
         year = 2000,
       editor = {{Alloin}, Danielle and {Olsen}, Knut and {Galaz}, Gaspar},
       series = {Astronomical Society of the Pacific Conference Series},
       volume = {221},
        month = jan,
        pages = {199},
          doi = {10.48550/arXiv.astro-ph/0006253},
archivePrefix = {arXiv},
       eprint = {astro-ph/0006253},
 primaryClass = {astro-ph},
       adsurl = {https://ui.adsabs.harvard.edu/abs/2000ASPC..221..199O}
}

@ARTICLE{Junais_lines,
   author = {{Junais} and {Boissier}, S. and {Epinat}, B. and {Amram}, P. and {Madore}, B.~F. and {Boselli}, A. and {Koda}, J. and {Gil de Paz}, A. and {Mu{\~n}os Mateos}, J.~C. and {Chemin}, L.},
    title = "{First spectroscopic study of ionised gas emission lines in the extreme low surface brightness galaxy Malin 1}",
  journal = {A\&A},
   year = 2020,
    month = may,
   volume = {637},
    eid = {A21},
    pages = {A21},
    doi = {10.1051/0004-6361/201937330},
archivePrefix = {arXiv},
   eprint = {2003.09492},
 primaryClass = {astro-ph.GA},
   adsurl = {https://ui.adsabs.harvard.edu/abs/2020A\&A...637A..21J}
}

@ARTICLE{Moustakas2010,
       author = {{Moustakas}, John and {Kennicutt}, Jr., Robert C. and {Tremonti}, Christy A. and {Dale}, Daniel A. and {Smith}, John-David T. and {Calzetti}, Daniela},
        title = "{Optical Spectroscopy and Nebular Oxygen Abundances of the Spitzer/SINGS Galaxies}",
      journal = {ApJS},
         year = 2010,
        month = oct,
       volume = {190},
       number = {2},
        pages = {233-266},
          doi = {10.1088/0067-0049/190/2/233},
archivePrefix = {arXiv},
       eprint = {1007.4547},
 primaryClass = {astro-ph.CO},
       adsurl = {https://ui.adsabs.harvard.edu/abs/2010ApJS..190..233M}
}

@software{Earl2024,
  author       = {Nicholas Earl and
                  Ricky O'Steen and
                  Erik Tollerud and
                  brechmos and
                  Wolfgang Kerzendorf and
                  Ivo Busko and
                  P. L. Lim and
                  shaileshahuja and
                  Dan D'Avella and
                  Thomas Robitaille and
                  Adam Ginsburg and
                  Derek Homeier and
                  Brigitta Sipőcz and
                  Jesse Averbukh and
                  Brian Cherinka and
                  James Tocknell and
                  Sara Ogaz and
                  Robel Geda and
                  James Davies and
                  Kyle Conroy and
                  Hans Moritz Günther and
                  Kyle Barbary and
                  Kelle Cruz and
                  Jonathan Foster and
                  Michael Droettboom and
                  Duy Nguyen and
                  E. M. Bray and
                  Andy Casey and
                  Henry Ferguson},
  title        = {astropy/specutils: v2.3.0},
  month        = feb,
  year         = 2026,
  publisher    = {Zenodo},
  version      = {v2.3.0},
  doi          = {10.5281/zenodo.18473282},
  url          = {https://doi.org/10.5281/zenodo.18473282},
  swhid        = {swh:1:dir:7325c4b0234eb3071a79320fbcf54990365036dc
                   ;origin=https://doi.org/10.5281/zenodo.1421356;vis
                   it=swh:1:snp:dc14798f240356f3262c33329f9668795e5e1
                   db4;anchor=swh:1:rel:dd98653928a7d47027d178869f6a4
                   cf356feb8f4;path=astropy-specutils-4d8dbea}
}

@article{Astropy,
title = {Astropy: A community Python package for astronomy},
author = {Robitaille, Thomas P. and Tollerud, Erik J. and Perry Greenfield and Michael Droettboom and Erik Bray and Tom Aldcroft and Matt Davis and Adam Ginsburg and Price-Whelan, Adrian M. and Kerzendorf, Wolfgang E. and Alexander Conley and Neil Crighton and Kyle Barbary and Demitri Muna and Ferguson, Henry C and Fr\'ed\'eric Grollier and Parikh, Madhura M. and Nair, Prasanth H. and G\"unther, Hans M. and Christoph Deil and Casey, Andrew R.},
year = 2013,
doi = "10.1051/0004-6361/201322068",
language = {English},
volume = {558},
journal = {A\&A},
issn = {0004-6361},
publisher = {EDP Sciences}
}

@ARTICLE{Gavazzi2011,
       author = {{Gavazzi}, G. and {Savorgnan}, G. and {Fumagalli}, Mattia},
        title = "{The complete census of optically selected AGNs in the Coma supercluster: the dependence of AGN activity on the local environment}",
      journal = {A\&A},
         year = 2011,
        month = oct,
       volume = {534},
          eid = {A31},
        pages = {A31},
          doi = {10.1051/0004-6361/201117461},
archivePrefix = {arXiv},
       eprint = {1107.3702},
 primaryClass = {astro-ph.CO},
       adsurl = {https://ui.adsabs.harvard.edu/abs/2011A\&A...534A..31G}
}

@ARTICLE{Vandokkum2015,
       author = {{van Dokkum}, Pieter G. and {Abraham}, Roberto and {Merritt}, Allison and {Zhang}, Jielai and {Geha}, Marla and {Conroy}, Charlie},
        title = "{Forty-seven Milky Way-sized, Extremely Diffuse Galaxies in the Coma Cluster}",
      journal = {ApJL},
         year = 2015,
        month = jan,
       volume = {798},
       number = {2},
          eid = {L45},
        pages = {L45},
          doi = {10.1088/2041-8205/798/2/L45},
archivePrefix = {arXiv},
       eprint = {1410.8141},
 primaryClass = {astro-ph.GA},
       adsurl = {https://ui.adsabs.harvard.edu/abs/2015ApJ...798L..45V}
}

@ARTICLE{Schlegel1998,
       author = {{Schlegel}, David J. and {Finkbeiner}, Douglas P. and {Davis}, Marc},
        title = "{Maps of Dust Infrared Emission for Use in Estimation of Reddening and Cosmic Microwave Background Radiation Foregrounds}",
      journal = {ApJ},
         year = 1998,
        month = jun,
       volume = {500},
       number = {2},
        pages = {525-553},
          doi = {10.1086/305772},
archivePrefix = {arXiv},
       eprint = {astro-ph/9710327},
 primaryClass = {astro-ph},
       adsurl = {https://ui.adsabs.harvard.edu/abs/1998ApJ...500..525S}
}

@inproceedings{MTObjects,
  title={Improved detection of faint extended astronomical objects through statistical attribute filtering},
  author={Teeninga, Paul and Moschini, Ugo and Trager, Scott C and Wilkinson, Michael HF},
  booktitle={International Symposium on Mathematical Morphology and Its Applications to Signal and Image Processing},
  pages={157--168},
  year={2015},
  organization={Springer}
}

@ARTICLE{Conroy2009,
   author = {{Conroy}, Charlie and {Gunn}, James E. and {White}, Martin},
    title = "{The Propagation of Uncertainties in Stellar Population Synthesis Modeling. I. The Relevance of Uncertain Aspects of Stellar Evolution and the Initial Mass Function to the Derived Physical Properties of Galaxies}",
  journal = {ApJ},
   year = 2009,
    month = jul,
   volume = {699},
   number = {1},
    pages = {486-506},
    doi = {10.1088/0004-637X/699/1/486},
archivePrefix = {arXiv},
   eprint = {0809.4261},
 primaryClass = {astro-ph},
   adsurl = {https://ui.adsabs.harvard.edu/abs/2009ApJ...699..486C}
}

@ARTICLE{Taylor2011,
   author = {{Taylor}, Edward N. and {Hopkins}, Andrew M. and {Baldry}, Ivan K. and {Brown}, Michael J.~I. and {Driver}, Simon P. and {Kelvin}, Lee S. and {Hill}, David T. and {Robotham}, Aaron S.~G. and {Bland-Hawthorn}, Joss and {Jones}, D.~H. and {Sharp}, R.~G. and {Thomas}, Daniel and {Liske}, Jochen and {Loveday}, Jon and {Norberg}, Peder and {Peacock}, J.~A. and {Bamford}, Steven P. and {Brough}, Sarah and {Colless}, Matthew and {Cameron}, Ewan and {Conselice}, Christopher J. and {Croom}, Scott M. and {Frenk}, C.~S. and {Gunawardhana}, Madusha and {Kuijken}, Konrad and {Nichol}, R.~C. and {Parkinson}, H.~R. and {Phillipps}, S. and {Pimbblet}, K.~A. and {Popescu}, C.~C. and {Prescott}, Matthew and {Sutherland}, W.~J. and {Tuffs}, R.~J. and {van Kampen}, Eelco and {Wijesinghe}, D.},
    title = "{Galaxy And Mass Assembly (GAMA): stellar mass estimates}",
  journal = {MNRAS},
   year = 2011,
    month = dec,
   volume = {418},
   number = {3},
    pages = {1587-1620},
    doi = {10.1111/j.1365-2966.2011.19536.x},
archivePrefix = {arXiv},
   eprint = {1108.0635},
 primaryClass = {astro-ph.CO},
   adsurl = {https://ui.adsabs.harvard.edu/abs/2011MNRAS.418.1587T}
}

@ARTICLE{Calzetti2000,
   author = {{Calzetti}, Daniela and {Armus}, Lee and {Bohlin}, Ralph C. and {Kinney}, Anne L. and {Koornneef}, Jan and {Storchi-Bergmann}, Thaisa},
    title = "{The Dust Content and Opacity of Actively Star-forming Galaxies}",
  journal = {ApJ},
   year = 2000,
    month = apr,
   volume = {533},
   number = {2},
    pages = {682-695},
    doi = {10.1086/308692},
archivePrefix = {arXiv},
   eprint = {astro-ph/9911459},
 primaryClass = {astro-ph},
   adsurl = {https://ui.adsabs.harvard.edu/abs/2000ApJ...533..682C}
}

@ARTICLE{Cardelli89,
   author = {{Cardelli}, Jason A. and {Clayton}, Geoffrey C. and {Mathis}, John S.},
    title = "{The Relationship between Infrared, Optical, and Ultraviolet Extinction}",
  journal = {ApJ},
   year = 1989,
    month = oct,
   volume = {345},
    pages = {245},
    doi = {10.1086/167900},
   adsurl = {https://ui.adsabs.harvard.edu/abs/1989ApJ...345..245C}
}

@ARTICLE{Martin2019,
   author = {{Martin}, G. and {Kaviraj}, S. and {Laigle}, C. and {Devriendt}, J.~E.~G. and {Jackson}, R.~A. and {Peirani}, S. and {Dubois}, Y. and {Pichon}, C. and {Slyz}, A.},
    title = "{The formation and evolution of low-surface-brightness galaxies}",
  journal = {MNRAS},
   year = 2019,
    month = may,
   volume = {485},
   number = {1},
    pages = {796-818},
    doi = {10.1093/MNRAS/stz356},
archivePrefix = {arXiv},
   eprint = {1902.04580},
 primaryClass = {astro-ph.GA},
   adsurl = {https://ui.adsabs.harvard.edu/abs/2019MNRAS.485..796M}
}

@inproceedings{Attention,
 author = {Vaswani, Ashish and Shazeer, Noam and Parmar, Niki and Uszkoreit, Jakob and Jones, Llion and Gomez, Aidan N and Kaiser, \L ukasz and Polosukhin, Illia},
 booktitle = {Advances in Neural Information Processing Systems},
 editor = {I. Guyon and U. Von Luxburg and S. Bengio and H. Wallach and R. Fergus and S. Vishwanathan and R. Garnett},
 pages = {},
 publisher = {Curran Associates, Inc.},
 title = {Attention is All you Need},
 url = {https://proceedings.neurips.cc/paper_files/paper/2017/file/3f5ee243547dee91fbd053c1c4a845aa-Paper.pdf},
 volume = {30},
 year = {2017}
}

@INPROCEEDINGS{MODS,
   author = {{Pogge}, R.~W. and {Atwood}, B. and {Brewer}, D.~F. and {Byard}, P.~L. and {Derwent}, M.~A. and {Gonzalez}, R. and {Martini}, P. and {Mason}, J.~A. and {O'Brien}, T.~P. and {Osmer}, P.~S. and {Pappalardo}, D.~P. and {Steinbrecher}, D.~P. and {Teiga}, E.~J. and {Zhelem}, R.},
    title = "{The multi-object double spectrographs for the Large Binocular Telescope}",
    booktitle = {Ground-based and Airborne Instrumentation for Astronomy III},
    year = 2010,
   editor = {{McLean}, Ian S. and {Ramsay}, Suzanne K. and {Takami}, Hideki},
   series = {Society of Photo-Optical Instrumentation Engineers (SPIE) Conference Series},
   volume = {7735},
    month = jul,
    eid = {77350A},
    pages = {77350A},
    doi = {10.1117/12.857215},
   adsurl = {https://ui.adsabs.harvard.edu/abs/2010SPIE.7735E..0AP}
}

@ARTICLE{DES,
   author = {{Abbott}, T.~M.~C. and {Abdalla}, F.~B. and {Allam}, S. and {Amara}, A. and {Annis}, J. and {Asorey}, J. and {Avila}, S. and {Ballester}, O. and {Banerji}, M. and {Barkhouse}, W. and {Baruah}, L. and {Baumer}, M. and {Bechtol}, K. and {Becker}, M.~R. and {Benoit-L{\'e}vy}, A. and {Bernstein}, G.~M. and {Bertin}, E. and {Blazek}, J. and {Bocquet}, S. and {Brooks}, D. and {Brout}, D. and {Buckley-Geer}, E. and {Burke}, D.~L. and {Busti}, V. and {Campisano}, R. and {Cardiel-Sas}, L. and {Carnero Rosell}, A. and {Carrasco Kind}, M. and {Carretero}, J. and {Castander}, F.~J. and {Cawthon}, R. and {Chang}, C. and {Chen}, X. and {Conselice}, C. and {Costa}, G. and {Crocce}, M. and {Cunha}, C.~E. and {D'Andrea}, C.~B. and {da Costa}, L.~N. and {Das}, R. and {Daues}, G. and {Davis}, T.~M. and {Davis}, C. and {De Vicente}, J. and {DePoy}, D.~L. and {DeRose}, J. and {Desai}, S. and {Diehl}, H.~T. and {Dietrich}, J.~P. and {Dodelson}, S. and {Doel}, P. and {Drlica-Wagner}, A. and {Eifler}, T.~F. and {Elliott}, A.~E. and {Evrard}, A.~E. and {Farahi}, A. and {Fausti Neto}, A. and {Fernandez}, E. and {Finley}, D.~A. and {Flaugher}, B. and {Foley}, R.~J. and {Fosalba}, P. and {Friedel}, D.~N. and {Frieman}, J. and {Garc{\'\i}a-Bellido}, J. and {Gaztanaga}, E. and {Gerdes}, D.~W. and {Giannantonio}, T. and {Gill}, M.~S.~S. and {Glazebrook}, K. and {Goldstein}, D.~A. and {Gower}, M. and {Gruen}, D. and {Gruendl}, R.~A. and {Gschwend}, J. and {Gupta}, R.~R. and {Gutierrez}, G. and {Hamilton}, S. and {Hartley}, W.~G. and {Hinton}, S.~R. and {Hislop}, J.~M. and {Hollowood}, D. and {Honscheid}, K. and {Hoyle}, B. and {Huterer}, D. and {Jain}, B. and {James}, D.~J. and {Jeltema}, T. and {Johnson}, M.~W.~G. and {Johnson}, M.~D. and {Kacprzak}, T. and {Kent}, S. and {Khullar}, G. and {Klein}, M. and {Kovacs}, A. and {Koziol}, A.~M.~G. and {Krause}, E. and {Kremin}, A. and {Kron}, R. and {Kuehn}, K. and {Kuhlmann}, S. and {Kuropatkin}, N. and {Lahav}, O. and {Lasker}, J. and {Li}, T.~S. and {Li}, R.~T. and {Liddle}, A.~R. and {Lima}, M. and {Lin}, H. and {L{\'o}pez-Reyes}, P. and {MacCrann}, N. and {Maia}, M.~A.~G. and {Maloney}, J.~D. and {Manera}, M. and {March}, M. and {Marriner}, J. and {Marshall}, J.~L. and {Martini}, P. and {McClintock}, T. and {McKay}, T. and {McMahon}, R.~G. and {Melchior}, P. and {Menanteau}, F. and {Miller}, C.~J. and {Miquel}, R. and {Mohr}, J.~J. and {Morganson}, E. and {Mould}, J. and {Neilsen}, E. and {Nichol}, R.~C. and {Nogueira}, F. and {Nord}, B. and {Nugent}, P. and {Nunes}, L. and {Ogando}, R.~L.~C. and {Old}, L. and {Pace}, A.~B. and {Palmese}, A. and {Paz-Chinch{\'o}n}, F. and {Peiris}, H.~V. and {Percival}, W.~J. and {Petravick}, D. and {Plazas}, A.~A. and {Poh}, J. and {Pond}, C. and {Porredon}, A. and {Pujol}, A. and {Refregier}, A. and {Reil}, K. and {Ricker}, P.~M. and {Rollins}, R.~P. and {Romer}, A.~K. and {Roodman}, A. and {Rooney}, P. and {Ross}, A.~J. and {Rykoff}, E.~S. and {Sako}, M. and {Sanchez}, M.~L. and {Sanchez}, E. and {Santiago}, B. and {Saro}, A. and {Scarpine}, V. and {Scolnic}, D. and {Serrano}, S. and {Sevilla-Noarbe}, I. and {Sheldon}, E. and {Shipp}, N. and {Silveira}, M.~L. and {Smith}, M. and {Smith}, R.~C. and {Smith}, J.~A. and {Soares-Santos}, M. and {Sobreira}, F. and {Song}, J. and {Stebbins}, A. and {Suchyta}, E. and {Sullivan}, M. and {Swanson}, M.~E.~C. and {Tarle}, G. and {Thaler}, J. and {Thomas}, D. and {Thomas}, R.~C. and {Troxel}, M.~A. and {Tucker}, D.~L. and {Vikram}, V. and {Vivas}, A.~K. and {Walker}, A.~R. and {Wechsler}, R.~H. and {Weller}, J. and {Wester}, W. and {Wolf}, R.~C. and {Wu}, H. and {Yanny}, B. and {Zenteno}, A. and {Zhang}, Y. and {Zuntz}, J. and {DES Collaboration} and {Juneau}, S. and {Fitzpatrick}, M. and {Nikutta}, R.},
    title = "{The Dark Energy Survey: Data Release 1}",
  journal = {ApJS},
   year = 2018,
    month = dec,
   volume = {239},
   number = {2},
    eid = {18},
    pages = {18},
    doi = {10.3847/1538-4365/aae9f0},
archivePrefix = {arXiv},
   eprint = {1801.03181},
 primaryClass = {astro-ph.IM},
   adsurl = {https://ui.adsabs.harvard.edu/abs/2018ApJS..239...18A}
}

@ARTICLE{des_object_catolog,
   author = {{Sevilla-Noarbe}, I. and {Bechtol}, K. and {Carrasco Kind}, M. and {Carnero Rosell}, A. and {Becker}, M.~R. and {Drlica-Wagner}, A. and {Gruendl}, R.~A. and {Rykoff}, E.~S. and {Sheldon}, E. and {Yanny}, B. and {Alarcon}, A. and {Allam}, S. and {Amon}, A. and {Benoit-L{\'e}vy}, A. and {Bernstein}, G.~M. and {Bertin}, E. and {Burke}, D.~L. and {Carretero}, J. and {Choi}, A. and {Diehl}, H.~T. and {Everett}, S. and {Flaugher}, B. and {Gaztanaga}, E. and {Gschwend}, J. and {Harrison}, I. and {Hartley}, W.~G. and {Hoyle}, B. and {Jarvis}, M. and {Johnson}, M.~D. and {Kessler}, R. and {Kron}, R. and {Kuropatkin}, N. and {Leistedt}, B. and {Li}, T.~S. and {Menanteau}, F. and {Morganson}, E. and {Ogando}, R.~L.~C. and {Palmese}, A. and {Paz-Chinch{\'o}n}, F. and {Pieres}, A. and {Pond}, C. and {Rodriguez-Monroy}, M. and {Smith}, J. Allyn and {Stringer}, K.~M. and {Troxel}, M.~A. and {Tucker}, D.~L. and {de Vicente}, J. and {Wester}, W. and {Zhang}, Y. and {Abbott}, T.~M.~C. and {Aguena}, M. and {Annis}, J. and {Avila}, S. and {Bhargava}, S. and {Bridle}, S.~L. and {Brooks}, D. and {Brout}, D. and {Castander}, F.~J. and {Cawthon}, R. and {Chang}, C. and {Conselice}, C. and {Costanzi}, M. and {Crocce}, M. and {da Costa}, L.~N. and {Pereira}, M.~E.~S. and {Davis}, T.~M. and {Desai}, S. and {Dietrich}, J.~P. and {Doel}, P. and {Eckert}, K. and {Evrard}, A.~E. and {Ferrero}, I. and {Fosalba}, P. and {Garc{\'\i}a-Bellido}, J. and {Gerdes}, D.~W. and {Giannantonio}, T. and {Gruen}, D. and {Gutierrez}, G. and {Hinton}, S.~R. and {Hollowood}, D.~L. and {Honscheid}, K. and {Huff}, E.~M. and {Huterer}, D. and {James}, D.~J. and {Jeltema}, T. and {Kuehn}, K. and {Lahav}, O. and {Lidman}, C. and {Lima}, M. and {Lin}, H. and {Maia}, M.~A.~G. and {Marshall}, J.~L. and {Martini}, P. and {Melchior}, P. and {Miquel}, R. and {Mohr}, J.~J. and {Morgan}, R. and {Neilsen}, E. and {Plazas}, A.~A. and {Romer}, A.~K. and {Roodman}, A. and {Sanchez}, E. and {Scarpine}, V. and {Schubnell}, M. and {Serrano}, S. and {Smith}, M. and {Suchyta}, E. and {Tarle}, G. and {Thomas}, D. and {To}, C. and {Varga}, T.~N. and {Wechsler}, R.~H. and {Weller}, J. and {Wilkinson}, R.~D. and {DES Collaboration}},
    title = "{Dark Energy Survey Year 3 Results: Photometric Data Set for Cosmology}",
  journal = {ApJS},
   year = 2021,
    month = jun,
   volume = {254},
   number = {2},
    eid = {24},
    pages = {24},
    doi = {10.3847/1538-4365/abeb66},
archivePrefix = {arXiv},
   eprint = {2011.03407},
 primaryClass = {astro-ph.CO},
   adsurl = {https://ui.adsabs.harvard.edu/abs/2021ApJS..254...24S}
}

@ARTICLE{Thuruthipilly2024A,
   author = {{Thuruthipilly}, H. and {Junais} and {Pollo}, A. and {Sureshkumar}, U. and {Grespan}, M. and {Sawant}, P. and {Ma{\l}ek}, K. and {Zadrozny}, A.},
    title = "{Shedding light on low-surface-brightness galaxies in dark energy surveys with transformer models}",
  journal = {A\&A},
   year = 2024,
    month = feb,
   volume = {682},
    eid = {A4},
    pages = {A4},
    doi = {10.1051/0004-6361/202347649},
archivePrefix = {arXiv},
   eprint = {2310.13543},
 primaryClass = {astro-ph.GA},
   adsurl = {https://ui.adsabs.harvard.edu/abs/2024A\&A...682A...4T}
}

@ARTICLE{Hareesh+26,
       author = {{Thuruthipilly}, Hareesh and {Lisiecki}, Krzysztof and {Junais} and {Ma{\l}ek}, Katarzyna and {Pollo}, Agnieszka and {Pearson}, William J. and {Vanzanella}, Antonio and {Pal}, Saptarshi and {Figueira}, Miguel and {Dabhade}, Pratik and {Durkalec}, Anna and {Cotter}, Aidan P. and {Sureshkumar}, Unnikrishnan and {Hazra}, Nandini and {Matera}, Patryk and {Dey}, Subhrata and {Vr{\'a}bel}, Michal and {Dutta}, Anirban and {Willems}, Henry and {Principi Cavaterra}, Nicola and {Dobrowolska}, Natalia and {Knop}, Wojciech},
        title = "{From DES to KiDS: Domain adaptation for cross-survey detections of low-surface-brightness galaxies}",
      journal = {A\&A},
         year = 2026,
        month = jul,
       volume = {711},
          eid = {A187},
        pages = {A187},
          doi = {10.1051/0004-6361/202659547},
archivePrefix = {arXiv},
       eprint = {2605.13842},
 primaryClass = {astro-ph.GA},
       adsurl = {https://ui.adsabs.harvard.edu/abs/2026A\&A...711A.187T}
}

@ARTICLE{LSST,
   author = {{Ivezi{\'c}}, {\v{Z}}eljko and {Kahn}, Steven M. and {Tyson}, J. Anthony and {Abel}, Bob and {Acosta}, Emily and {Allsman}, Robyn and {Alonso}, David and {AlSayyad}, Yusra and {Anderson}, Scott F. and {Andrew}, John and {Angel}, James Roger P. and {Angeli}, George Z. and {Ansari}, Reza and {Antilogus}, Pierre and {Araujo}, Constanza and {Armstrong}, Robert and {Arndt}, Kirk T. and {Astier}, Pierre and {Aubourg}, {\'E}ric and {Auza}, Nicole and {Axelrod}, Tim S. and {Bard}, Deborah J. and {Barr}, Jeff D. and {Barrau}, Aurelian and {Bartlett}, James G. and {Bauer}, Amanda E. and {Bauman}, Brian J. and {Baumont}, Sylvain and {Bechtol}, Ellen and {Bechtol}, Keith and {Becker}, Andrew C. and {Becla}, Jacek and {Beldica}, Cristina and {Bellavia}, Steve and {Bianco}, Federica B. and {Biswas}, Rahul and {Blanc}, Guillaume and {Blazek}, Jonathan and {Blandford}, Roger D. and {Bloom}, Josh S. and {Bogart}, Joanne and {Bond}, Tim W. and {Booth}, Michael T. and {Borgland}, Anders W. and {Borne}, Kirk and {Bosch}, James F. and {Boutigny}, Dominique and {Brackett}, Craig A. and {Bradshaw}, Andrew and {Brandt}, William Nielsen and {Brown}, Michael E. and {Bullock}, James S. and {Burchat}, Patricia and {Burke}, David L. and {Cagnoli}, Gianpietro and {Calabrese}, Daniel and {Callahan}, Shawn and {Callen}, Alice L. and {Carlin}, Jeffrey L. and {Carlson}, Erin L. and {Chandrasekharan}, Srinivasan and {Charles-Emerson}, Glenaver and {Chesley}, Steve and {Cheu}, Elliott C. and {Chiang}, Hsin-Fang and {Chiang}, James and {Chirino}, Carol and {Chow}, Derek and {Ciardi}, David R. and {Claver}, Charles F. and {Cohen-Tanugi}, Johann and {Cockrum}, Joseph J. and {Coles}, Rebecca and {Connolly}, Andrew J. and {Cook}, Kem H. and {Cooray}, Asantha and {Covey}, Kevin R. and {Cribbs}, Chris and {Cui}, Wei and {Cutri}, Roc and {Daly}, Philip N. and {Daniel}, Scott F. and {Daruich}, Felipe and {Daubard}, Guillaume and {Daues}, Greg and {Dawson}, William and {Delgado}, Francisco and {Dellapenna}, Alfred and {de Peyster}, Robert and {de Val-Borro}, Miguel and {Digel}, Seth W. and {Doherty}, Peter and {Dubois}, Richard and {Dubois-Felsmann}, Gregory P. and {Durech}, Josef and {Economou}, Frossie and {Eifler}, Tim and {Eracleous}, Michael and {Emmons}, Benjamin L. and {Fausti Neto}, Angelo and {Ferguson}, Henry and {Figueroa}, Enrique and {Fisher-Levine}, Merlin and {Focke}, Warren and {Foss}, Michael D. and {Frank}, James and {Freemon}, Michael D. and {Gangler}, Emmanuel and {Gawiser}, Eric and {Geary}, John C. and {Gee}, Perry and {Geha}, Marla and {Gessner}, Charles J.~B. and {Gibson}, Robert R. and {Gilmore}, D. Kirk and {Glanzman}, Thomas and {Glick}, William and {Goldina}, Tatiana and {Goldstein}, Daniel A. and {Goodenow}, Iain and {Graham}, Melissa L. and {Gressler}, William J. and {Gris}, Philippe and {Guy}, Leanne P. and {Guyonnet}, Augustin and {Haller}, Gunther and {Harris}, Ron and {Hascall}, Patrick A. and {Haupt}, Justine and {Hernandez}, Fabio and {Herrmann}, Sven and {Hileman}, Edward and {Hoblitt}, Joshua and {Hodgson}, John A. and {Hogan}, Craig and {Howard}, James D. and {Huang}, Dajun and {Huffer}, Michael E. and {Ingraham}, Patrick and {Innes}, Walter R. and {Jacoby}, Suzanne H. and {Jain}, Bhuvnesh and {Jammes}, Fabrice and {Jee}, M. James and {Jenness}, Tim and {Jernigan}, Garrett and {Jevremovi{\'c}}, Darko and {Johns}, Kenneth and {Johnson}, Anthony S. and {Johnson}, Margaret W.~G. and {Jones}, R. Lynne and {Juramy-Gilles}, Claire and {Juri{\'c}}, Mario and {Kalirai}, Jason S. and {Kallivayalil}, Nitya J. and {Kalmbach}, Bryce and {Kantor}, Jeffrey P. and {Karst}, Pierre and {Kasliwal}, Mansi M. and {Kelly}, Heather and {Kessler}, Richard and {Kinnison}, Veronica and {Kirkby}, David and {Knox}, Lloyd and {Kotov}, Ivan V. and {Krabbendam}, Victor L. and {Krughoff}, K. Simon and {Kub{\'a}nek}, Petr and {Kuczewski}, John and {Kulkarni}, Shri and {Ku}, John and {Kurita}, Nadine R. and {Lage}, Craig S. and {Lambert}, Ron and {Lange}, Travis and {Langton}, J. Brian and {Le Guillou}, Laurent and {Levine}, Deborah and {Liang}, Ming and {Lim}, Kian-Tat and {Lintott}, Chris J. and {Long}, Kevin E. and {Lopez}, Margaux and {Lotz}, Paul J. and {Lupton}, Robert H. and {Lust}, Nate B. and {MacArthur}, Lauren A. and {Mahabal}, Ashish and {Mandelbaum}, Rachel and {Markiewicz}, Thomas W. and {Marsh}, Darren S. and {Marshall}, Philip J. and {Marshall}, Stuart and {May}, Morgan and {McKercher}, Robert and {McQueen}, Michelle and {Meyers}, Joshua and {Migliore}, Myriam and {Miller}, Michelle and {Mills}, David J.},
    title = "{LSST: From Science Drivers to Reference Design and Anticipated Data Products}",
  journal = {ApJ},
   year = 2019,
    month = mar,
   volume = {873},
   number = {2},
    eid = {111},
    pages = {111},
    doi = {10.3847/1538-4357/ab042c},
archivePrefix = {arXiv},
   eprint = {0805.2366},
 primaryClass = {astro-ph},
   adsurl = {https://ui.adsabs.harvard.edu/abs/2019ApJ...873..111I}
}

@article{Chan18,
    author = {Chan, T K and Kereš, D and Wetzel, A and Hopkins, P F and Faucher-Giguère, C-A and El-Badry, K and Garrison-Kimmel, S and Boylan-Kolchin, M},
    title = {The origin of ultra diffuse galaxies: stellar feedback and quenching},
    journal = {MNRAS},
    volume = {478},
    number = {1},
    pages = {906-925},
    year = {2018},
    month = {05},
    issn = {0035-8711},
    doi = {10.1093/MNRAS/sty1153},
    url = {https://doi.org/10.1093/MNRAS/sty1153},
    eprint = {https://academic.oup.com/MNRAS/article-pdf/478/1/906/25013008/sty1153.pdf},
}

@article{Gannon2024,
    author = {Gannon, Jonah S and Ferré-Mateu, Anna and Forbes, Duncan A and Brodie, Jean P and Buzzo, Maria Luisa and Romanowsky, Aaron J},
    title = {A Catalogue and analysis of ultra-diffuse galaxy spectroscopic properties},
    journal = {MNRAS},
    volume = {531},
    number = {1},
    pages = {1856-1869},
    year = {2024},
    month = {05},
    issn = {0035-8711},
    doi = {10.1093/MNRAS/stae1287},
    url = {https://doi.org/10.1093/MNRAS/stae1287},
    eprint = {https://academic.oup.com/MNRAS/article-pdf/531/1/1856/57908506/stae1287.pdf},
}

@article{Greco2018,
doi = {10.3847/1538-4357/aab842},
url = {https://dx.doi.org/10.3847/1538-4357/aab842},
year = {2018},
month = {apr},
publisher = {The American Astronomical Society},
volume = {857},
number = {2},
pages = {104},
author = {Greco, Johnny P. and Greene, Jenny E. and Strauss, Michael A. and Macarthur, Lauren A. and Flowers, Xzavier and Goulding, Andy D. and Huang, Song and Kim, Ji Hoon and Komiyama, Yutaka and Leauthaud, Alexie and Leisman, Lukas and Lupton, Robert H. and Sifón, Cristóbal and Wang, Shiang-Yu},
title = {Illuminating Low Surface Brightness Galaxies with the Hyper Suprime-Cam Survey},
journal = {ApJ}
}

@ARTICLE{Tanoglidis2018,
   author = {{Tanoglidis}, D. and {Drlica-Wagner}, A. and {Wei}, K. and {Li}, T.~S. and {S{\'a}nchez}, J. and {Zhang}, Y. and {Peter}, A.~H.~G. and {Feldmeier-Krause}, A. and {Prat}, J. and {Casey}, K. and {Palmese}, A. and {S{\'a}nchez}, C. and {DeRose}, J. and {Conselice}, C. and {Gagnon}, L. and {Abbott}, T.~M.~C. and {Aguena}, M. and {Allam}, S. and {Avila}, S. and {Bechtol}, K. and {Bertin}, E. and {Bhargava}, S. and {Brooks}, D. and {Burke}, D.~L. and {Rosell}, A. Carnero and {Kind}, M. Carrasco and {Carretero}, J. and {Chang}, C. and {Costanzi}, M. and {da Costa}, L.~N. and {De Vicente}, J. and {Desai}, S. and {Diehl}, H.~T. and {Doel}, P. and {Eifler}, T.~F. and {Everett}, S. and {Evrard}, A.~E. and {Flaugher}, B. and {Frieman}, J. and {Garc{\'\i}a-Bellido}, J. and {Gerdes}, D.~W. and {Gruendl}, R.~A. and {Gschwend}, J. and {Gutierrez}, G. and {Hartley}, W.~G. and {Hollowood}, D.~L. and {Huterer}, D. and {James}, D.~J. and {Krause}, E. and {Kuehn}, K. and {Kuropatkin}, N. and {Maia}, M.~A.~G. and {March}, M. and {Marshall}, J.~L. and {Menanteau}, F. and {Miquel}, R. and {Ogando}, R.~L.~C. and {Paz-Chinch{\'o}n}, F. and {Romer}, A.~K. and {Roodman}, A. and {Sanchez}, E. and {Scarpine}, V. and {Serrano}, S. and {Sevilla-Noarbe}, I. and {Smith}, M. and {Suchyta}, E. and {Tarle}, G. and {Thomas}, D. and {Tucker}, D.~L. and {Walker}, A.~R. and {DES Collaboration}},
    title = "{Shadows in the Dark: Low-surface-brightness Galaxies Discovered in the Dark Energy Survey}",
  journal = {ApJS},
   year = 2021,
    month = feb,
   volume = {252},
   number = {2},
    eid = {18},
    pages = {18},
    doi = {10.3847/1538-4365/abca89},
archivePrefix = {arXiv},
   eprint = {2006.04294},
 primaryClass = {astro-ph.GA},
   adsurl = {https://ui.adsabs.harvard.edu/abs/2021ApJS..252...18T}
}

@ARTICLE{Janowiecki2019,
       author = {{Janowiecki}, Steven and {Jones}, Michael G. and {Leisman}, Lukas and {Webb}, Andrew},
        title = "{The environment of H I-bearing ultra-diffuse galaxies in the ALFALFA survey}",
      journal = {MNRAS},
         year = 2019,
        month = nov,
       volume = {490},
       number = {1},
        pages = {566-577},
          doi = {10.1093/MNRAS/stz1868},
archivePrefix = {arXiv},
       eprint = {1906.11543},
 primaryClass = {astro-ph.GA},
       adsurl = {https://ui.adsabs.harvard.edu/abs/2019MNRAS.490..566J}
}

@article{ Yiping2026,
	author = {{Su, Yiping} and {Yi, Zhenping} and {Li, Sihan} and {Liu, Meng} and {Kong, Xiaoming} and {Bu, Yude} and {Li, Dong}},
	title = {Searching for ultra-diffuse galaxies in the Dark Energy Survey using an object-detection algorithm},
	DOI= "10.1051/0004-6361/202556863",
	url= "https://doi.org/10.1051/0004-6361/202556863",
	journal = {A\&A},
	year = 2026,
	volume = 706,
	pages = "A35",
}

@software{larry_bradley_2025_14889440,
  author       = {Larry Bradley and
                  Brigitta Sip{\H o}cz and
                  Thomas Robitaille and
                  Erik Tollerud and
                  Z\`e Vin{\'{\i}}cius and
                  Christoph Deil and
                  Kyle Barbary and
                  Tom J Wilson and
                  Ivo Busko and
                  Axel Donath and
                  Hans Moritz G{\"u}nther and
                  Mihai Cara and
                  P. L. Lim and
                  Sebastian Me{\ss}linger and
                  Zach Burnett and
                  Simon Conseil and
                  Michael Droettboom and
                  Azalee Bostroem and
                  E. M. Bray and
                  Lars Andersen Bratholm and
                  William Jamieson and
                  Adam Ginsburg and
                  Geert Barentsen and
                  Matt Craig and
                  Sergio Pascual and
                  Shivangee Rathi and
                  Marshall Perrin and
                  Brett M. Morris},
  title        = {astropy/photutils: 2.2.0},
  month        = feb,
  year         = 2025,
  publisher    = {Zenodo},
  version      = {2.2.0},
  doi          = {10.5281/zenodo.14889440},
  url          = {https://doi.org/10.5281/zenodo.14889440},
  swhid        = {swh:1:dir:11159107f27a28985192ed1118b1f2055709d093
                   ;origin=https://doi.org/10.5281/zenodo.596036;visi
                   t=swh:1:snp:ae8c4a55d349d43e53cfe9ce92e678fcfe840f
                   3b;anchor=swh:1:rel:0117f67e8888adcdfc85308287dd9c
                   854b466389;path=astropy-photutils-ffb96c5
                  },
}

@ARTICLE{Jenny,
   author = {{Kadowaki}, Jennifer and {Zaritsky}, Dennis and {Donnerstein}, R.~L. and {RS}, Pranjal and {Karunakaran}, Ananthan and {Spekkens}, Kristine},
    title = "{On the Properties of Spectroscopically Confirmed Ultra-diffuse Galaxies across Environments}",
  journal = {ApJ},
   year = 2021,
    month = dec,
   volume = {923},
   number = {2},
    eid = {257},
    pages = {257},
    doi = {10.3847/1538-4357/ac2948},
archivePrefix = {arXiv},
   eprint = {2110.00015},
 primaryClass = {astro-ph.GA},
   adsurl = {https://ui.adsabs.harvard.edu/abs/2021ApJ...923..257K}
}

@article{Zaritsky_2023,
doi = {10.3847/1538-4365/acdd71},
url = {https://doi.org/10.3847/1538-4365/acdd71},
year = {2023},
month = {jul},
publisher = {The American Astronomical Society},
volume = {267},
number = {2},
pages = {27},
author = {Zaritsky, Dennis and Donnerstein, Richard and Dey, Arjun and Karunakaran, Ananthan and Kadowaki, Jennifer and Khim, Donghyeon J. and Spekkens, Kristine and Zhang, Huanian},
title = {Systematically Measuring Ultra-diffuse Galaxies (SMUDGes). V. The Complete SMUDGes Catalog and the Nature of Ultradiffuse Galaxies},
journal = {ApJS}
}

@ARTICLE{Bellazzini2017,
   author = {{Bellazzini}, M. and {Belokurov}, V. and {Magrini}, L. and {Fraternali}, F. and {Testa}, V. and {Beccari}, G. and {Marchetti}, A. and {Carini}, R.},
    title = "{Redshift, metallicity and size of two extended dwarf Irregular galaxies: a link between dwarf Irregulars and ultra diffuse galaxies?}",
  journal = {MNRAS},
   year = 2017,
    month = may,
   volume = {467},
   number = {3},
    pages = {3751-3758},
    doi = {10.1093/MNRAS/stx236},
archivePrefix = {arXiv},
   eprint = {1701.07632},
 primaryClass = {astro-ph.GA},
   adsurl = {https://ui.adsabs.harvard.edu/abs/2017MNRAS.467.3751B}
}

@article{Sipgi,
  author = {Gargiulo, A and Fumana, M and Bisogni, S and Franzetti, P and Cassarà, L P and Garilli, B and Scodeggio, M and Vietri, G},
  title = {sipgi: an interactive pipeline for spectroscopic data reduction},
  journal = {MNRAS},
  volume = {514},
  number = {2},
  pages = {2902-2914},
  year = {2022},
  month = {04},
  issn = {0035-8711},
  doi = {10.1093/MNRAS/stac1065},
  url = {https://doi.org/10.1093/MNRAS/stac1065},
  eprint = {https://academic.oup.com/MNRAS/article-pdf/514/2/2902/44166546/stac1065.pdf},
}

@ARTICLE{Garilli2010_EZ,
   author = {{Garilli}, B. and {Fumana}, M. and {Franzetti}, P. and {Paioro}, L. and {Scodeggio}, M. and {Le F{\`e}vre}, O. and {Paltani}, S. and {Scaramella}, R.},
    title = "{EZ: A Tool For Automatic Redshift Measurement}",
  journal = {PASP},
   year = 2010,
    month = jul,
   volume = {122},
   number = {893},
    pages = {827},
    doi = {10.1086/654903},
archivePrefix = {arXiv},
   eprint = {1005.2825},
 primaryClass = {astro-ph.IM},
   adsurl = {https://ui.adsabs.harvard.edu/abs/2010PASP..122..827G}
}

@ARTICLE{Guzzo2014_VIPERS,
   author = {{Guzzo}, L. and {Scodeggio}, M. and {Garilli}, B. and {Granett}, B.~R. and {Fritz}, A. and {Abbas}, U. and {Adami}, C. and {Arnouts}, S. and {Bel}, J. and {Bolzonella}, M. and {Bottini}, D. and {Branchini}, E. and {Cappi}, A. and {Coupon}, J. and {Cucciati}, O. and {Davidzon}, I. and {De Lucia}, G. and {de la Torre}, S. and {Franzetti}, P. and {Fumana}, M. and {Hudelot}, P. and {Ilbert}, O. and {Iovino}, A. and {Krywult}, J. and {Le Brun}, V. and {Le F{\`e}vre}, O. and {Maccagni}, D. and {Ma{\l}ek}, K. and {Marulli}, F. and {McCracken}, H.~J. and {Paioro}, L. and {Peacock}, J.~A. and {Polletta}, M. and {Pollo}, A. and {Schlagenhaufer}, H. and {Tasca}, L.~A.~M. and {Tojeiro}, R. and {Vergani}, D. and {Zamorani}, G. and {Zanichelli}, A. and {Burden}, A. and {Di Porto}, C. and {Marchetti}, A. and {Marinoni}, C. and {Mellier}, Y. and {Moscardini}, L. and {Nichol}, R.~C. and {Percival}, W.~J. and {Phleps}, S. and {Wolk}, M.},
    title = "{The VIMOS Public Extragalactic Redshift Survey (VIPERS). An unprecedented view of galaxies and large-scale structure at 0.5 < z < 1.2}",
  journal = {A\&A},
   year = 2014,
    month = jun,
   volume = {566},
    eid = {A108},
    pages = {A108},
    doi = {10.1051/0004-6361/201321489},
archivePrefix = {arXiv},
   eprint = {1303.2623},
 primaryClass = {astro-ph.CO},
   adsurl = {https://ui.adsabs.harvard.edu/abs/2014A\&A...566A.108G}
}

@ARTICLE{LeFevre2005_VVDS,
   author = {{Le F{\`e}vre}, O. and {Vettolani}, G. and {Garilli}, B. and {Tresse}, L. and {Bottini}, D. and {Le Brun}, V. and {Maccagni}, D. and {Picat}, J.~P. and {Scaramella}, R. and {Scodeggio}, M. and {Zanichelli}, A. and {Adami}, C. and {Arnaboldi}, M. and {Arnouts}, S. and {Bardelli}, S. and {Bolzonella}, M. and {Cappi}, A. and {Charlot}, S. and {Ciliegi}, P. and {Contini}, T. and {Foucaud}, S. and {Franzetti}, P. and {Gavignaud}, I. and {Guzzo}, L. and {Ilbert}, O. and {Iovino}, A. and {McCracken}, H.~J. and {Marano}, B. and {Marinoni}, C. and {Mathez}, G. and {Mazure}, A. and {Meneux}, B. and {Merighi}, R. and {Paltani}, S. and {Pell{\`o}}, R. and {Pollo}, A. and {Pozzetti}, L. and {Radovich}, M. and {Zamorani}, G. and {Zucca}, E. and {Bondi}, M. and {Bongiorno}, A. and {Busarello}, G. and {Lamareille}, F. and {Mellier}, Y. and {Merluzzi}, P. and {Ripepi}, V. and {Rizzo}, D.},
    title = "{The VIMOS VLT deep survey. First epoch VVDS-deep survey: 11 564 spectra with 17.5 {\ensuremath{\leq}} IAB {\ensuremath{\leq}} 24, and the redshift distribution over 0 {\ensuremath{\leq}} z {\ensuremath{\leq}} 5}",
  journal = {A\&A},
   year = 2005,
    month = sep,
   volume = {439},
   number = {3},
    pages = {845-862},
    doi = {10.1051/0004-6361:20041960},
archivePrefix = {arXiv},
   eprint = {astro-ph/0409133},
 primaryClass = {astro-ph},
   adsurl = {https://ui.adsabs.harvard.edu/abs/2005A\&A...439..845L}
}

@ARTICLE{Lilly2007_zcosmos,
   author = {{Lilly}, S.~J. and {Le F{\`e}vre}, O. and {Renzini}, A. and {Zamorani}, G. and {Scodeggio}, M. and {Contini}, T. and {Carollo}, C.~M. and {Hasinger}, G. and {Kneib}, J. -P. and {Iovino}, A. and {Le Brun}, V. and {Maier}, C. and {Mainieri}, V. and {Mignoli}, M. and {Silverman}, J. and {Tasca}, L.~A.~M. and {Bolzonella}, M. and {Bongiorno}, A. and {Bottini}, D. and {Capak}, P. and {Caputi}, K. and {Cimatti}, A. and {Cucciati}, O. and {Daddi}, E. and {Feldmann}, R. and {Franzetti}, P. and {Garilli}, B. and {Guzzo}, L. and {Ilbert}, O. and {Kampczyk}, P. and {Kovac},, K. and {Lamareille}, F. and {Leauthaud}, A. and {Le Borgne}, J. -F. and {McCracken}, H.~J. and {Marinoni}, C. and {Pello}, R. and {Ricciardelli}, E. and {Scarlata}, C. and {Vergani}, D. and {Sanders}, D.~B. and {Schinnerer}, E. and {Scoville}, N. and {Taniguchi}, Y. and {Arnouts}, S. and {Aussel}, H. and {Bardelli}, S. and {Brusa}, M. and {Cappi}, A. and {Ciliegi}, P. and {Finoguenov}, A. and {Foucaud}, S. and {Franceschini}, A. and {Halliday}, C. and {Impey}, C. and {Knobel}, C. and {Koekemoer}, A. and {Kurk}, J. and {Maccagni}, D. and {Maddox}, S. and {Marano}, B. and {Marconi}, G. and {Meneux}, B. and {Mobasher}, B. and {Moreau}, C. and {Peacock}, J.~A. and {Porciani}, C. and {Pozzetti}, L. and {Scaramella}, R. and {Schiminovich}, D. and {Shopbell}, P. and {Smail}, I. and {Thompson}, D. and {Tresse}, L. and {Vettolani}, G. and {Zanichelli}, A. and {Zucca}, E.},
    title = "{zCOSMOS: A Large VLT/VIMOS Redshift Survey Covering 0 < z < 3 in the COSMOS Field}",
  journal = {ApJS},
   year = 2007,
    month = sep,
   volume = {172},
   number = {1},
    pages = {70-85},
    doi = {10.1086/516589},
archivePrefix = {arXiv},
   eprint = {astro-ph/0612291},
 primaryClass = {astro-ph},
   adsurl = {https://ui.adsabs.harvard.edu/abs/2007ApJS..172...70L}
}

@ARTICLE{Polletta2023,
   author = {{Polletta}, M. and {Nonino}, M. and {Frye}, B. and {Gargiulo}, A. and {Bisogni}, S. and {Garuda}, N. and {Thompson}, D. and {Lehnert}, M. and {Pascale}, M. and {Willner}, S.~P. and {Kamieneski}, P. and {Leimbach}, R. and {Cheng}, C. and {Coe}, D. and {Cohen}, S.~H. and {Conselice}, C.~J. and {Dai}, L. and {Diego}, J. and {Dole}, H. and {Driver}, S.~P. and {D'Silva}, J.~C.~J. and {Fontana}, A. and {Foo}, N. and {Furtak}, L.~J. and {Grogin}, N.~A. and {Harrington}, K. and {Hathi}, N.~P. and {Jansen}, R.~A. and {Kelly}, P. and {Koekemoer}, A.~M. and {Mancini}, C. and {Marshall}, M.~A. and {Pierel}, J.~D.~R. and {Pirzkal}, N. and {Robotham}, A. and {Rutkowski}, M.~J. and {Ryan}, R.~E. and {Snigula}, J.~M. and {Summers}, J. and {Tompkins}, S. and {Willmer}, C.~N.~A. and {Windhorst}, R.~A. and {Yan}, H. and {Yun}, M.~S. and {Zitrin}, A.},
    title = "{Spectroscopy of the supernova H0pe host galaxy at redshift 1.78}",
  journal = {A\&A},
   year = 2023,
    month = jul,
   volume = {675},
    eid = {L4},
    pages = {L4},
    doi = {10.1051/0004-6361/202346964},
archivePrefix = {arXiv},
   eprint = {2306.12385},
 primaryClass = {astro-ph.GA},
   adsurl = {https://ui.adsabs.harvard.edu/abs/2023A\&A...675L...4P}
}

@ARTICLE{Rabitz2017,
   author = {{Rabitz}, A. and {Lamer}, G. and {Schwope}, A. and {Takey}, A.},
    title = "{Distant clusters of galaxies in the 2XMM/SDSS footprint: follow-up observations with the LBT}",
  journal = {A\&A},
   year = 2017,
    month = nov,
   volume = {607},
    eid = {A56},
    pages = {A56},
    doi = {10.1051/0004-6361/201731128},
archivePrefix = {arXiv},
   eprint = {1709.02436},
 primaryClass = {astro-ph.CO},
   adsurl = {https://ui.adsabs.harvard.edu/abs/2017A\&A...607A..56R}
}

@ARTICLE{CIGALE_Burgarella2005,
   author = {{Burgarella}, D. and {Buat}, V. and {Iglesias-P{\'a}ramo}, J.},
    title = "{Star formation and dust attenuation properties in galaxies from a statistical ultraviolet-to-far-infrared analysis}",
  journal = {MNRAS},
   year = 2005,
    month = jul,
   volume = {360},
   number = {4},
    pages = {1413-1425},
    doi = {10.1111/j.1365-2966.2005.09131.x},
archivePrefix = {arXiv},
   eprint = {astro-ph/0504434},
 primaryClass = {astro-ph},
   adsurl = {https://ui.adsabs.harvard.edu/abs/2005MNRAS.360.1413B}
}

@ARTICLE{CIGALE_Noll2009,
   author = {{Noll}, S. and {Burgarella}, D. and {Giovannoli}, E. and {Buat}, V. and
   {Marcillac}, D. and {Mu{\~n}oz-Mateos}, J.~C.},
    title = "{Analysis of galaxy spectral energy distributions from far-UV to far-IR with CIGALE: studying a SINGS test sample}",
  journal = {A\&A},
   year = 2009,
    month = dec,
   volume = {507},
   number = {3},
    pages = {1793-1813},
    doi = {10.1051/0004-6361/200912497},
archivePrefix = {arXiv},
   eprint = {0909.5439},
 primaryClass = {astro-ph.CO},
   adsurl = {https://ui.adsabs.harvard.edu/abs/2009A\&A...507.1793N}
}

@ARTICLE{CIGALE_Boquien2019,
   author = {{Boquien}, M. and {Burgarella}, D. and {Roehlly}, Y. and {Buat}, V. and
   {Ciesla}, L. and {Corre}, D. and {Inoue}, A.~K. and {Salas}, H.},
    title = "{CIGALE: a python Code Investigating GALaxy Emission}",
  journal = {A\&A},
   year = 2019,
    month = feb,
   volume = {622},
    eid = {A103},
    pages = {A103},
    doi = {10.1051/0004-6361/201834156},
archivePrefix = {arXiv},
   eprint = {1811.03094},
 primaryClass = {astro-ph.GA},
   adsurl = {https://ui.adsabs.harvard.edu/abs/2019A\&A...622A.103B}
}

@INCOLLECTION{Samuel,
   author = {{Boissier}, Samuel},
    title = "{Star Formation in Galaxies}",
  publisher = {The Astronomical Society},
  booktitle = {Planets, Stars and Stellar Systems. Volume 6: Extragalactic Astronomy and Cosmology},
   year = 2013,
   editor = {{Oswalt}, Terry D. and {Keel}, William C.},
   volume = {6},
    pages = {141},
    doi = {10.1007/978-94-007-5609-0_3},
   adsurl = {https://ui.adsabs.harvard.edu/abs/2013pss6.book..141B}
}

@ARTICLE{CIGALE_Malek2018,
   author = {{Ma{\l}ek}, K. and {Buat}, V. and {Roehlly}, Y. and {Burgarella}, D. and {Hurley}, P.~D. and {Shirley}, R. and {Duncan}, K. and {Efstathiou}, A. and {Papadopoulos}, A. and {Vaccari}, M. and {Farrah}, D. and {Marchetti}, L. and {Oliver}, S.},
    title = "{HELP: modelling the spectral energy distributions of Herschel detected galaxies in the ELAIS N1 field}",
  journal = {A\&A},
   year = 2018,
    month = nov,
   volume = {620},
    eid = {A50},
    pages = {A50},
    doi = {10.1051/0004-6361/201833131},
archivePrefix = {arXiv},
   eprint = {1809.00529},
 primaryClass = {astro-ph.GA},
   adsurl = {https://ui.adsabs.harvard.edu/abs/2018A\&A...620A..50M}
}

@article{Zhong2008,
  author = {Zhong, G. H. and Liang, Y. C. and Liu, F. S. and Hammer, F. and Hu, J. Y. and Chen, X. Y. and Deng, L. C. and Zhang, B.},
  title = {A large sample of low surface brightness disc galaxies from the SDSS – I. The sample and the stellar populations},
  journal = {MNRAS},
  volume = {391},
  number = {2},
  pages = {986-999},
  year = {2008},
  month = {11},
  issn = {0035-8711},
  doi = {10.1111/j.1365-2966.2008.13972.x},
  url = {https://doi.org/10.1111/j.1365-2966.2008.13972.x},
  eprint = {https://academic.oup.com/MNRAS/article-pdf/391/2/986/5778848/MNRAS0391-0986.pdf},
}

@article{Rong_2020,
doi = {10.3847/2041-8213/aba8aa},
url = {https://dx.doi.org/10.3847/2041-8213/aba8aa},
year = {2020},
month = {aug},
publisher = {The American Astronomical Society},
volume = {899},
number = {1},
pages = {L12},
author = {Rong, Yu and Zhu, Kai and Johnston, Evelyn J. and Zhang, Hong-Xin and Cao, Tianwen and Puzia, Thomas H. and Galaz, Gaspar},
title = {Lessons on Star-forming Ultra-diffuse Galaxies from the Stacked Spectra of the Sloan Digital Sky Survey},
journal = {ApJL}
}

@article{LaMarca,
	author = {{La Marca} and {Iodice, Enrichetta} and {Cantiello, Michele} and {Forbes, Duncan A.} and {Rejkuba, Marina} and {Hilker, Michael} and {Arnaboldi, Magda} and {Greggio, Laura} and {Spiniello, Chiara} and {Mieske, Steffen} and {Venhola, Aku} and {Spavone, Marilena} and {D’Ago, Giuseppe} and {Raj, Maria Angela} and {Ragusa, Rossella} and {Mirabile, Marco} and {Rampazzo, Roberto} and {Peletier, Reynier} and {Paolillo, Maurizio} and {Challapa, Nelvy Choque} and {Schipani, Pietro}},
	title = {Galaxy populations in the Hydra I cluster from the VEGAS survey - II. The ultra-diffuse galaxy population},
	DOI= "10.1051/0004-6361/202142367",
	url= "https://doi.org/10.1051/0004-6361/202142367",
	journal = {A\&A},
	year = 2022,
	volume = 665,
	pages = "A105",
}

@article{Lim_2020,
doi = {10.3847/1538-4357/aba433},
url = {https://dx.doi.org/10.3847/1538-4357/aba433},
year = {2020},
month = {aug},
publisher = {The American Astronomical Society},
volume = {899},
number = {1},
pages = {69},
author = {Lim, Sungsoon and Côté, Patrick and Peng, Eric W. and Ferrarese, Laura and Roediger, Joel C. and Durrell, Patrick R. and Mihos, J. Christopher and Wang, Kaixiang and Gwyn, S. D. J. and Cuillandre, Jean-Charles and Liu, Chengze and Sánchez-Janssen, Rubén and Toloba, Elisa and Sales, Laura V. and Guhathakurta, Puragra and Lançon, Ariane and Puzia, Thomas H.},
title = {The Next Generation Virgo Cluster Survey (NGVS). XXX. Ultra-diffuse Galaxies and Their Globular Cluster Systems},
journal = {ApJ}
}

@article{Abazajian_2009,
doi = {10.1088/0067-0049/182/2/543},
url = {https://doi.org/10.1088/0067-0049/182/2/543},
year = {2009},
month = {may},
publisher = {The American Astronomical Society},
volume = {182},
number = {2},
pages = {543},
author = {Abazajian, Kevork N. and Adelman-McCarthy, Jennifer K. and Agüeros, Marcel A. and Allam, Sahar S. and Prieto, Carlos Allende and An, Deokkeun and Anderson, Kurt S. J. and Anderson, Scott F. and Annis, James and Bahcall, Neta A. and Bailer-Jones, C. A. L. and Barentine, J. C. and Bassett, Bruce A. and Becker, Andrew C. and Beers, Timothy C. and Bell, Eric F. and Belokurov, Vasily and Berlind, Andreas A. and Berman, Eileen F. and Bernardi, Mariangela and Bickerton, Steven J. and Bizyaev, Dmitry and Blakeslee, John P. and Blanton, Michael R. and Bochanski, John J. and Boroski, William N. and Brewington, Howard J. and Brinchmann, Jarle and Brinkmann, J. and Brunner, Robert J. and Budavári, Tamás and Carey, Larry N. and Carliles, Samuel and Carr, Michael A. and Castander, Francisco J. and Cinabro, David and Connolly, A. J. and Csabai, István and Cunha, Carlos E. and Czarapata, Paul C. and Davenport, James R. A. and de Haas, Ernst and Dilday, Ben and Doi, Mamoru and Eisenstein, Daniel J. and Evans, Michael L. and Evans, N. W. and Fan, Xiaohui and Friedman, Scott D. and Frieman, Joshua A. and Fukugita, Masataka and Gänsicke, Boris T. and Gates, Evalyn and Gillespie, Bruce and Gilmore, G. and Gonzalez, Belinda and Gonzalez, Carlos F. and Grebel, Eva K. and Gunn, James E. and Györy, Zsuzsanna and Hall, Patrick B. and Harding, Paul and Harris, Frederick H. and Harvanek, Michael and Hawley, Suzanne L. and Hayes, Jeffrey J. E. and Heckman, Timothy M. and Hendry, John S. and Hennessy, Gregory S. and Hindsley, Robert B. and Hoblitt, J. and Hogan, Craig J. and Hogg, David W. and Holtzman, Jon A. and Hyde, Joseph B. and Ichikawa, Shin-ichi and Ichikawa, Takashi and Im, Myungshin and Ivezić, Željko and Jester, Sebastian and Jiang, Linhua and Johnson, Jennifer A. and Jorgensen, Anders M. and Jurić, Mario and Kent, Stephen M. and Kessler, R. and Kleinman, S. J. and Knapp, G. R. and Konishi, Kohki and Kron, Richard G. and Krzesinski, Jurek and Kuropatkin, Nikolay and Lampeitl, Hubert and Lebedeva, Svetlana and Lee, Myung Gyoon and Lee, Young Sun and Leger, R. French and Lépine, Sébastien and Li, Nolan and Lima, Marcos and Lin, Huan and Long, Daniel C. and Loomis, Craig P. and Loveday, Jon and Lupton, Robert H. and Magnier, Eugene and Malanushenko, Olena and Malanushenko, Viktor and Mandelbaum, Rachel and Margon, Bruce and Marriner, John P. and Martínez-Delgado, David and Matsubara, Takahiko and McGehee, Peregrine M. and McKay, Timothy A. and Meiksin, Avery and Morrison, Heather L. and Mullally, Fergal and Munn, Jeffrey A. and Murphy, Tara and Nash, Thomas and Nebot, Ada and Neilsen, Eric H. and Newberg, Heidi Jo and Newman, Peter R. and Nichol, Robert C. and Nicinski, Tom and Nieto-Santisteban, Maria and Nitta, Atsuko and Okamura, Sadanori and Oravetz, Daniel J. and Ostriker, Jeremiah P. and Owen, Russell and Padmanabhan, Nikhil and Pan, Kaike and Park, Changbom and Pauls, George and Peoples, John and Percival, Will J. and Pier, Jeffrey R. and Pope, Adrian C. and Pourbaix, Dimitri and Price, Paul A. and Purger, Norbert and Quinn, Thomas and Raddick, M. Jordan and Fiorentin, Paola Re and Richards, Gordon T. and Richmond, Michael W. and Riess, Adam G. and Rix, Hans-Walter and Rockosi, Constance M. and Sako, Masao and Schlegel, David J. and Schneider, Donald P. and Scholz, Ralf-Dieter and Schreiber, Matthias R. and Schwope, Axel D. and Seljak, Uroš and Sesar, Branimir and Sheldon, Erin and Shimasaku, Kazu and Sibley, Valena C. and Simmons, A. E. and Sivarani, Thirupathi and Smith, J. Allyn and Smith, Martin C. and Smolčić, Vernesa and Snedden, Stephanie A. and Stebbins, Albert and Steinmetz, Matthias and Stoughton, Chris and Strauss, Michael A. and SubbaRao, Mark and Suto, Yasushi and Szalay, Alexander S. and Szapudi, István and Szkody, Paula and Tanaka, Masayuki and Tegmark, Max and Teodoro, Luis F. A. and Thakar, Aniruddha R. and Tremonti, Christy A. and Tucker, Douglas L. and Uomoto, Alan and Vanden Berk, Daniel E. and Vandenberg, Jan and Vidrih, S. and Vogeley, Michael S. and Voges, Wolfgang and Vogt, Nicole P. and Wadadekar, Yogesh and Watters, Shannon and Weinberg, David H. and West, Andrew A. and White, Simon D. M. and Wilhite, Brian C. and Wonders, Alainna C. and Yanny, Brian and Yocum, D. R. and York, Donald G. and Zehavi, Idit and Zibetti, Stefano and Zucker, Daniel B.},
title = {SEVENTH DR OF THE SDSS},
journal = {ApJS}
}

@article{Mihos_2015,
doi = {10.1088/2041-8205/809/2/L21},
url = {https://dx.doi.org/10.1088/2041-8205/809/2/L21},
year = {2015},
month = {aug},
publisher = {The American Astronomical Society},
volume = {809},
number = {2},
pages = {L21},
author = {Mihos, J. Christopher and Durrell, Patrick R. and Ferrarese, Laura and Feldmeier, John J. and Côté, Patrick and Peng, Eric W. and Harding, Paul and Liu, Chengze and Gwyn, Stephen and Cuillandre, Jean-Charles},
title = {GALAXIES AT THE EXTREMES: ULTRA-DIFFUSE GALAXIES IN THE VIRGO CLUSTER},
journal = {ApJL}
}

@article{Koda_2015,
doi = {10.1088/2041-8205/807/1/L2},
url = {https://dx.doi.org/10.1088/2041-8205/807/1/L2},
year = {2015},
month = {jun},
publisher = {The American Astronomical Society},
volume = {807},
number = {1},
pages = {L2},
author = {Koda, Jin and Yagi, Masafumi and Yamanoi, Hitomi and Komiyama, Yutaka},
title = {APPROXIMATELY A THOUSAND ULTRA-DIFFUSE GALAXIES IN THE COMA CLUSTER},
journal = {ApJL}
}

@article{vanderburg2016udg,
  author = {van der Burg, R. F. J. and Muzzin, A. and Hoekstra, H.},
  title = {The abundance and spatial distribution of ultra‑diffuse galaxies in nearby galaxy clusters},
  journal = {A\&A},
  volume = {590},
  pages = {A20},
  year = {2016},
  doi = {...}
}

@ARTICLE{Kado-Fong,
   author = {{Kado-Fong}, Erin and {Greene}, Jenny E. and {Huang}, Song and {Goulding}, Andy},
    title = "{Ultra-diffuse Galaxies as Extreme Star-forming Environments. I. Mapping Star Formation in H I-rich UDGs}",
  journal = {ApJ},
   year = 2022,
    month = dec,
   volume = {941},
   number = {1},
    eid = {11},
    pages = {11},
    doi = {10.3847/1538-4357/ac9964},
archivePrefix = {arXiv},
   eprint = {2209.05492},
 primaryClass = {astro-ph.GA},
   adsurl = {https://ui.adsabs.harvard.edu/abs/2022ApJ...941...11K}
}

@ARTICLE{Leroy,
   author = {{Leroy}, Adam K. and {Walter}, Fabian and {Brinks}, Elias and {Bigiel}, Frank and {de Blok}, W.~J.~G. and {Madore}, Barry and {Thornley}, M.~D.},
    title = "{The Star Formation Efficiency in Nearby Galaxies: Measuring Where Gas Forms Stars Effectively}",
  journal = {ApJ},
   year = 2008,
    month = dec,
   volume = {136},
   number = {6},
    pages = {2782-2845},
    doi = {10.1088/0004-6256/136/6/2782},
archivePrefix = {arXiv},
   eprint = {0810.2556},
 primaryClass = {astro-ph},
   adsurl = {https://ui.adsabs.harvard.edu/abs/2008AJ....136.2782L}
}

@article{ Xu,
	author = {{Xu} and {Ramos-Ceja, Miriam E.} and {Pacaud, Florian} and {Reiprich, Thomas H.} and {Erben, Thomas}},
	title = {Catalog of X-ray-selected extended galaxy clusters from the ROSAT All-Sky Survey (RXGCC)⋆⋆⋆},
	DOI= "10.1051/0004-6361/202140908",
	url= "https://doi.org/10.1051/0004-6361/202140908",
	journal = {A\&A},
	year = 2022,
	volume = 658,
	pages = "A59",
}

@ARTICLE{Cardona-Barrero,
   author = {{Cardona-Barrero}, Salvador and {Di Cintio}, Arianna and {Brook}, Christopher B.~A. and {Ruiz-Lara}, Tomas and {Beasley}, Michael A. and {Falc{\'o}n-Barroso}, Jesus and {Macci{\`o}}, Andrea V.},
    title = "{NIHAO XXIV: rotation- or pressure-supported systems? Simulated Ultra Diffuse Galaxies show a broad distribution in their stellar kinematics}",
  journal = {MNRAS},
   year = 2020,
    month = oct,
   volume = {497},
   number = {4},
    pages = {4282-4292},
    doi = {10.1093/MNRAS/staa2094},
archivePrefix = {arXiv},
   eprint = {2004.09535},
 primaryClass = {astro-ph.GA},
   adsurl = {https://ui.adsabs.harvard.edu/abs/2020MNRAS.497.4282C}
}

@ARTICLE{Di_cintio2017,
   author = {{Di Cintio}, Arianna and {Brook}, Chris B. and {Dutton}, Aaron A. and {Macci{\`o}}, Andrea V. and {Obreja}, Aura and {Dekel}, Avishai},
    title = "{NIHAO - XI. Formation of ultra-diffuse galaxies by outflows}",
  journal = {MNRAS},
   year = 2017,
    month = mar,
   volume = {466},
   number = {1},
    pages = {L1-L6},
    doi = {10.1093/MNRASl/slw210},
archivePrefix = {arXiv},
   eprint = {1608.01327},
 primaryClass = {astro-ph.GA},
   adsurl = {https://ui.adsabs.harvard.edu/abs/2017MNRAS.466L...1D}
}

@article{euclid2023,
author = {{Euclid Collaboration} and Scaramella, R.      and Amiaux, J. and others},
  title = "{Euclid preparation. XXXI. The Euclid mission and science objectives}",
  journal = {A\&A},
  volume = {677},
  pages = {A11},
  year = {2023},
  doi = {10.1051/0004-6361/202346347}
}

@article{Prole_2019,
  author = {Prole, D J and van der Burg, R F J and Hilker, M and Davies, J I},
  title = {Observational properties of ultra-diffuse galaxies in low-density environments: field UDGs are predominantly blue and star forming},
  journal = {MNRAS},
  volume = {488},
  number = {2},
  pages = {2143-2157},
  year = {2019},
  month = {07},
  issn = {0035-8711},
  doi = {10.1093/MNRAS/stz1843},
  url = {https://doi.org/10.1093/MNRAS/stz1843},
  eprint = {https://academic.oup.com/MNRAS/article-pdf/488/2/2143/28967628/stz1843.pdf},
}

@article{ Marleau_2021,
	author = {{Marleau} and {Habas, Rebecca} and {Poulain, Mélina} and {Duc, Pierre-Alain} and {Müller, Oliver} and {Lim, Sungsoon} and {Durrell, Patrick R.} and {Sánchez-Janssen, Rubén} and {Paudel, Sanjaya} and {Lammim Ahad, Syeda} and {Chougule, Abhishek} and {Bílek, Michal} and {Fensch, Jérémy}},
	title = {Ultra diffuse galaxies in the MATLAS low-to-moderate density fields⋆},
	DOI= "10.1051/0004-6361/202141432",
	url= "https://doi.org/10.1051/0004-6361/202141432",
	journal = {A\&A},
	year = 2021,
	volume = 654,
	pages = "A105",
}

@article{Cohen_2018,
doi = {10.3847/1538-4357/aae7c8},
url = {https://dx.doi.org/10.3847/1538-4357/aae7c8},
year = {2018},
month = {nov},
publisher = {The American Astronomical Society},
volume = {868},
number = {2},
pages = {96},
author = {Cohen, Yotam and Dokkum, Pieter van and Danieli, Shany and Romanowsky, Aaron J. and Abraham, Roberto and Merritt, Allison and Zhang, Jielai and Mowla, Lamiya and Kruijssen, J. M. Diederik and Conroy, Charlie and Wasserman, Asher},
title = {The Dragonfly Nearby Galaxies Survey. V. HST/ACS Observations of 23 Low Surface Brightness Objects in the Fields of NGC 1052, NGC 1084, M96, and NGC 4258},
journal = {ApJ}
}

@article{Leisman_2017,
doi = {10.3847/1538-4357/aa7575},
url = {https://dx.doi.org/10.3847/1538-4357/aa7575},
year = {2017},
month = {jun},
publisher = {The American Astronomical Society},
volume = {842},
number = {2},
pages = {133},
author = {Leisman, Lukas and Haynes, Martha P. and Janowiecki, Steven and Hallenbeck, Gregory and Józsa, Gyula and Giovanelli, Riccardo and Adams, Elizabeth A. K. and Neira, David Bernal and Cannon, John M. and Janesh, William F. and Rhode, Katherine L. and Salzer, John J.},
title = {(Almost) Dark Galaxies in the ALFALFA Survey: Isolated H i-bearing Ultra-diffuse Galaxies},
journal = {ApJ}
}

@article{Blanton_2005,
doi = {10.1086/431416},
url = {https://dx.doi.org/10.1086/431416},
year = {2005},
month = {sep},
publisher = {},
volume = {631},
number = {1},
pages = {208},
author = {Blanton, Michael R. and Lupton, Robert H. and Schlegel, David J. and Strauss, Michael A. and Brinkmann, J. and Fukugita, Masataka and Loveday, Jon},
title = {The Properties and Luminosity Function of Extremely Low Luminosity Galaxies*},
journal = {ApJ}
}

@ARTICLE{CIGALE_Buat2014,
   author = {{Buat}, V. and {Heinis}, S. and {Boquien}, M. and {Burgarella}, D. and {Charmandaris}, V. and {Boissier}, S. and {Boselli}, A. and {Le Borgne}, D. and {Morrison}, G.},
    title = "{Ultraviolet to infrared emission of z > 1 galaxies: Can we derive reliable star formation rates and stellar masses?}",
  journal = {A\&A},
   year = 2014,
    month = jan,
   volume = {561},
    eid = {A39},
    pages = {A39},
    doi = {10.1051/0004-6361/201322081},
archivePrefix = {arXiv},
   eprint = {1310.7712},
 primaryClass = {astro-ph.CO},
   adsurl = {https://ui.adsabs.harvard.edu/abs/2014A\&A...561A..39B}
}

@article{Driver2005,
  author = {Driver, S. P. and Liske, J. and Cross, N. J. G. and De Propris, R. and Allen, P. D.},
  title = {The Millennium Galaxy Catalogue: the space density and surface-brightness distribution(s) of galaxies},
  journal = {MNRAS},
  volume = {360},
  number = {1},
  pages = {81-103},
  year = {2005},
  month = {06},
  issn = {0035-8711},
  doi = {10.1111/j.1365-2966.2005.08990.x},
  url = {https://doi.org/10.1111/j.1365-2966.2005.08990.x},
  eprint = {https://academic.oup.com/MNRAS/article-pdf/360/1/81/5968223/360-1-81.pdf},
}

@ARTICLE{CIGALE_Ciesla2016,
   author = {{Ciesla}, L. and {Boselli}, A. and {Elbaz}, D. and {Boissier}, S. and {Buat}, V. and {Charmandaris}, V. and {Schreiber}, C. and {B{\'e}thermin}, M. and {Baes}, M. and {Boquien}, M. and {De Looze}, I. and {Fern{\'a}ndez-Ontiveros}, J.~A. and {Pappalardo}, C. and {Spinoglio}, L. and {Viaene}, S.},
    title = "{The imprint of rapid star formation quenching on the spectral energy distributions of galaxies}",
  journal = {A\&A},
   year = 2016,
    month = jan,
   volume = {585},
    eid = {A43},
    pages = {A43},
    doi = {10.1051/0004-6361/201527107},
archivePrefix = {arXiv},
   eprint = {1510.07657},
 primaryClass = {astro-ph.GA},
   adsurl = {https://ui.adsabs.harvard.edu/abs/2016A\&A...585A..43C}
}

@ARTICLE{BC2003,
   author = {{Bruzual}, G. and {Charlot}, S.},
    title = "{Stellar population synthesis at the resolution of 2003}",
  journal = {MNRAS},
   year = 2003,
    month = oct,
   volume = {344},
   number = {4},
    pages = {1000-1028},
    doi = {10.1046/j.1365-8711.2003.06897.x},
archivePrefix = {arXiv},
   eprint = {astro-ph/0309134},
 primaryClass = {astro-ph},
   adsurl = {https://ui.adsabs.harvard.edu/abs/2003MNRAS.344.1000B}
}

@ARTICLE{Chabrier2003,
   author = {{Chabrier}, Gilles},
    title = "{Galactic Stellar and Substellar Initial Mass Function}",
  journal = {PASP},
   year = 2003,
    month = jul,
   volume = {115},
   number = {809},
    pages = {763-795},
    doi = {10.1086/376392},
archivePrefix = {arXiv},
   eprint = {astro-ph/0304382},
 primaryClass = {astro-ph},
   adsurl = {https://ui.adsabs.harvard.edu/abs/2003PASP..115..763C}
}

@ARTICLE{CF2000,
   author = {{Charlot}, St{\'e}phane and {Fall}, S. Michael},
    title = "{A Simple Model for the Absorption of Starlight by Dust in Galaxies}",
  journal = {ApJ},
   year = 2000,
    month = aug,
   volume = {539},
   number = {2},
    pages = {718-731},
    doi = {10.1086/309250},
archivePrefix = {arXiv},
   eprint = {astro-ph/0003128},
 primaryClass = {astro-ph},
   adsurl = {https://ui.adsabs.harvard.edu/abs/2000ApJ...539..718C}
}

@article{Junais23,
	author ={{Junais} and {Małek, K.} and {Boissier, S.} and {Pearson, W. J.} and {Pollo, A.} and {Boselli, A.} and {Boquien, M.} and {Donevski, D.} and {Goto, T.} and {Hamed, M.} and {Kim, S. J.} and {Koda, J.} and {Matsuhara, H.} and {Riccio, G.} and {Romano, M.}},
	title = {Variation in optical and infrared properties of galaxies in relation to their surface brightness⋆},
	DOI= "10.1051/0004-6361/202346528",
	url= "https://doi.org/10.1051/0004-6361/202346528",
	journal = {A\&A},
	year = 2023,
	volume = 676,
	pages = "A41"
}

@ARTICLE{Rahman,
   author = {{Rahman}, Nurur and {Howell}, Justin H. and {Helou}, George and {Mazzarella}, Joseph M. and {Buckalew}, Brent},
    title = "{Exploring Infrared Properties of Giant Low Surface Brightness Galaxies}",
  journal = {ApJ},
   year = 2007,
    month = jul,
   volume = {663},
   number = {2},
    pages = {908-923},
    doi = {10.1086/518554},
archivePrefix = {arXiv},
   eprint = {0704.1483},
 primaryClass = {astro-ph},
   adsurl = {https://ui.adsabs.harvard.edu/abs/2007ApJ...663..908R}
}

@article{ Junais,
	author = {{Junais} and {Weilbacher, P. M.} and {Epinat, B.} and {Boissier, S.} and {Galaz, G.} and {Johnston, E. J.} and {Puzia, T. H.} and {Amram, P.} and {Małek, K.}},
	title = {MUSE observations of the giant low surface brightness galaxy Malin 1: Numerous HII regions, star formation rate, metallicity, and dust attenuation⋆},
	DOI= "10.1051/0004-6361/202347669",
	url= "https://doi.org/10.1051/0004-6361/202347669",
	journal = {A\&A},
	year = 2024,
	volume = 681,
	pages = "A100",
}

@BOOK{Osterbrock,
   author = {{Osterbrock}, Donald E.},
   publisher = {},
    title = "{Astrophysics of gaseous nebulae and active galactic nuclei}",
   year = 1989,
   adsurl = {https://ui.adsabs.harvard.edu/abs/1989agna.book.....O}
}

@ARTICLE{Shields,
       author = {{Shields}, Tristen and {Rieke}, Marcia and {Hainline}, Kevin and {Helton}, Jakob M. and {Bunker}, Andrew J. and {Carreira}, Courtney and {Curtis-Lake}, Emma and {Eisenstein}, Daniel J. and {Johnson}, Benjamin D. and {Rinaldi}, Pierluigi and {Robertson}, Brant and {Williams}, Christina C. and {Willmer}, Christopher N.~A. and {Sun}, Yang},
        title = "{JADES: low surface brightness galaxies at 0.4 < z < 0.8 in GOODS-S}",
      journal = {MNRAS},
         year = 2026,
        month = mar,
       volume = {546},
       number = {4},
          eid = {stag202},
        pages = {stag202},
          doi = {10.1093/MNRAS/stag202},
archivePrefix = {arXiv},
       eprint = {2511.17738},
 primaryClass = {astro-ph.GA},
       adsurl = {https://ui.adsabs.harvard.edu/abs/2026MNRAS.546ag202S}
}

@ARTICLE{McGaugh,
       author = {{McGaugh}, S.~S. and {Bothun}, G.~D.},
        title = "{Structural Characteristics and Stellar Composition of Low Surface Brightness Disk Galaxies}",
      journal = {ApJ},
         year = 1994,
        month = feb,
       volume = {107},
        pages = {530},
          doi = {10.1086/116874},
archivePrefix = {arXiv},
       eprint = {astro-ph/9311003},
 primaryClass = {astro-ph},
       adsurl = {https://ui.adsabs.harvard.edu/abs/1994AJ....107..530M}
}

@ARTICLE{Greene2022,
       author = {{Greene}, Jenny E. and {Greco}, Johnny P. and {Goulding}, Andy D. and {Huang}, Song and {Kado-Fong}, Erin and {Danieli}, Shany and {Li}, Jiaxuan and {Kim}, Ji Hoon and {Komiyama}, Yutaka and {Leauthaud}, Alexie and {MacArthur}, Lauren A. and {Sif{\'o}n}, Crist{\'o}bal},
        title = "{The Nature of Low-surface-brightness Galaxies in the Hyper Suprime-Cam Survey}",
      journal = {ApJ},
         year = 2022,
        month = jul,
       volume = {933},
       number = {2},
          eid = {150},
        pages = {150},
          doi = {10.3847/1538-4357/ac7238},
archivePrefix = {arXiv},
       eprint = {2204.11883},
 primaryClass = {astro-ph.GA},
       adsurl = {https://ui.adsabs.harvard.edu/abs/2022ApJ...933..150G}
}

@ARTICLE{MISTRAL_reduction,
       author = {{Schmitt}, J. and {Adami}, C. and {Dennefeld}, M. and {Agneray}, F. and {Basa}, S. and {Brunei}, J.~C. and {Buat}, V. and {Burgarella}, D. and {Carvalho}, C. and {Castagnoli}, G. and {Grosso}, N. and {Huppert}, F. and {Moreau}, C. and {Moreau}, F. and {Moreau}, L. and {Muslimov}, E. and {Pascal}, S. and {Perruchot}, S. and {Russeil}, D. and {Beuzit}, J.~L. and {Dolon}, F. and {Ferrari}, M. and {Hamelin}, B. and {Le Van Suu}, A. and {Aravind}, K. and {Gotz}, D. and {Jehin}, E. and {LeFloc'h}, E. and {Palmerio}, J. and {Saccardi}, A. and {Schneider}, B. and {Sch{\"u}ssler}, F. and {Turpin}, D. and {Vergani}, S.~D.},
        title = "{Multi-purpose InSTRument for Astronomy at Low-resolution: MISTRAL at the Observatoire de Haute-Provence<xref rid=``FN2'' ref-type=``fn''/>}",
      journal = {A\&A},
         year = 2024,
        month = jul,
       volume = {687},
          eid = {A198},
        pages = {A198},
          doi = {10.1051/0004-6361/202449254},
archivePrefix = {arXiv},
       eprint = {2404.03705},
 primaryClass = {astro-ph.IM},
       adsurl = {https://ui.adsabs.harvard.edu/abs/2024A\&A...687A.198S}
}

@article{Yuan_2018,
 title={Spatially resolved star formation and dust attenuation in Mrk 848: Comparison of the integral field spectra and the UV-to-IR SED},
 volume={613},
 ISSN={1432-0746},
 url={http://dx.doi.org/10.1051/0004-6361/201731865},
 DOI={10.1051/0004-6361/201731865},
 journal={A\&A},
 publisher={EDP Sciences},
 author={Yuan, Fang-Ting and Argudo-Fernández, María and Shen, Shiyin and Hao, Lei and Jiang, Chunyan and Yin, Jun and Boquien, Médéric and Lin, Lihwai},
 year={2018},
 month=may, pages={A13} }

@ARTICLE{Hinz,
   author = {{Hinz}, J.~L. and {Rieke}, M.~J. and {Rieke}, G.~H. and {Willmer}, C.~N.~A. and {Misselt}, K. and {Engelbracht}, C.~W. and {Blaylock}, M. and {Pickering}, T.~E.},
    title = "{Spitzer Observations of Low-Luminosity Isolated and Low Surface Brightness Galaxies}",
  journal = {ApJ},
   year = 2007,
    month = jul,
   volume = {663},
   number = {2},
    pages = {895-907},
    doi = {10.1086/518817},
archivePrefix = {arXiv},
   eprint = {0704.2059},
 primaryClass = {astro-ph},
   adsurl = {https://ui.adsabs.harvard.edu/abs/2007ApJ...663..895H}
}

@ARTICLE{MISTRAL,
   author = {{Paiella}, A. and {Cacciotti}, F. and {Isopi}, G. and {Barbavara}, E. and {Battistelli}, E.~S. and {de Bernardis}, P. and {Capalbo}, V. and {Carbone}, A. and {Carretti}, E. and {Ciccalotti}, E. and {Columbro}, F. and {Coppolecchia}, A. and {Cruciani}, A. and {D'Alessandro}, G. and {De Petris}, M. and {Govoni}, F. and {Lamagna}, L. and {Levati}, E. and {Marongiu}, P. and {Mascia}, A. and {Masi}, S. and {Molinari}, E. and {Murgia}, M. and {Navarrini}, A. and {Novelli}, A. and {Occhiuzzi}, A. and {Orlati}, A. and {Pappalardo}, E. and {Pettinari}, G. and {Piacentini}, F. and {Pisanu}, T. and {Poppi}, S. and {Porceddu}, I. and {Ritacco}, A. and {Schirru}, M.~R. and {Vargiu}, G.},
    title = "{The MISTRAL Instrument and the Characterization of Its Detector Array}",
  journal = {J. Low Temp. Phys.},
   year = 2024,
    month = nov,
   volume = {217},
   number = {3-4},
    pages = {436-445},
    doi = {10.1007/s10909-024-03210-1},
   adsurl = {https://ui.adsabs.harvard.edu/abs/2024JLTP..217..436P}
}

@article{Zaritsky_2019,
doi = {10.3847/1538-4365/aaefe9},
url = {https://doi.org/10.3847/1538-4365/aaefe9},
year = {2018},
month = {dec},
publisher = {The American Astronomical Society},
volume = {240},
number = {1},
pages = {1},
author = {Zaritsky, Dennis and Donnerstein, Richard and Dey, Arjun and Kadowaki, Jennifer and Zhang, Huanian and Karunakaran, Ananthan and Martínez-Delgado, David and Rahman, Mubdi and Spekkens, Kristine},
title = {Systematically Measuring Ultra-diffuse Galaxies (SMUDGes). I. Survey Description and First Results in the Coma Galaxy Cluster and Environs},
journal = {ApJS}
}

@article{cosmos2020,
doi = {10.3847/1538-4365/ac3078},
url = {https://doi.org/10.3847/1538-4365/ac3078},
year = {2022},
month = {jan},
publisher = {The American Astronomical Society},
volume = {258},
number = {1},
pages = {11},
author = {Weaver, J. R. and Kauffmann, O. B. and Ilbert, O. and McCracken, H. J. and Moneti, A. and Toft, S. and Brammer, G. and Shuntov, M. and Davidzon, I. and Hsieh, B. C. and Laigle, C. and Anastasiou, A. and Jespersen, C. K. and Vinther, J. and Capak, P. and Casey, C. M. and McPartland, C. J. R. and Milvang-Jensen, B. and Mobasher, B. and Sanders, D. B. and Zalesky, L. and Arnouts, S. and Aussel, H. and Dunlop, J. S. and Faisst, A. and Franx, M. and Furtak, L. J. and Fynbo, J. P. U. and Gould, K. M. L. and Greve, T. R. and Gwyn, S. and Kartaltepe, J. S. and Kashino, D. and Koekemoer, A. M. and Kokorev, V. and Le Fèvre, O. and Lilly, S. and Masters, D. and Magdis, G. and Mehta, V. and Peng, Y. and Riechers, D. A. and Salvato, M. and Sawicki, M. and Scarlata, C. and Scoville, N. and Shirley, R. and Silverman, J. D. and Sneppen, A. and Smolc̆ić, V. and Steinhardt, C. and Stern, D. and Tanaka, M. and Taniguchi, Y. and Teplitz, H. I. and Vaccari, M. and Wang, W.-H. and Zamorani, G.},
title = {COSMOS2020: A Panchromatic View of the Universe to z ∼ 10 from Two Complementary Catalogs},
journal = {ApJS}
}

@INPROCEEDINGS{Mistral_proposal,
   author = {{Bernaud}, E. and {Junais} and {Boissier}, S. and {Adami}, C. and {Thuruthipilly}, H. and {Ilbert}, O. and {Gavazzi}, R. and {Dubois}, C. and {Mercier}, W.},
    title = "{Measuring the redshift of ultra-diffuse galaxies candidates at OHP with MISTRAL}",
  booktitle = {SF2A-2025: Proceedings of the Annual meeting of the French Society of Astronomy and Astrophysics. Eds.: A. Siebert},
   year = 2025,
   editor = {{Siebert}, A. and {Bailli{\'e}}, K. and {B{\'e}thermin}, M. and {Cantalloube}, F. and {Josselin}, E. and {Lagarde}, N. and {Malzac}, J. and {Richard}, J. and {Selliez}, L. and {Venot}, O.},
    month = dec,
    pages = {391-393},
   adsurl = {https://ui.adsabs.harvard.edu/abs/2025sf2a.conf..391B}
}

@article{Du_2020,
doi = {10.3847/1538-3881/ab6efb},
url = {https://doi.org/10.3847/1538-3881/ab6efb},
year = {2020},
month = {mar},
publisher = {The American Astronomical Society},
volume = {159},
number = {4},
pages = {138},
author = {Du, Wei and Cheng, Cheng and Zheng, Zheng and Wu, Hong},
title = {Stellar Mass and Stellar Mass-to-light Ratio–Color Relations for Low Surface Brightness Galaxies},
journal = {AJ}
}

@article{Ebrov_2025,
 title={Photometric stellar masses for galaxies in DESI Legacy Imaging Surveys},
 ISSN={1432-0746},
 url={http://dx.doi.org/10.1051/0004-6361/202453448},
 DOI={10.1051/0004-6361/202453448},
 journal={Astronomy \&amp; Astrophysics},
 publisher={EDP Sciences},
 author={Ebrová, Ivana and Bílek, Michal and Eliášek, Jiří},
 year={2025},
 month=nov }

@ARTICLE{Kroupa,
   author = {{Kroupa}, Pavel},
    title = "{On the variation of the initial mass function}",
  journal = {MNRAS},
   year = 2001,
    month = apr,
   volume = {322},
   number = {2},
    pages = {231-246},
    doi = {10.1046/j.1365-8711.2001.04022.x},
archivePrefix = {arXiv},
   eprint = {astro-ph/0009005},
 primaryClass = {astro-ph},
   adsurl = {https://ui.adsabs.harvard.edu/abs/2001MNRAS.322..231K}
}

@BOOK{Press2007,
       author = {{Press}, William H. and {Teukolsky}, Saul A. and {Vetterling}, William T. and {Flannery}, Brian P.},
       publisher = {},
        title = "{Numerical Recipes 3rd Edition: The Art of Scientific Computing}",
         year = 2007,
       adsurl = {https://ui.adsabs.harvard.edu/abs/2007nras.book.....P}
}

@ARTICLE{Kennicutt94,
   author = {{Kennicutt}, Jr., Robert C. and {Tamblyn}, Peter and {Congdon}, Charles E.},
    title = "{Past and Future Star Formation in Disk Galaxies}",
  journal = {ApJ},
   year = 1994,
    month = nov,
   volume = {435},
    pages = {22},
    doi = {10.1086/174790},
   adsurl = {https://ui.adsabs.harvard.edu/abs/1994ApJ...435...22K}
}

@ARTICLE{Karunakaran,
   author = {{Karunakaran}, Ananthan and {Motiwala}, Khadeejah and {Spekkens}, Kristine and {Zaritsky}, Dennis and {Donnerstein}, Richard L. and {Dey}, Arjun},
    title = "{Systematically Measuring Ultradiffuse Galaxies. VII. The H I Survey Overview}",
  journal = {ApJ},
   year = 2024,
    month = nov,
   volume = {975},
   number = {1},
    eid = {91},
    pages = {91},
    doi = {10.3847/1538-4357/ad77cf},
archivePrefix = {arXiv},
   eprint = {2408.07119},
 primaryClass = {astro-ph.GA},
   adsurl = {https://ui.adsabs.harvard.edu/abs/2024ApJ...975...91K}
}

@ARTICLE{DECALs,
   author = {{Dey}, Arjun and {Schlegel}, David J. and {Lang}, Dustin and {Blum}, Robert and {Burleigh}, Kaylan and {Fan}, Xiaohui and {Findlay}, Joseph R. and {Finkbeiner}, Doug and {Herrera}, David and {Juneau}, St{\'e}phanie and {Landriau}, Martin and {Levi}, Michael and {McGreer}, Ian and {Meisner}, Aaron and {Myers}, Adam D. and {Moustakas}, John and {Nugent}, Peter and {Patej}, Anna and {Schlafly}, Edward F. and {Walker}, Alistair R. and {Valdes}, Francisco and {Weaver}, Benjamin A. and {Y{\`e}che}, Christophe and {Zou}, Hu and {Zhou}, Xu and {Abareshi}, Behzad and {Abbott}, T.~M.~C. and {Abolfathi}, Bela and {Aguilera}, C. and {Alam}, Shadab and {Allen}, Lori and {Alvarez}, A. and {Annis}, James and {Ansarinejad}, Behzad and {Aubert}, Marie and {Beechert}, Jacqueline and {Bell}, Eric F. and {BenZvi}, Segev Y. and {Beutler}, Florian and {Bielby}, Richard M. and {Bolton}, Adam S. and {Brice{\~n}o}, C{\'e}sar and {Buckley-Geer}, Elizabeth J. and {Butler}, Karen and {Calamida}, Annalisa and {Carlberg}, Raymond G. and {Carter}, Paul and {Casas}, Ricard and {Castander}, Francisco J. and {Choi}, Yumi and {Comparat}, Johan and {Cukanovaite}, Elena and {Delubac}, Timoth{\'e}e and {DeVries}, Kaitlin and {Dey}, Sharmila and {Dhungana}, Govinda and {Dickinson}, Mark and {Ding}, Zhejie and {Donaldson}, John B. and {Duan}, Yutong and {Duckworth}, Christopher J. and {Eftekharzadeh}, Sarah and {Eisenstein}, Daniel J. and {Etourneau}, Thomas and {Fagrelius}, Parker A. and {Farihi}, Jay and {Fitzpatrick}, Mike and {Font-Ribera}, Andreu and {Fulmer}, Leah and {G{\"a}nsicke}, Boris T. and {Gaztanaga}, Enrique and {George}, Koshy and {Gerdes}, David W. and {Gontcho}, Satya Gontcho A. and {Gorgoni}, Claudio and {Green}, Gregory and {Guy}, Julien and {Harmer}, Diane and {Hernandez}, M. and {Honscheid}, Klaus and {Huang}, Lijuan Wendy and {James}, David J. and {Jannuzi}, Buell T. and {Jiang}, Linhua and {Joyce}, Richard and {Karcher}, Armin and {Karkar}, Sonia and {Kehoe}, Robert and {Kneib}, Jean-Paul and {Kueter-Young}, Andrea and {Lan}, Ting-Wen and {Lauer}, Tod R. and {Le Guillou}, Laurent and {Le Van Suu}, Auguste and {Lee}, Jae Hyeon and {Lesser}, Michael and {Perreault Levasseur}, Laurence and {Li}, Ting S. and {Mann}, Justin L. and {Marshall}, Robert and {Mart{\'\i}nez-V{\'a}zquez}, C.~E. and {Martini}, Paul and {du Mas des Bourboux}, H{\'e}lion and {McManus}, Sean and {Meier}, Tobias Gabriel and {M{\'e}nard}, Brice and {Metcalfe}, Nigel and {Mu{\~n}oz-Guti{\'e}rrez}, Andrea and {Najita}, Joan and {Napier}, Kevin and {Narayan}, Gautham and {Newman}, Jeffrey A. and {Nie}, Jundan and {Nord}, Brian and {Norman}, Dara J. and {Olsen}, Knut A.~G. and {Paat}, Anthony and {Palanque-Delabrouille}, Nathalie and {Peng}, Xiyan and {Poppett}, Claire L. and {Poremba}, Megan R. and {Prakash}, Abhishek and {Rabinowitz}, David and {Raichoor}, Anand and {Rezaie}, Mehdi and {Robertson}, A.~N. and {Roe}, Natalie A. and {Ross}, Ashley J. and {Ross}, Nicholas P. and {Rudnick}, Gregory and {Safonova}, Sasha and {Saha}, Abhijit and {S{\'a}nchez}, F. Javier and {Savary}, Elodie and {Schweiker}, Heidi and {Scott}, Adam and {Seo}, Hee-Jong and {Shan}, Huanyuan and {Silva}, David R. and {Slepian}, Zachary and {Soto}, Christian and {Sprayberry}, David and {Staten}, Ryan and {Stillman}, Coley M. and {Stupak}, Robert J. and {Summers}, David L. and {Sien Tie}, Suk and {Tirado}, H. and {Vargas-Maga{\~n}a}, Mariana and {Vivas}, A. Katherina and {Wechsler}, Risa H. and {Williams}, Doug and {Yang}, Jinyi and {Yang}, Qian and {Yapici}, Tolga and {Zaritsky}, Dennis and {Zenteno}, A. and {Zhang}, Kai and {Zhang}, Tianmeng and {Zhou}, Rongpu and {Zhou}, Zhimin},
    title = "{Overview of the DESI Legacy Imaging Surveys}",
  journal = {ApJ},
   year = 2019,
    month = may,
   volume = {157},
   number = {5},
    eid = {168},
    pages = {168},
    doi = {10.3847/1538-3881/ab089d},
archivePrefix = {arXiv},
   eprint = {1804.08657},
 primaryClass = {astro-ph.IM},
   adsurl = {https://ui.adsabs.harvard.edu/abs/2019AJ....157..168D}
}

@article{Li_2025,
 title={A negative stellar mass−gaseous metallicity gradient relation of dwarf galaxies modulated by stellar feedback},
 volume={698},
 ISSN={1432-0746},
 url={http://dx.doi.org/10.1051/0004-6361/202452978},
 DOI={10.1051/0004-6361/202452978},
 journal={A\&A},
 publisher={EDP Sciences},
 author={Li, Tie and Zhang, Hong-Xin and Lyu, Wenhe and Tang, Yimeng and Yao, Yao and Wang, Enci and Rong, Yu and Chen, Guangwen and Kong, Xu and Bian, Fuyan and Gu, Qiusheng and Johnston, Evelyn J. and Li, Xin and Mao, Shude and Shi, Yong and Wang, Junfeng and Wang, Xin and Yu, Xiaoling and Zheng, Zhiyuan},
 year={2025},
 month=jun,
 pages={A208}
 }

@article{sprayberry1995,
       author = {{Sprayberry}, D. and {Impey}, C.~D. and {Bothun}, G.~D. and {Irwin}, M.~J.},
        title = "{Properties of the Class of Giant Low Surface Brightness Spiral Galaxies}",
      journal = {ApJ},
         year = 1995,
        month = feb,
       volume = {109},
        pages = {558},
          doi = {10.1086/117300},
       adsurl = {https://ui.adsabs.harvard.edu/abs/1995AJ....109..558S}
}

@article{2025BERNAUD,
       author = {{Bernaud}, E. and {Boissier}, S. and {Junais} and {Ma{\l}ek}, K. and {Hugot}, E. and {Galaz}, G.},
        title = "{Characterizing the outer disk of extended-UV galaxies in the optical dotmain with deep surveys}",
      journal = {A\&A},
         year = 2025,
        month = aug,
       volume = {700},
          eid = {A56},
        pages = {A56},
          doi = {10.1051/0004-6361/202555273},
archivePrefix = {arXiv},
       eprint = {2506.11568},
 primaryClass = {astro-ph.GA},
       adsurl = {https://ui.adsabs.harvard.edu/abs/2025A\&A...700A..56B}
}

@ARTICLE{Saburova2021,
       author = {{Saburova}, Anna S. and {Chilingarian}, Igor V. and {Kasparova}, Anastasia V. and {Sil'chenko}, Olga K. and {Grishin}, Kirill A. and {Katkov}, Ivan Yu and {Uklein}, Roman I.},
        title = "{Observational insights on the origin of giant low surface brightness galaxies}",
      journal = {MNRAS},
         year = 2021,
        month = may,
       volume = {503},
       number = {1},
        pages = {830-849},
          doi = {10.1093/MNRAS/stab374},
archivePrefix = {arXiv},
       eprint = {2011.01238},
 primaryClass = {astro-ph.GA},
       adsurl = {https://ui.adsabs.harvard.edu/abs/2021MNRAS.503..830S}
}

@article{Peng_2002,
doi = {10.1086/340952},
url = {https://doi.org/10.1086/340952},
year = {2002},
month = {jul},
publisher = {},
volume = {124},
number = {1},
pages = {266},
author = {Peng, Chien Y. and Ho, Luis C. and Impey, Chris D. and Rix, Hans-Walter},
title = {Detailed Structural Decomposition of Galaxy
Images*},
journal = {AJ}
}

@ARTICLE{2023MNRAS.523.3991Z,
       author = {{Zhu}, Qirong and {P{\'e}rez-Monta{\~n}o}, Luis Enrique and {Rodriguez-Gomez}, Vicente and {Cervantes Sodi}, Bernardo and {Zjupa}, Jolanta and {Marinacci}, Federico and {Vogelsberger}, Mark and {Hernquist}, Lars},
        title = "{Giant low surface brightness galaxies in TNG100}",
      journal = {MNRAS},
         year = 2023,
        month = aug,
       volume = {523},
       number = {3},
        pages = {3991-4014},
          doi = {10.1093/MNRAS/stad1655},
       adsurl = {https://ui.adsabs.harvard.edu/abs/2023MNRAS.523.3991Z}
}

@ARTICLE{Beijersbergen1999A,
       author = {{Beijersbergen}, M. and {de Blok}, W.~J.~G. and {van der Hulst}, J.~M.},
        title = "{Surface photometry of bulge dominated low surface brightness galaxies}",
      journal = {A\&A},
         year = 1999,
        month = nov,
       volume = {351},
        pages = {903-919},
          doi = {10.48550/arXiv.astro-ph/9909483},
archivePrefix = {arXiv},
       eprint = {astro-ph/9909483},
 primaryClass = {astro-ph},
       adsurl = {https://ui.adsabs.harvard.edu/abs/1999A\&A...351..903B}
}

@ARTICLE{Trujillo2020,
       author = {{Trujillo}, Ignacio and {Chamba}, Nushkia and {Knapen}, Johan H.},
        title = "{A physically motivated definition for the size of galaxies in an era of ultradeep imaging}",
      journal = {MNRAS},
         year = 2020,
        month = mar,
       volume = {493},
       number = {1},
        pages = {87-105},
          doi = {10.1093/MNRAS/staa236},
archivePrefix = {arXiv},
       eprint = {2001.02689},
 primaryClass = {astro-ph.GA},
       adsurl = {https://ui.adsabs.harvard.edu/abs/2020MNRAS.493...87T}
}

@article{ Junais2020,
	author = {{Junais} and {Ruiz Cejudo, Ignacio} and {Guerra Arencibia, Sergio} and {Trujillo, Ignacio} and {Alarcon, Miguel R.} and {Serra-Ricart, Miquel} and {Knapen, Johan H.} and {Duc, Pierre-Alain}},
	title = {Deep imaging of the galaxy Malin 2 shows new faint structures and a candidate satellite dwarf galaxy},
	DOI= "10.1051/0004-6361/202556569",
	url= "https://doi.org/10.1051/0004-6361/202556569",
	journal = {A\&A},
	year = 2025,
	volume = 702,
	pages = "A136",
}

@ARTICLE{Carrick2015,
       author = {{Carrick}, Jonathan and {Turnbull}, Stephen J. and {Lavaux}, Guilhem and {Hudson}, Michael J.},
        title = "{Cosmological parameters from the comparison of peculiar velocities with predictions from the 2M++ density field}",
      journal = {MNRAS},
         year = 2015,
        month = jun,
       volume = {450},
       number = {1},
        pages = {317-332},
          doi = {10.1093/MNRAS/stv547},
archivePrefix = {arXiv},
       eprint = {1504.04627},
 primaryClass = {astro-ph.CO},
       adsurl = {https://ui.adsabs.harvard.edu/abs/2015MNRAS.450..317C}
}

@article{Lavaux2011,
    author = {Lavaux, Guilhem and Hudson, Michael J.},
    title = {The 2M++ galaxy redshift catalogue},
    journal = {MNRAS},
    volume = {416},
    number = {4},
    pages = {2840-2856},
    year = {2011},
    month = {10},
    issn = {0035-8711},
    doi = {10.1111/j.1365-2966.2011.19233.x},
    url = {https://doi.org/10.1111/j.1365-2966.2011.19233.x},
    eprint = {https://academic.oup.com/MNRAS/article-pdf/416/4/2840/17330470/MNRAS0416-2840.pdf}
}
\begin{appendix}
\onecolumn
\section{Observations}
\label{Appendix_A}
\begin{table*}[ht]
\centering
\begin{threeparttable}
\caption{Observation details for the sample targets.}
\begin{tabular}{llllllll}
\hline
ID & Catalogued names & RAJ2000 & DECJ2000 & PA & Date & Instrument & $t_\mathrm{exp}$ (s)\\
\hline
\textit{(1)}&\textit{(2)}&\textit{(3)}&\textit{(4)}&\textit{(5)}&\textit{(6)}&\textit{(7)}&\textit{(8)}\\
\hline
\\
LBT-1     & DES-63527438  & 02:00:27.20    & $-$04:33:44.41   & 297.48   & 2025-01-03 & MODS1/2 & 3600 ($1200\times3$)\\
LBT-2     & DES-112062114 & 02:23:18.67    & $-$02:03:25.12   & 315.5    & 2024-11-07 & MODS1   & 4800 ($1200\times4$)\\
LBT-3     & DES-116944307 & 02:25:49.08    & $-$00:23:34.77   & 300.78   & 2024-12-03 & MODS1   & 2400 ($1200\times2$)\\
LBT-4     & DES-142528427 & 00:09:48.73    & $-$05:23:26.41   & 5.46     & 2024-12-05 & MODS1   & 2400 ($1200\times2$)\\
LBT-5     & DES-150607882 & 02:20:33.76    & $-$09:12:27.63   & 42.13    & 2025-01-03 & MODS1/2 & 3600 ($1200\times3$)\\
LBT-6     & DES-159109120 & 23:44:38.03    & $-$00:41:25.37   & 340.65   & 2024-12-05 & MODS1   & 2400 ($1200\times2$)\\
LBT-7     & DES-177521435 & 00:06:42.67    & $+$04:34:18.78   & 294.97   & 2024-11-06 & MODS1   & 3600 ($1200\times3$)\\
LBT-8     & DES-208463768 & 21:37:07.43    & $-$00:31:04.31   & 9.74     & 2024-12-05 & MODS1   & 3600 ($1200\times3$)\\
LBT-9     & DES-217352898 & 01:23:06.29    & $+$04:54:36.86   & 332.41   & 2024-12-03 & MODS1   & 3600 ($1200\times3$)\\
LBT-10    & DES-261775966 & 01:55:29.22    & $-$00:53:51.68   & 64.96    & 2025-01-02 & MODS1/2 & 3600 ($1200\times3$)\\
LBT-11    & DES-262073487 & 01:54:18.36    & $+$00:22:33.89   & 328.93   & 2025-01-02 & MODS1/2 & 3600 ($1200\times3$)\\
LBT-12    & DES-331218165 & 02:57:14.60    & $-$02:17:37.14   & 320.62   & 2024-10-01 & MODS2   & 3600 ($1200\times3$)\\
LBT-13    & DES-332393749 & 02:56:05.66    & $+$02:48:29.14   & 315.54   & 2024-12-04 & MODS1   & 3600 ($1200\times3$)\\
OHP-1     & COSMOS20-306274     & 09:59:51.40    & +01:42:24.01 & 4.0    &  2024-04-15 &MISTRAL   & 2700\\
OHP-2     & SMDG-1028240+365101 & 10:28:23.97   & +36:51:00.79  & 76.0   &  2025-02-02 & MISTRAL   & 3600 ($1800\times2$)\\
OHP-3     & SMDG-1042235+641153 & 10:42:23.54    & +64:11:53.48 & 35.0   &  2025-02-02 & MISTRAL   & 3600 ($1800\times2$)\\
OHP-4     & SMDG-1225058+325042 & 12:25:05.80    & +32:50:42.07 & 42.0   &  2025-02-04 & MISTRAL   & 3600 ($1800\times2$)\\
\\
\hline
\label{tab:obs_details}
\end{tabular}
\begin{tablenotes}
\footnotesize
\item \textbf{Notes.} (1) ID used in this work. (2) Official catalogued names of the sources, as listed in DES~\citep{Thuruthipilly2024A}, COSMOS20~\citep{cosmos2020}, and SMDG~\citep{Zaritsky_2023}. (3) Right ascension. (4) Declination. (5) PA of the long slit. (6) Observation date in the ISO 8601 format. (7) Instrument's name. (8) Exposure time in seconds (s).
\end{tablenotes}
\end{threeparttable}
\end{table*}

\onecolumn
\section{Samples}
\label{Appendix_B}
The LBT and OHP samples are shown in Fig.~\ref {showing sample}. 
In the image on the left, the RGB cutout of the sample listed in Table~\ref{tab:obs_details} is shown.
The RGB cutout has been created by superimposing the \textit{g}, \textit{r}, and \textit{z}-band $33''$  radius cutout from DES-DR1.
The right side of the image~\ref{showing sample} shows the spectra associated with each galaxy in the rest frame; relevant emission lines are plotted and labelled. 
\begin{figure*}[ht]
\centering
\begin{minipage}{\textwidth}
\centering
\begin{minipage}[t]{0.27\textwidth}
\centering
\subcaption{
\\
\vspace{1em}
LBT-1}
\includegraphics[width=\linewidth]{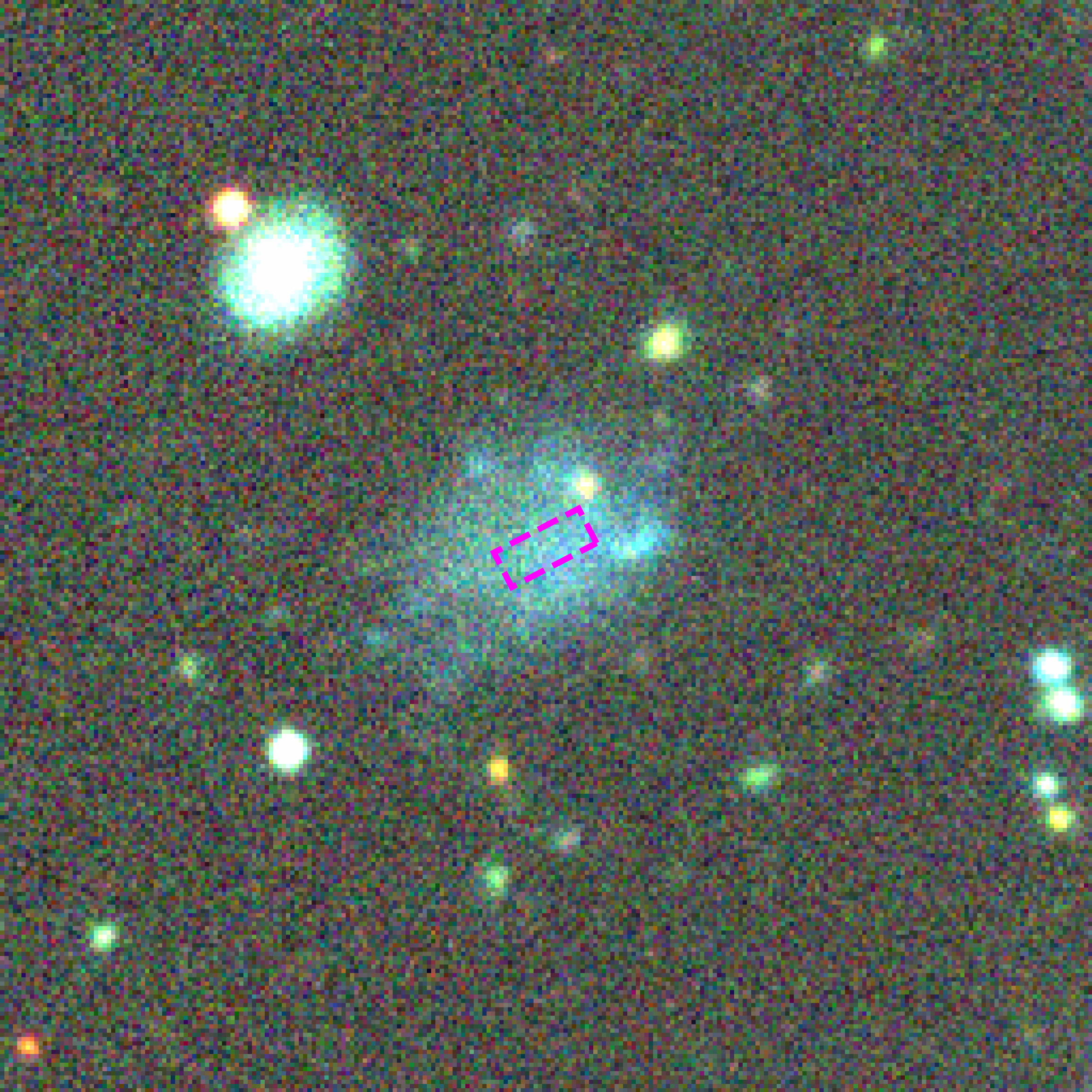}\vspace{-2.5em}
\label{435 cutout}
\end{minipage}
\hspace{0.001\textwidth}
\begin{minipage}[t]{0.71\textwidth}
\centering
\subcaption{
\\
\vspace{1em}
LBT-1
}
\includegraphics[width=\linewidth]{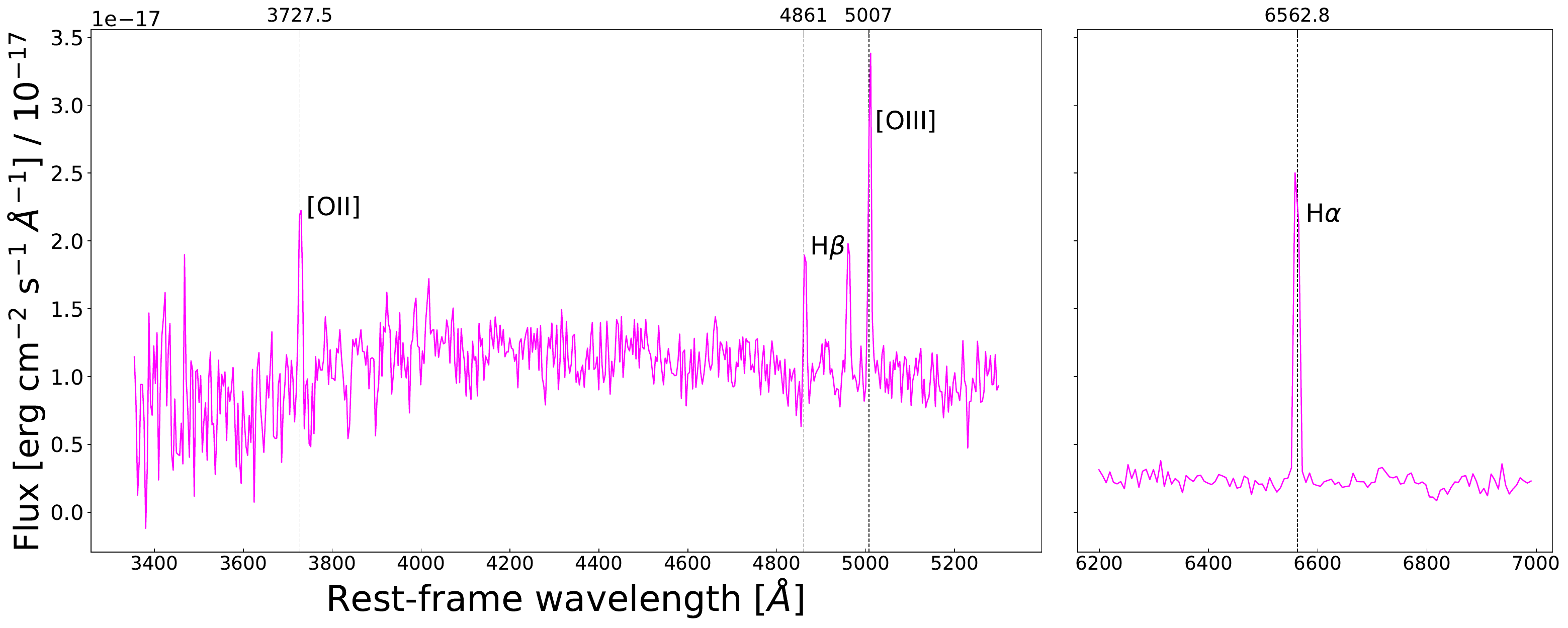}
\label{435 spectrum}\vspace{-2.5em}
\end{minipage}
\end{minipage}
\label{example}
\end{figure*}
\begin{figure*}[ht]
\centering
\begin{minipage}{\textwidth}
\centering
\begin{minipage}[t]{0.27\textwidth}
\centering
\subcaption*{LBT-2}
\includegraphics[width=\linewidth]{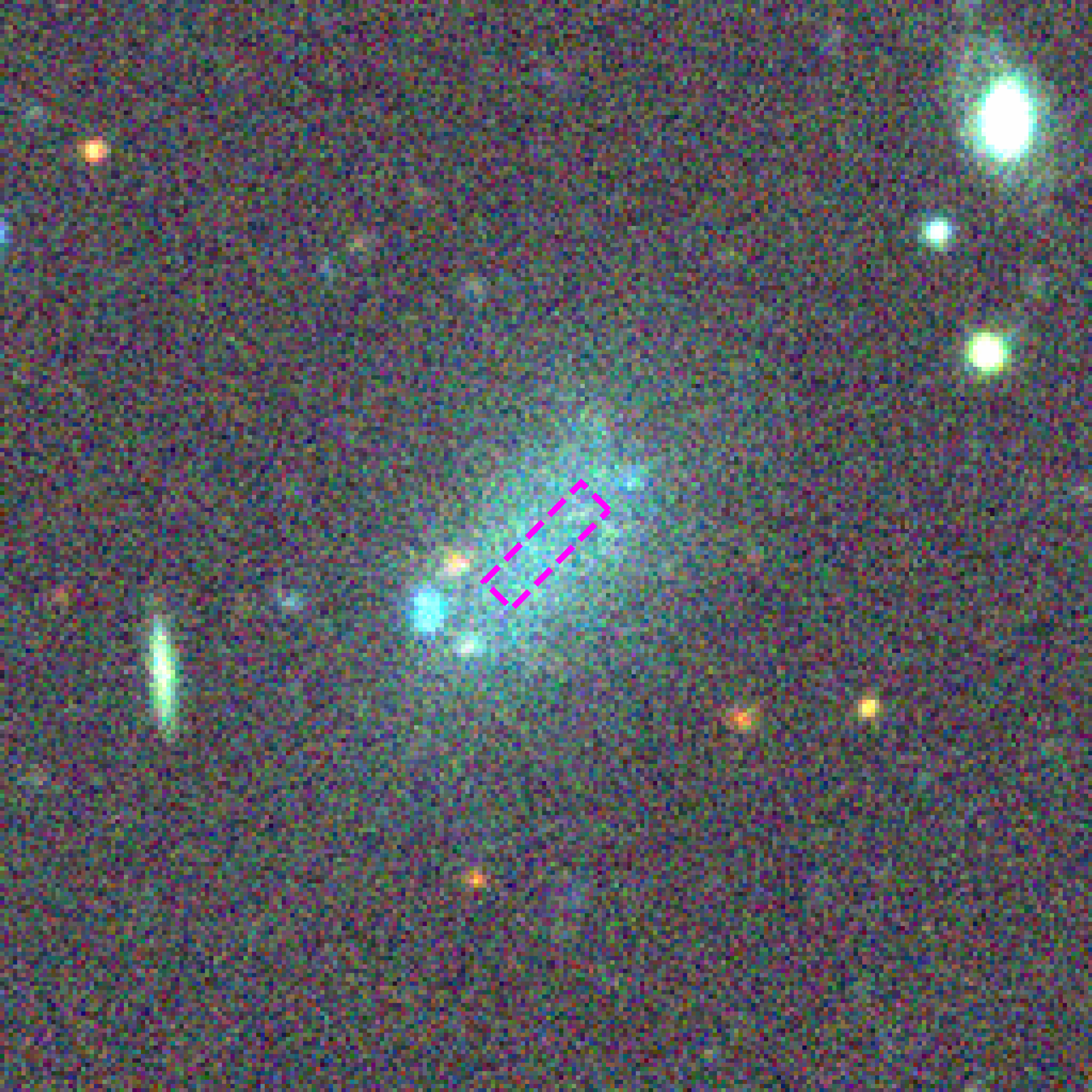}\vspace{-2.5em}
\label{435 cutout}
\end{minipage}
\hspace{0.001\textwidth}
\begin{minipage}[t]{0.71\textwidth}
\centering
\subcaption*{LBT-2}
\includegraphics[width=\linewidth]{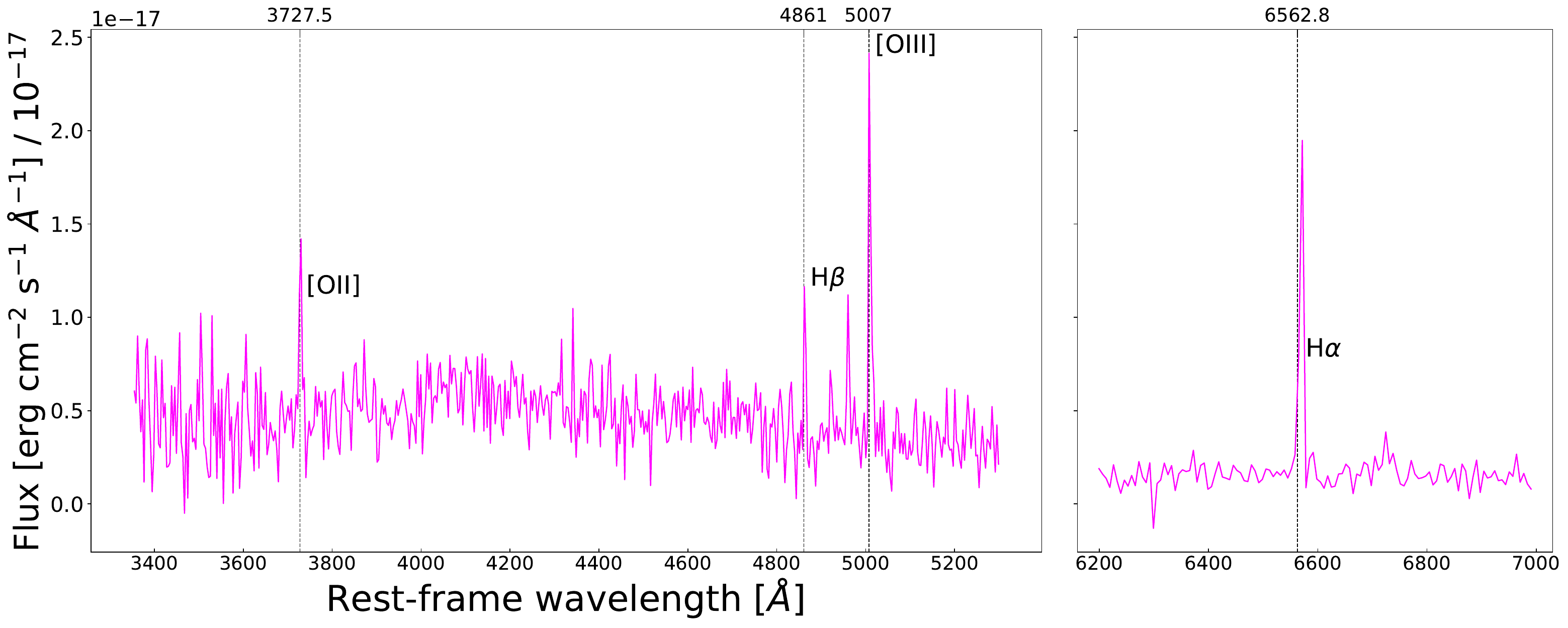}
\label{435 spectrum}\vspace{-2.5em}
\end{minipage}
\end{minipage}
\label{example}
\end{figure*}

\begin{figure*}[ht]
\centering
\begin{minipage}{\textwidth}
\centering
\begin{minipage}[t]{0.27\textwidth}
\centering
\subcaption*{\textbf{LBT-3}}
\includegraphics[width=\linewidth]{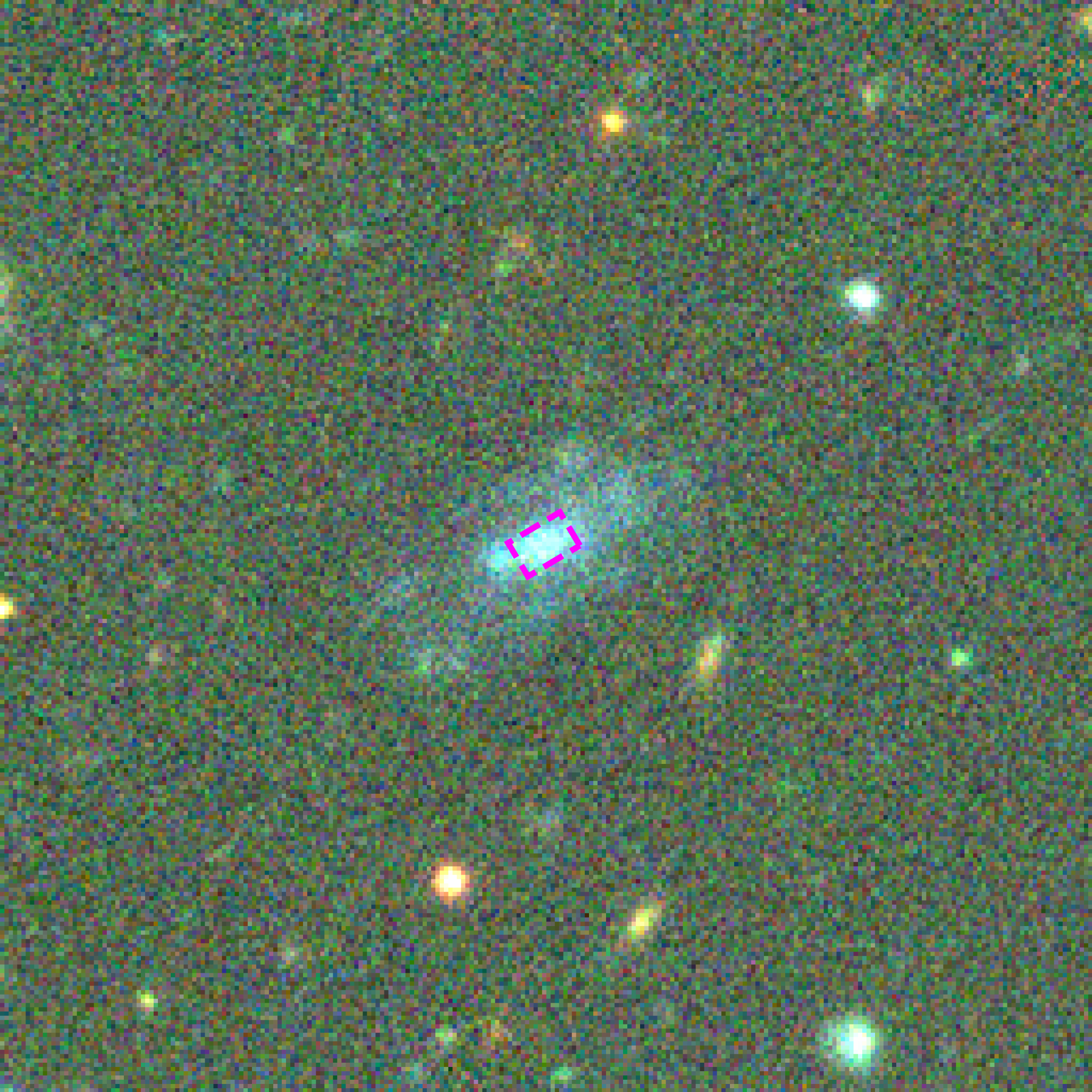}\vspace{-2.5em}
\label{435 cutout}
\end{minipage}
\hspace{0.001\textwidth}
\begin{minipage}[t]{0.71\textwidth}
\centering
\subcaption*{\textbf{LBT-3}}
\includegraphics[width=\linewidth]{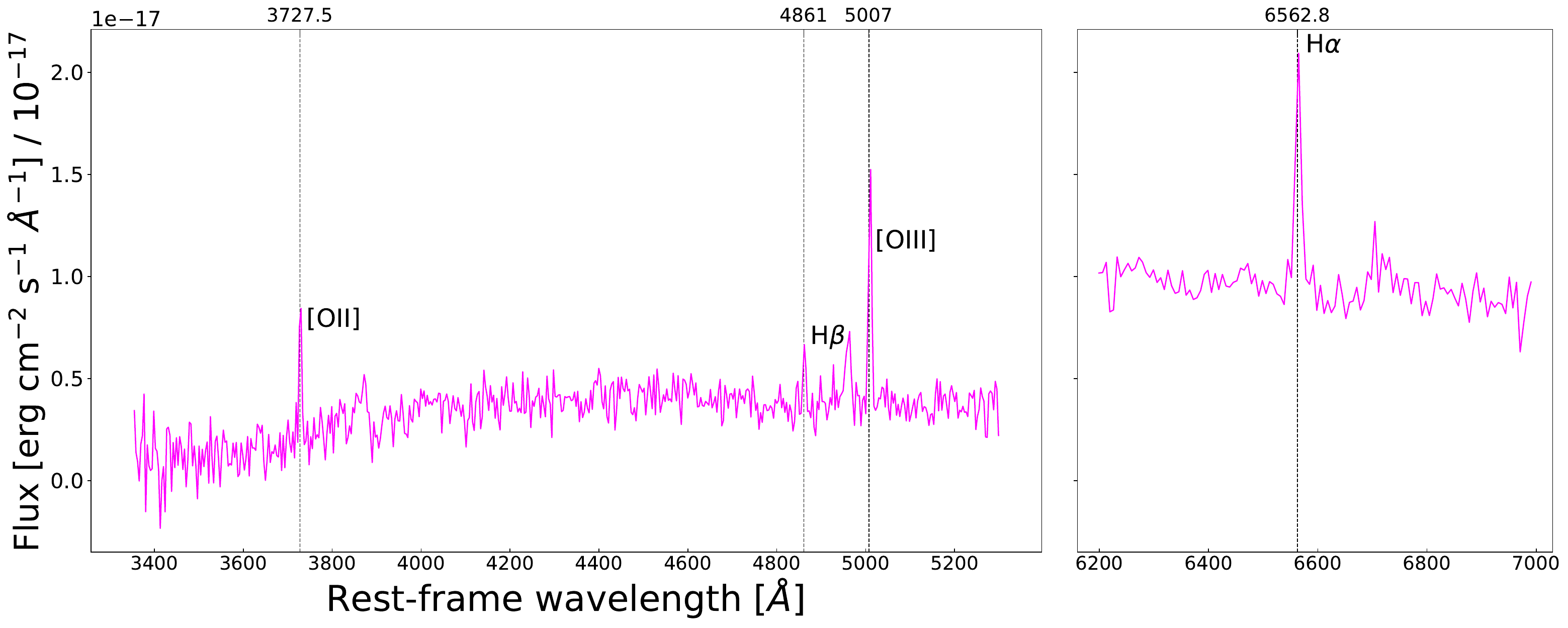}
\label{435 spectrum}\vspace{-2.5em}
\end{minipage}
\end{minipage}
\vspace{1em}
\setcounter{figure}{1}
\caption{(a) LBT and OHP-UDG candidate RGB cutouts. The $60''$ cutouts were made using the \textit{g}, \textit{r}, and \textit{z} bands. A rectangle is drawn on the cutout to mimic the PA and the width of the long slit used to observe the galaxies. The width of the slit is $2.4''$ and $1.9''$ for the LBT and OHP samples, respectively, while the length represents the size of the extraction box used during the reduction process of the 2D spectra to produce the final 1D spectra. (b) LBT and OHP UDG candidate spectra in the rest frame after data reduction. The extracted spectra were used to estimate the redshift and confirm the ultra-diffuse nature of the sources. Prominent emission lines are marked in the plot. Flux estimation was performed for the H$\alpha$ and H$\beta$ (only for ten of the LBT-UDG candidates) recombination lines. The sources in bold are confirmed UDGs, according to Table~\ref{tab:Halpha_flxs}.}
\label{showing sample}
\end{figure*}

\begin{figure*}[ht]
\centering
\begin{minipage}{\textwidth}
\centering
\begin{minipage}[t]{0.27\textwidth}
\centering
\subcaption*{\textbf{LBT-4}}
\includegraphics[width=\linewidth]{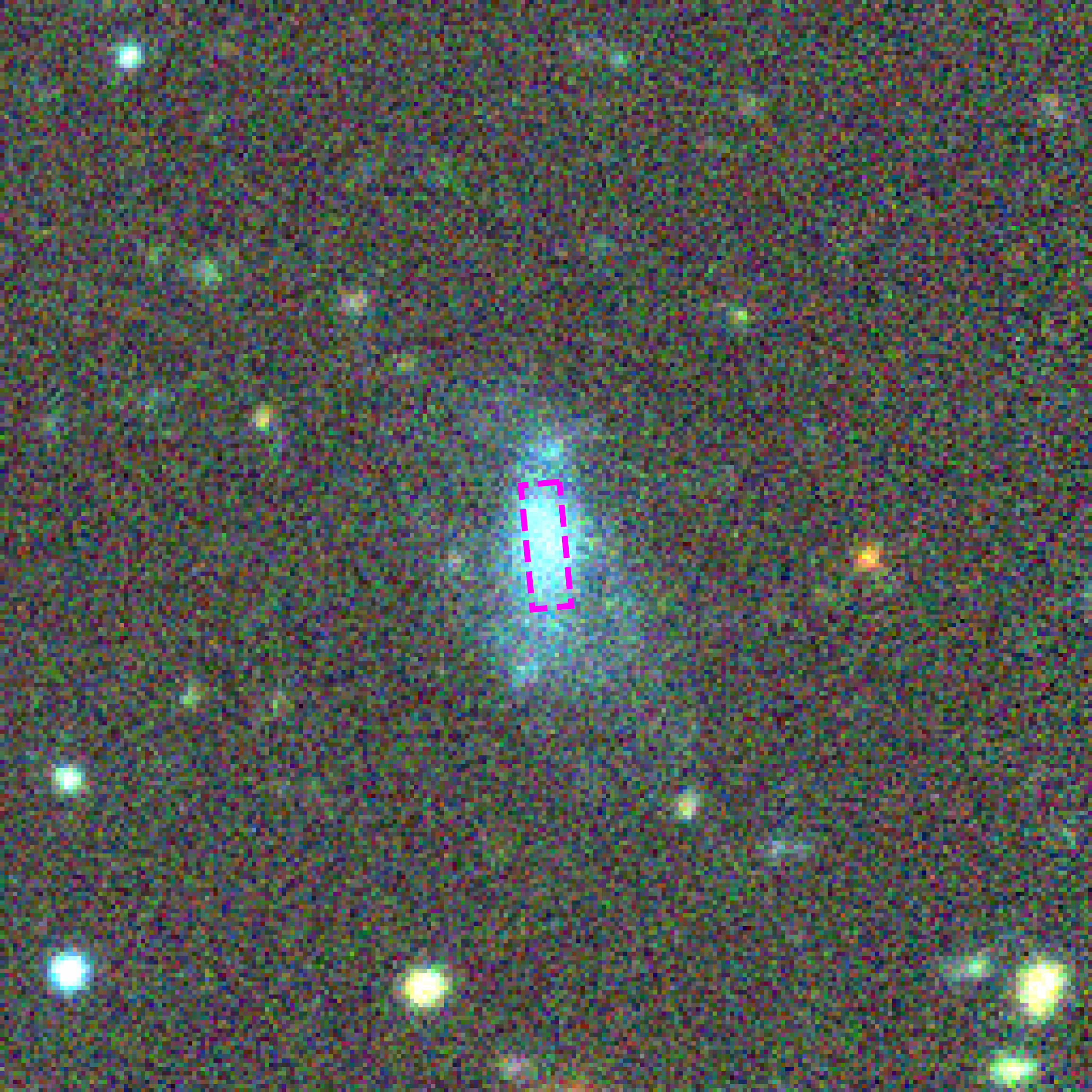}\vspace{-2.5em}
\label{435 cutout}
\end{minipage}
\hspace{0.001\textwidth}
\begin{minipage}[t]{0.71\textwidth}
\centering
\subcaption*{\textbf{LBT-4}}
\includegraphics[width=\linewidth]{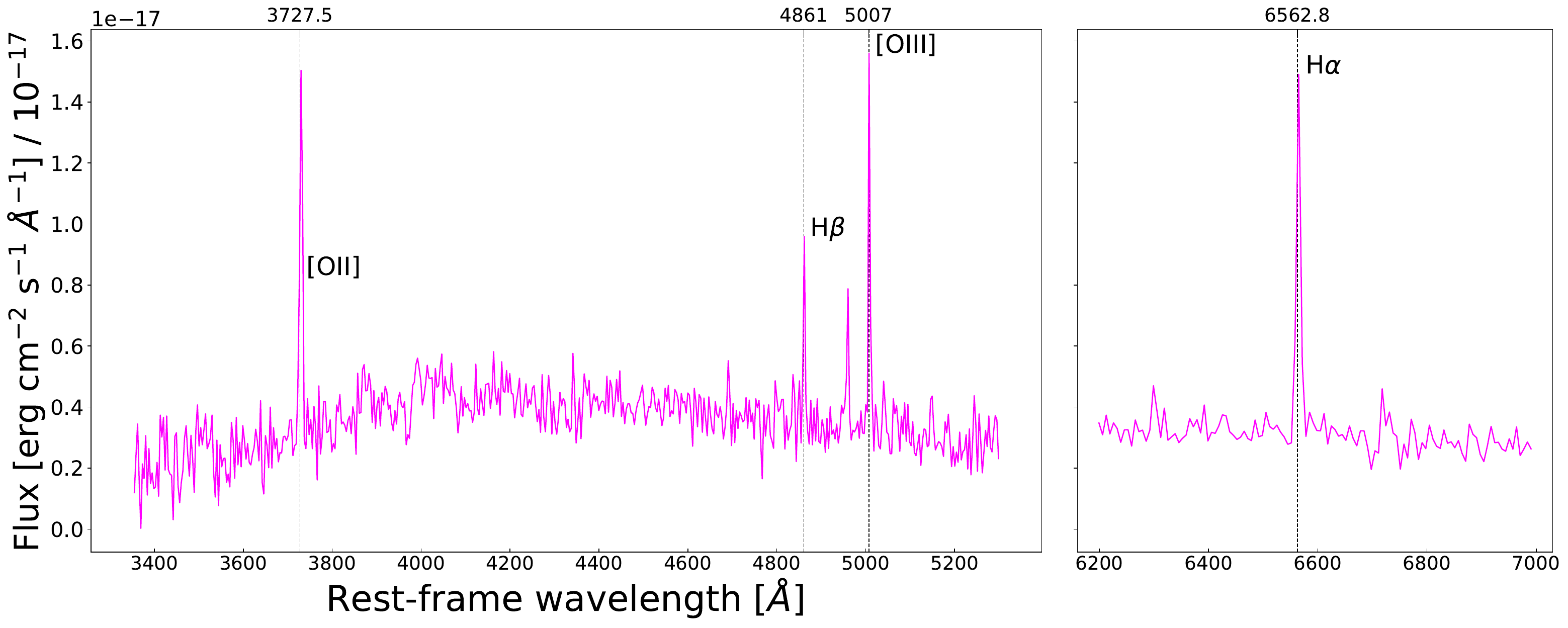}
\label{435 spectrum}\vspace{-2.5em}
\end{minipage}
\end{minipage}
\label{example}
\end{figure*}

\begin{figure*}[ht]
\centering
\begin{minipage}{\textwidth}
\centering
\begin{minipage}[t]{0.27\textwidth}
\centering
\subcaption*{LBT-5}
\includegraphics[width=\linewidth]{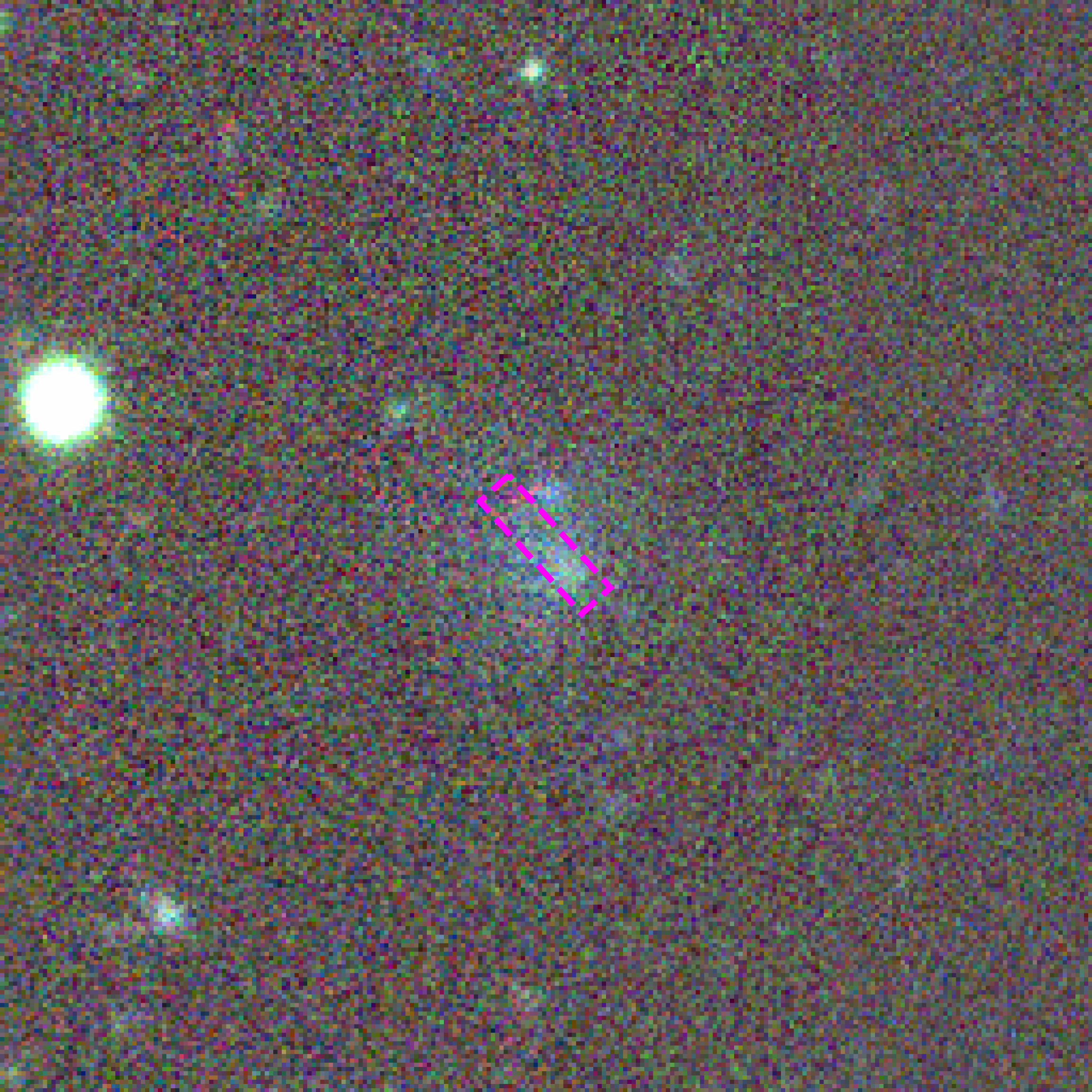}\vspace{-2.5em}
\label{435 cutout}
\end{minipage}
\hspace{0.001\textwidth}
\begin{minipage}[t]{0.71\textwidth}
\centering
\subcaption*{LBT-5}
\includegraphics[width=\linewidth]{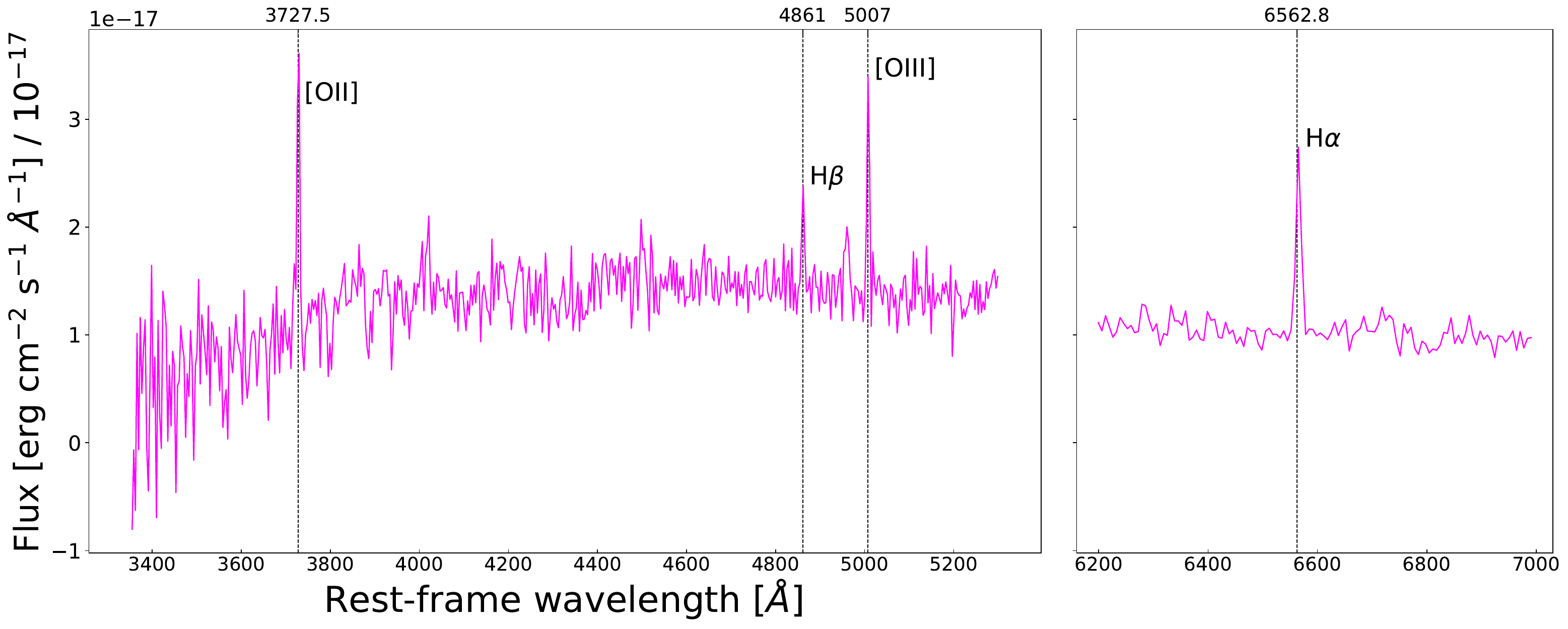}\vspace{-2.5em}
\label{435 spectrum}
\end{minipage}
\end{minipage}
\label{example}
\end{figure*}

\begin{figure*}[ht]
\centering
\begin{minipage}{\textwidth}
\centering
\begin{minipage}[t]{0.27\textwidth}
\centering
\subcaption*{\textbf{LBT-6}}
\includegraphics[width=\linewidth]{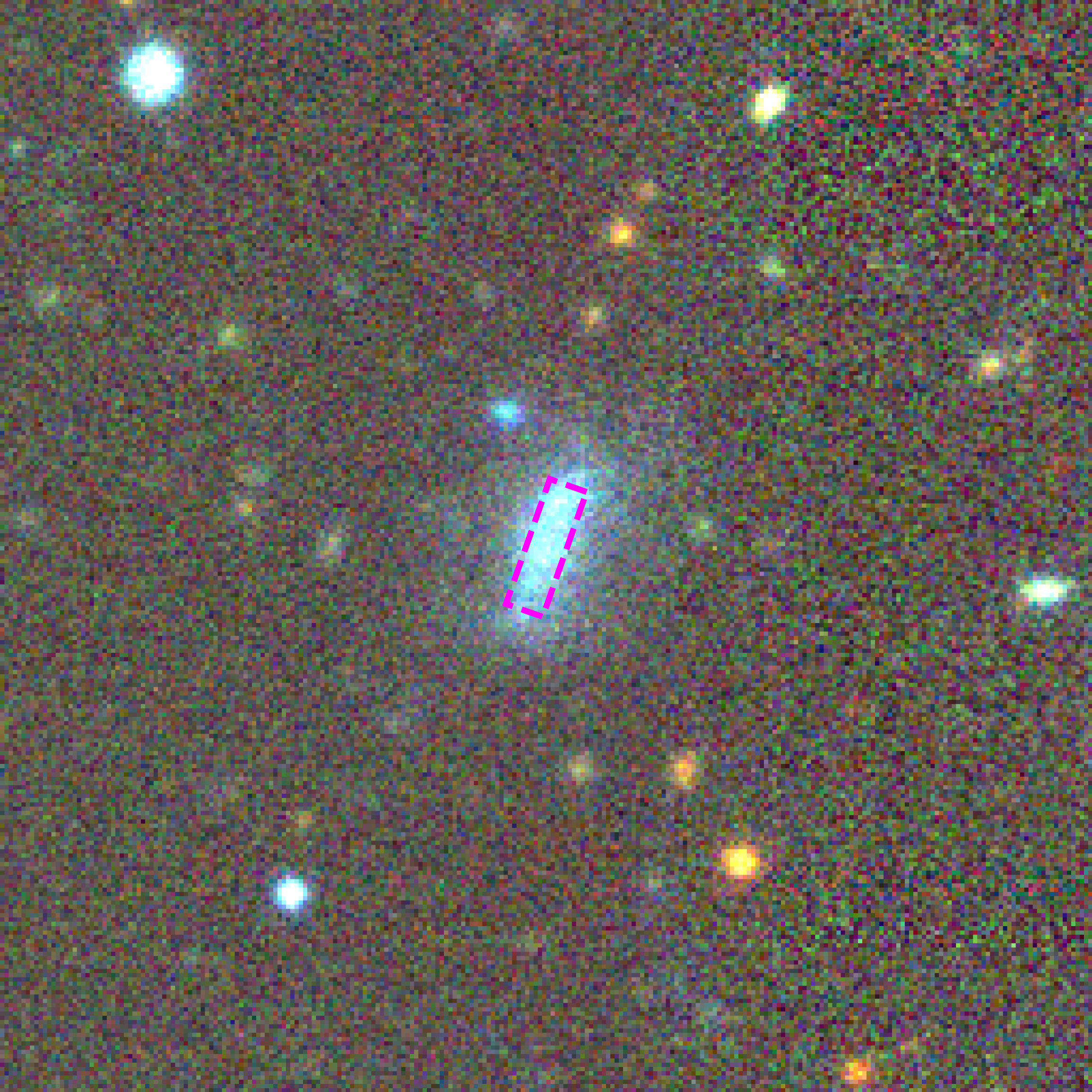}\vspace{-2.5em}
\end{minipage}
\hspace{0.001\textwidth}
\begin{minipage}[t]{0.71\textwidth}
\centering
\subcaption*{\textbf{LBT-6}}
\includegraphics[width=\linewidth]{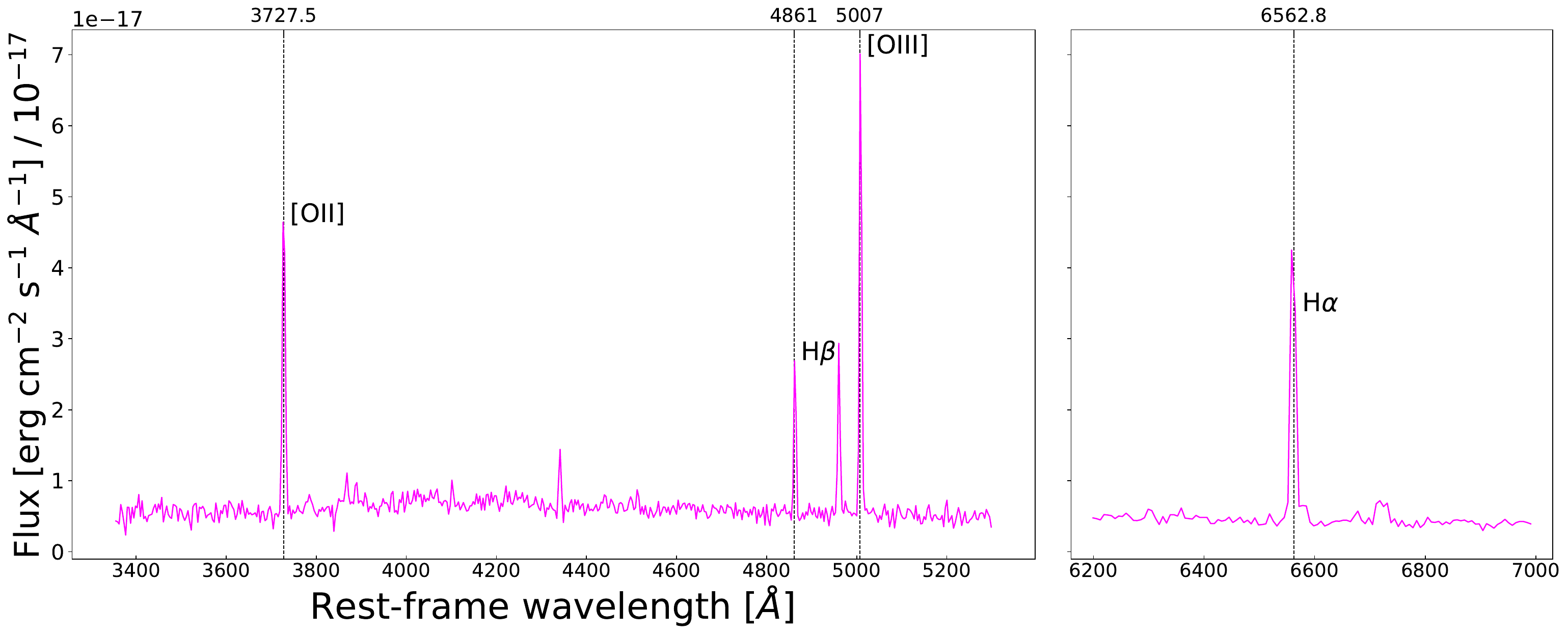}\vspace{-2.5em}
\label{435 spectrum}
\end{minipage}
\end{minipage}
\label{example}
\end{figure*}

\begin{figure*}[ht]
\centering
\begin{minipage}{\textwidth}
\centering
\begin{minipage}[t]{0.27\textwidth}
\centering
\subcaption*{\textbf{LBT-7}}
\includegraphics[width=\linewidth]{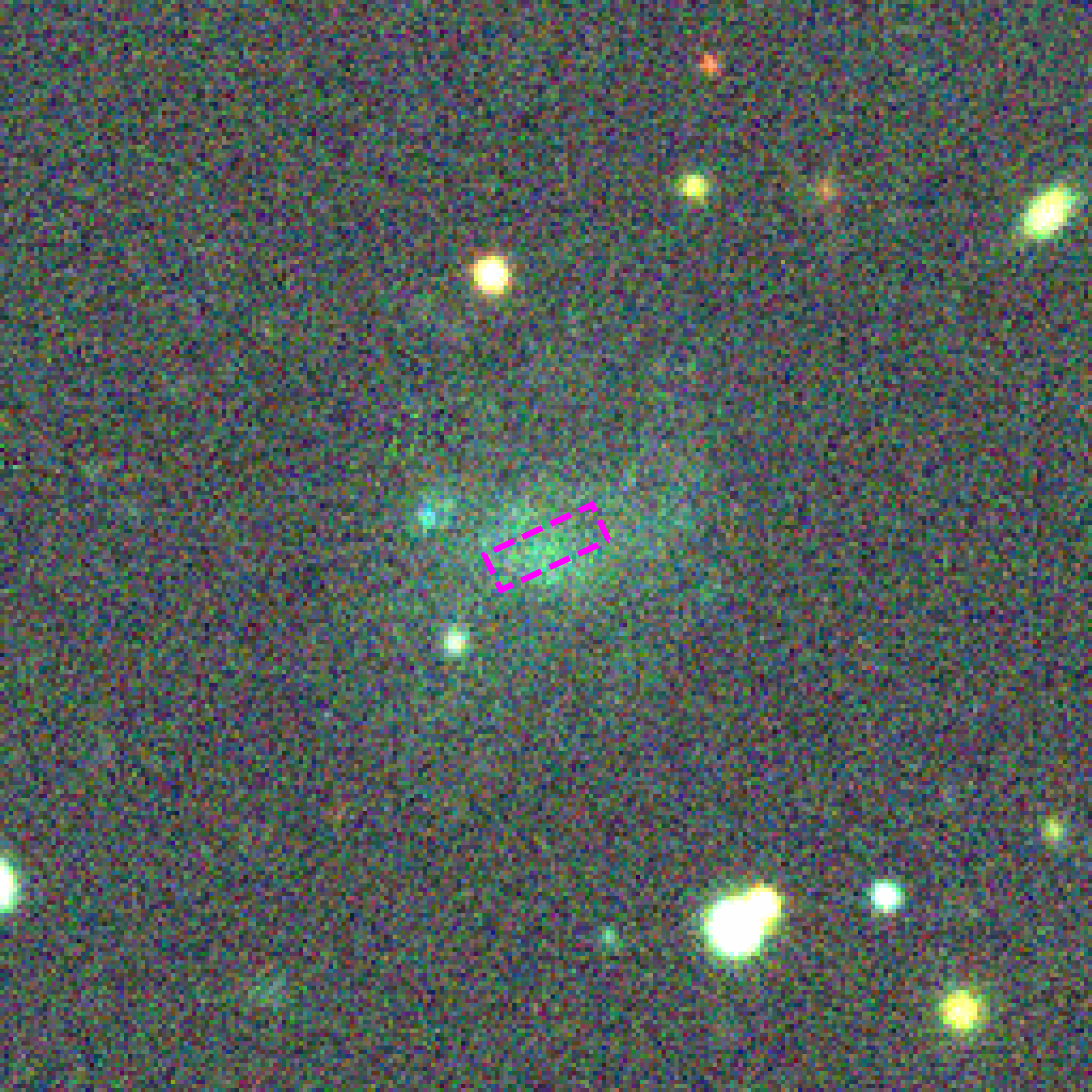}\vspace{-2.5em}
\end{minipage}
\hspace{0.001\textwidth}
\begin{minipage}[t]{0.71\textwidth}
\centering
\subcaption*{\textbf{LBT-7}}
\includegraphics[width=\linewidth]{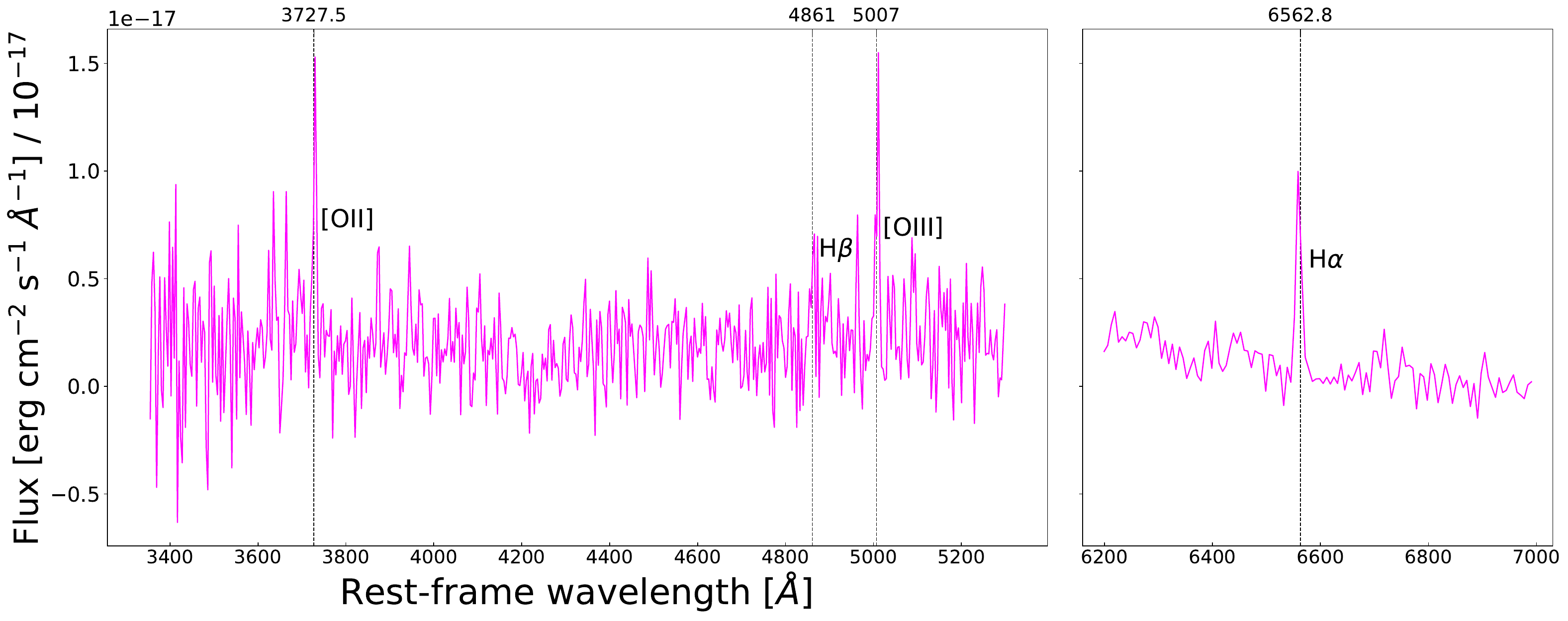}\vspace{-2.5em}
\label{435 spectrum}
\end{minipage}
\end{minipage}
\vspace{2em}
\caption*{\raggedright \textbf{Fig.~B.1.} continued.}
\label{example}
\end{figure*}

\begin{figure*}[ht]
\centering
\begin{minipage}[t]{0.27\textwidth}
\centering
\subcaption*{\textbf{LBT-8}}
\includegraphics[width=\linewidth]{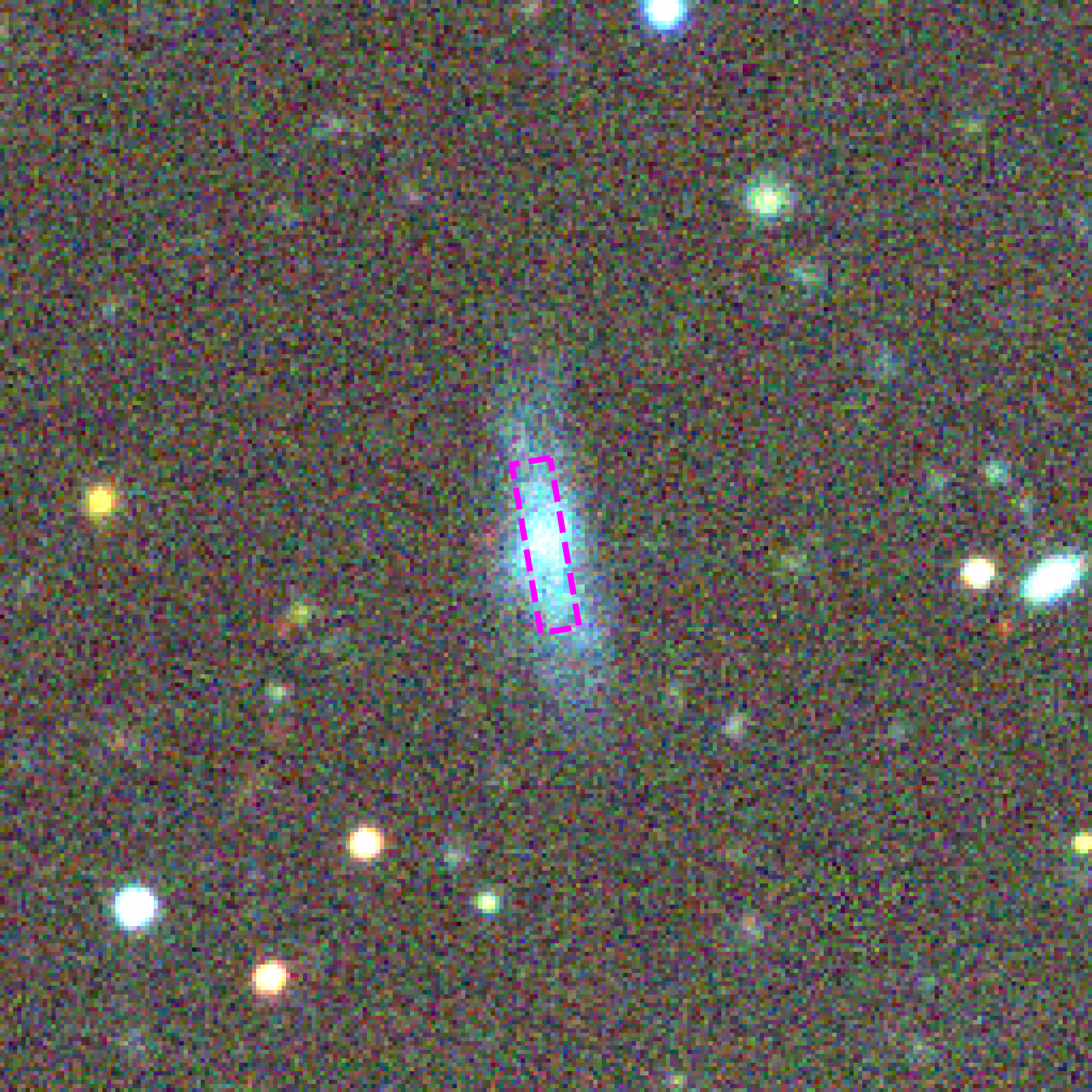}\vspace{-2.5em}
\label{435 cutout}
\end{minipage}
\hspace{0.001\textwidth}
\begin{minipage}[t]{0.71\textwidth}
\centering
\subcaption*{\textbf{LBT-8}}
\includegraphics[width=\linewidth]{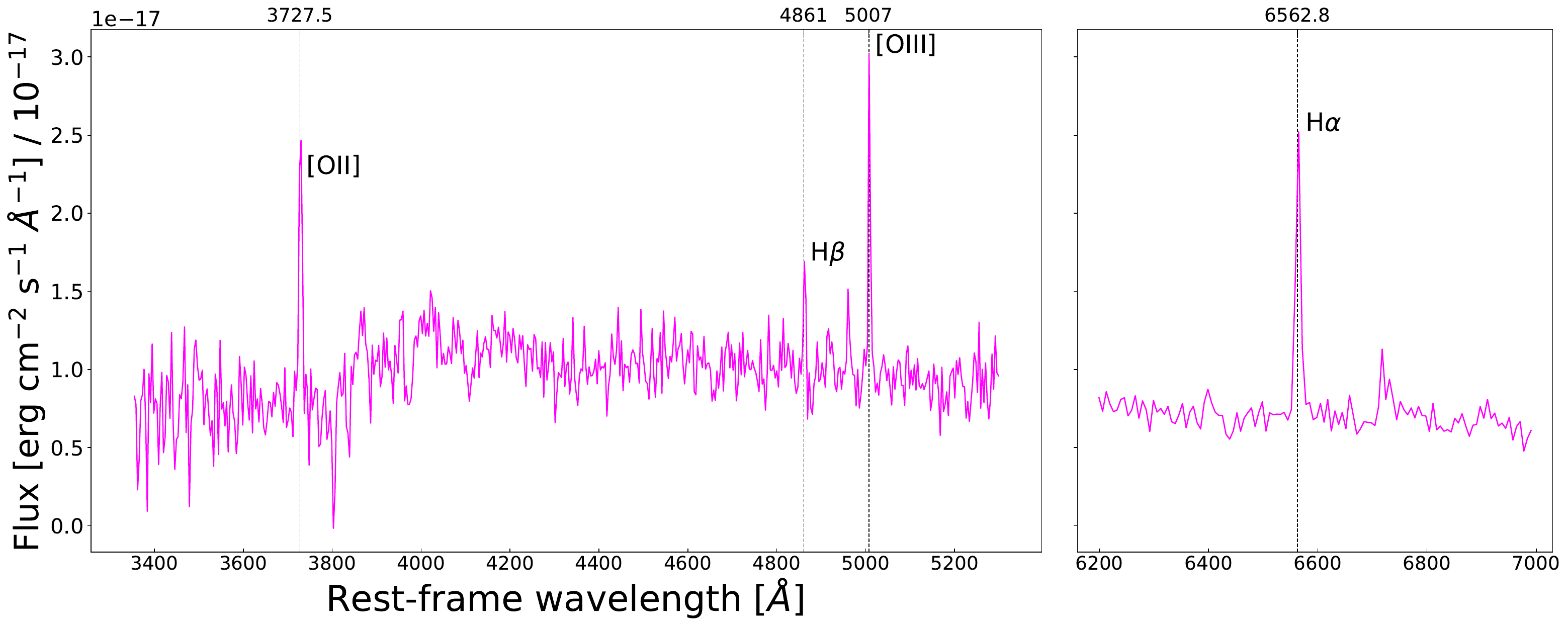}\vspace{-2.5em}
\label{435 spectrum}
\end{minipage}
\label{example}
\end{figure*}

\begin{figure*}[ht]
\centering
\begin{minipage}[t]{0.27\textwidth}
\centering
\subcaption*{LBT-9}
\includegraphics[width=\linewidth]{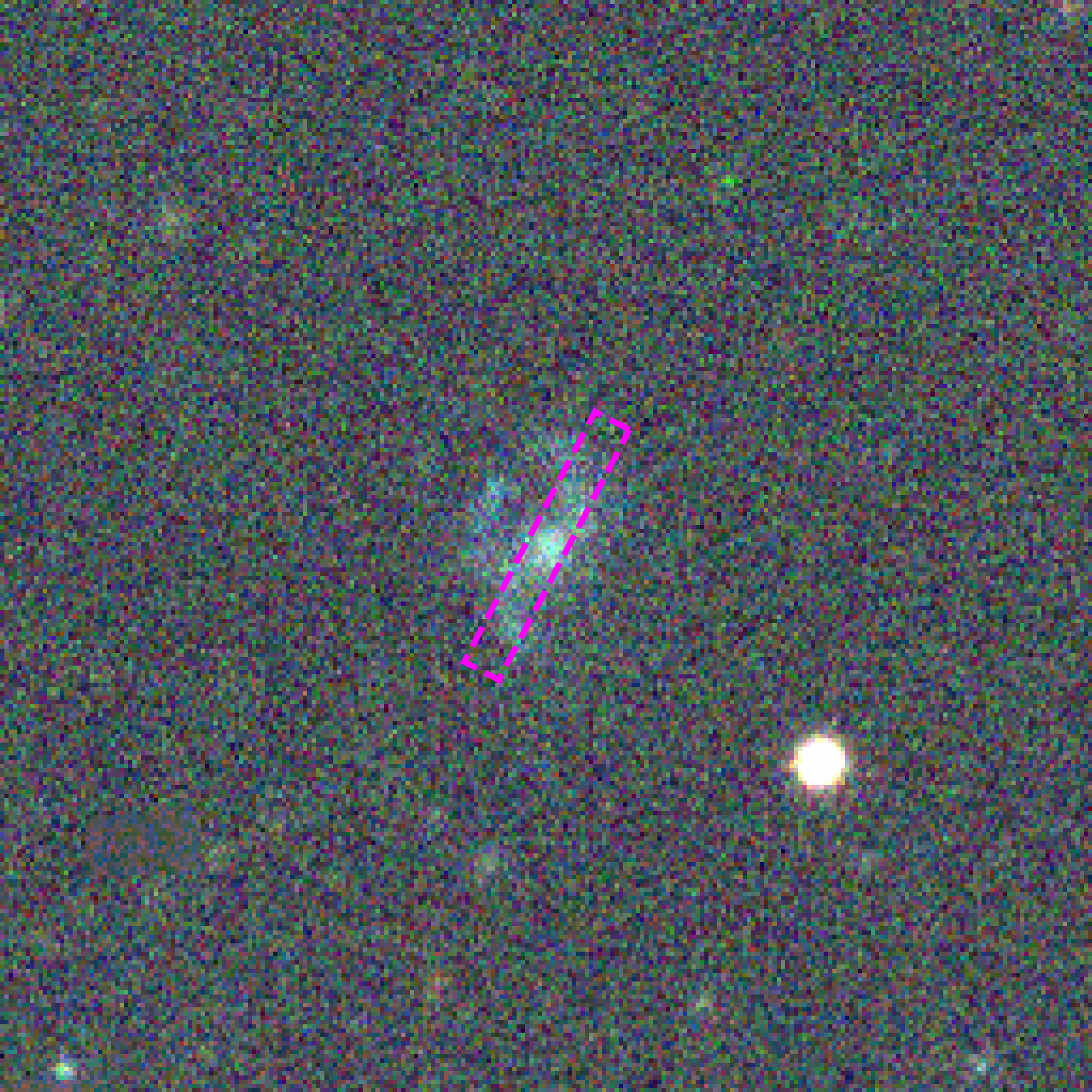}\vspace{-2.5em}
\label{435 cutout}
\end{minipage}
\hspace{0.001\textwidth}
\begin{minipage}[t]{0.71\textwidth}
\centering
\subcaption*{LBT-9}
\includegraphics[width=\linewidth]{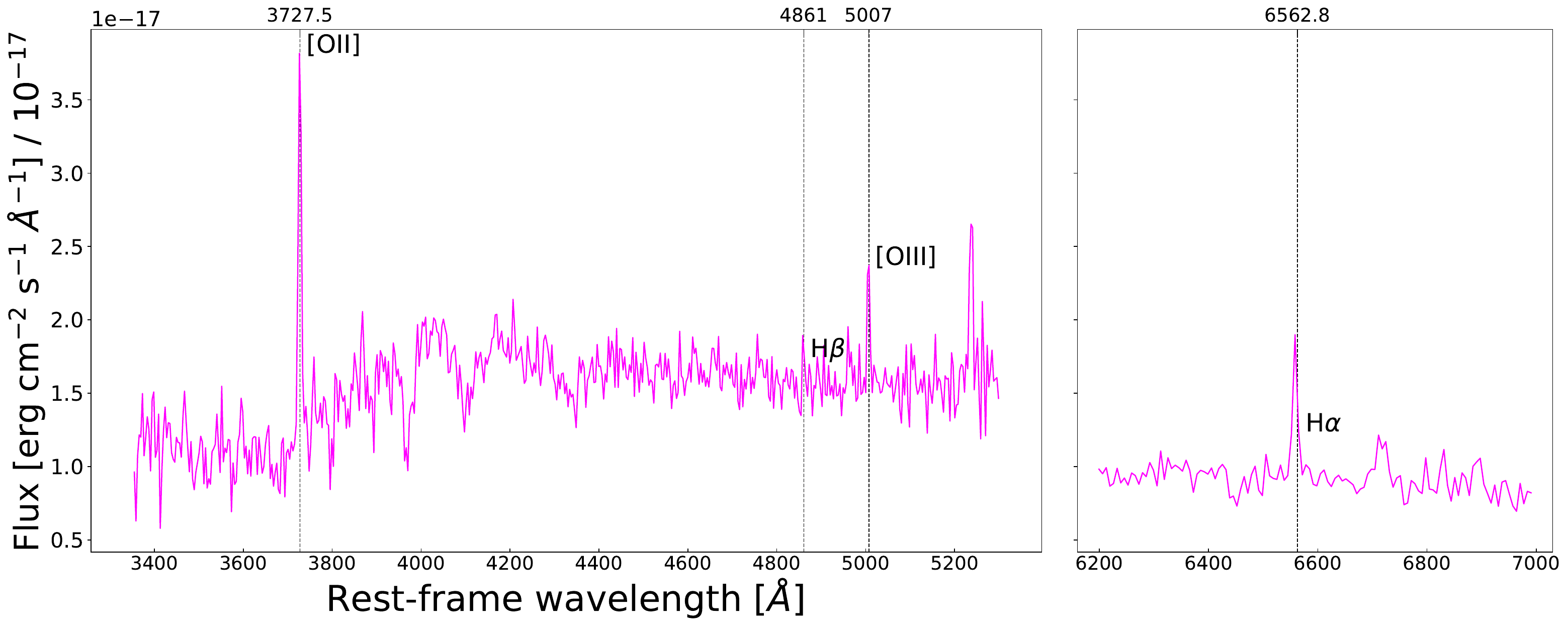}\vspace{-2.5em}
\label{435 spectrum}
\end{minipage}
\label{example}
\end{figure*}

\begin{figure*}[ht]
\centering
\begin{minipage}[t]{0.27\textwidth}
\centering
\subcaption*{\textbf{LBT-10}}
\includegraphics[width=\linewidth]{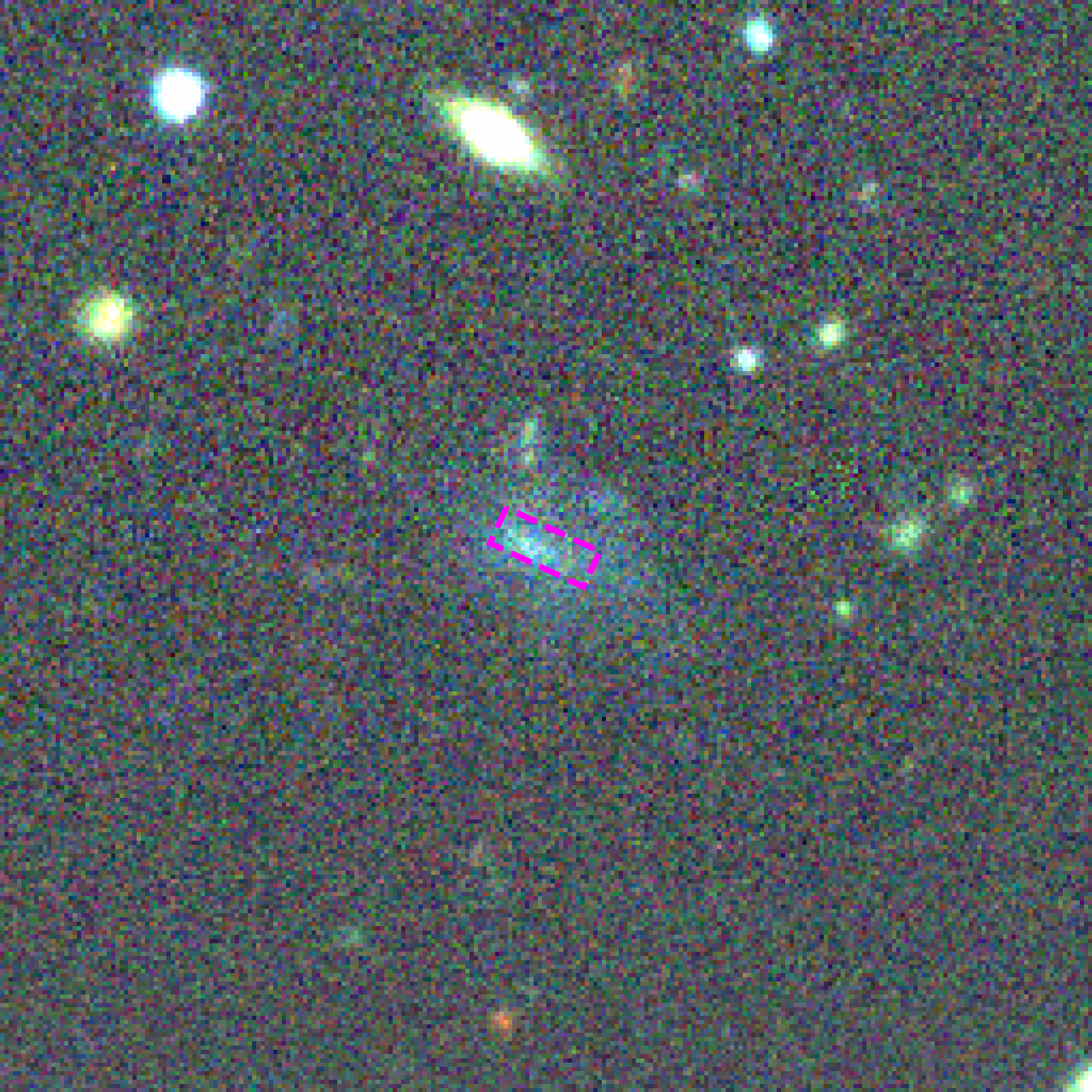}\vspace{-2.5em}
\label{435 cutout}
\end{minipage}
\hspace{0.001\textwidth}
\begin{minipage}[t]{0.71\textwidth}
\centering
\subcaption*{\textbf{LBT-10}}
\includegraphics[width=\linewidth]{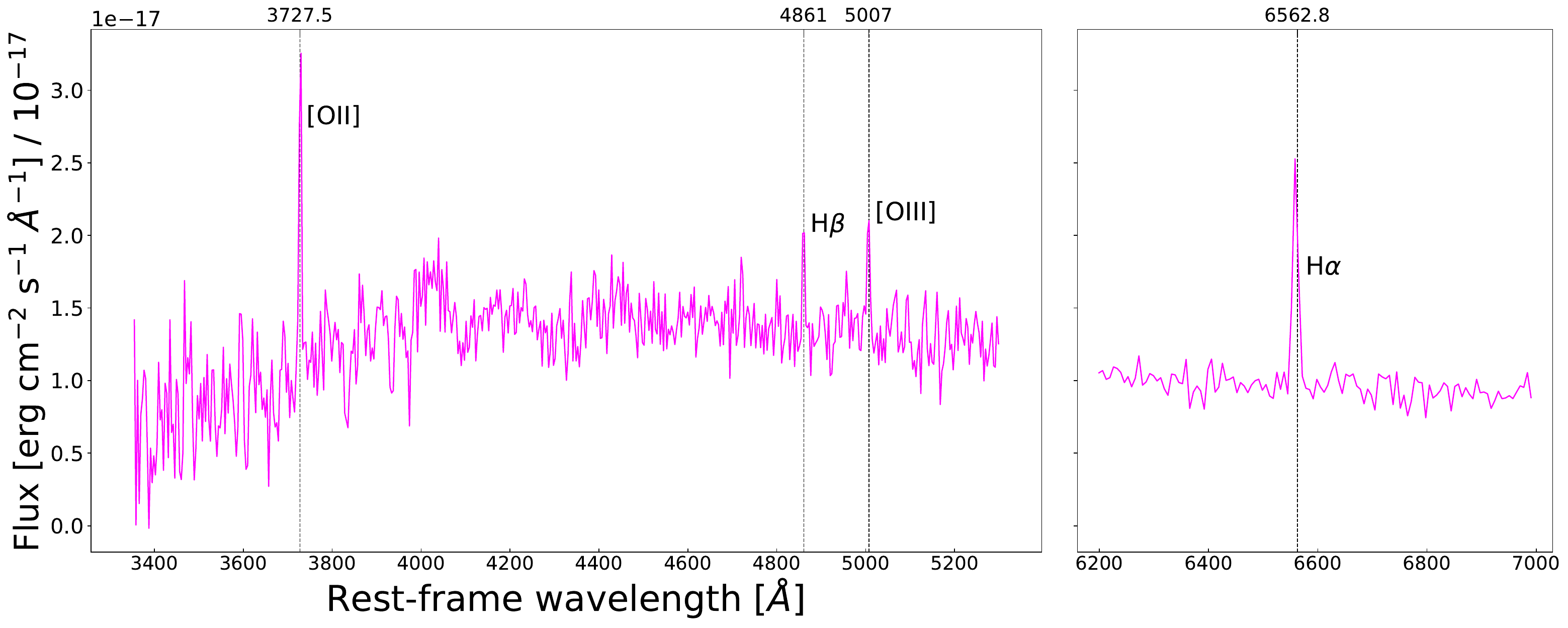}\vspace{-2.5em}
\label{435 spectrum}
\end{minipage}
\label{example}
\end{figure*}

\begin{figure*}[ht]
\centering
\begin{minipage}[t]{0.27\textwidth}
\centering
\subcaption*{\textbf{LBT-11}}
\includegraphics[width=\linewidth]{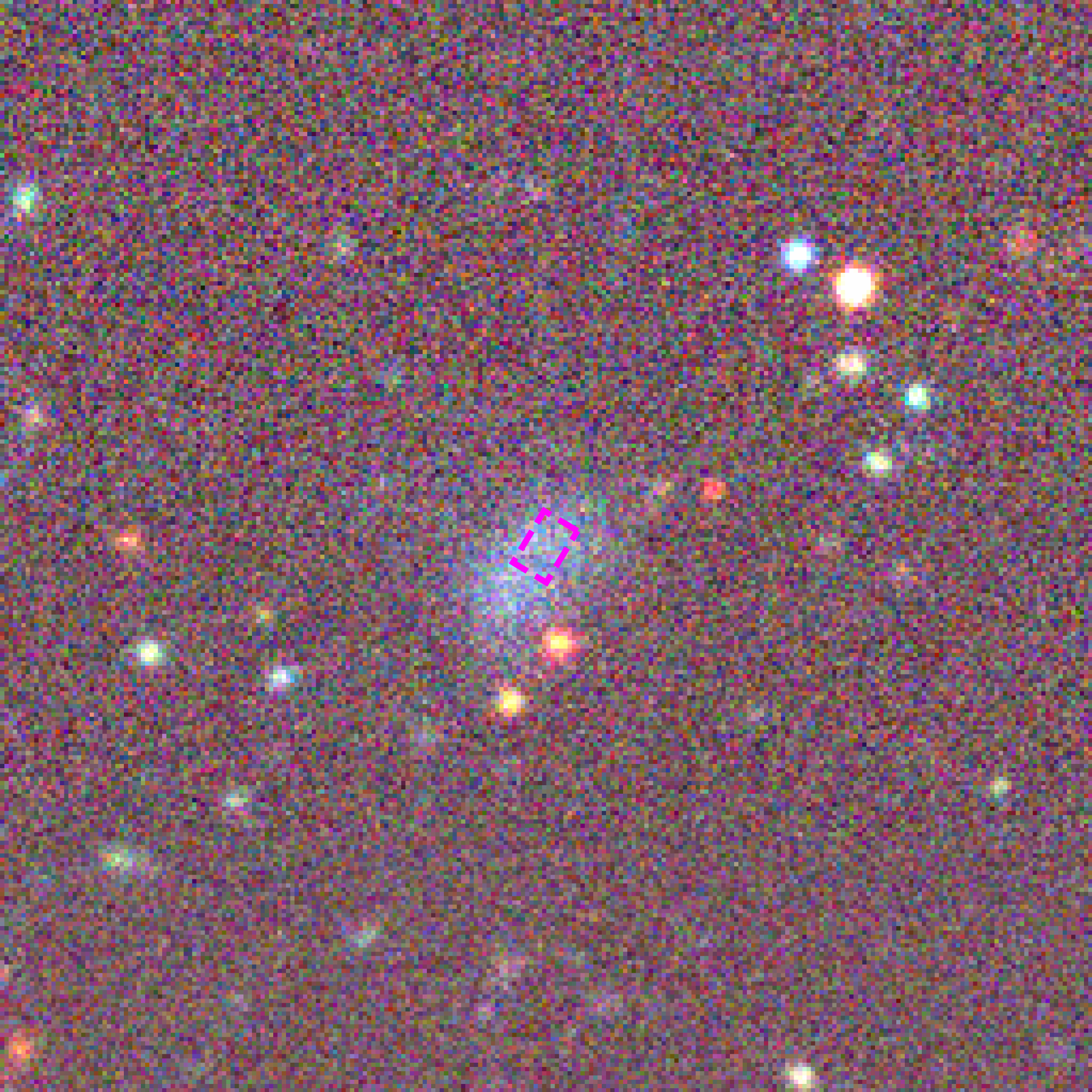}\vspace{-2.5em}
\label{435 cutout}
\end{minipage}
\hspace{0.001\textwidth}
\begin{minipage}[t]{0.71\textwidth}
\centering
\subcaption*{\textbf{LBT-11}}
\includegraphics[width=\linewidth]{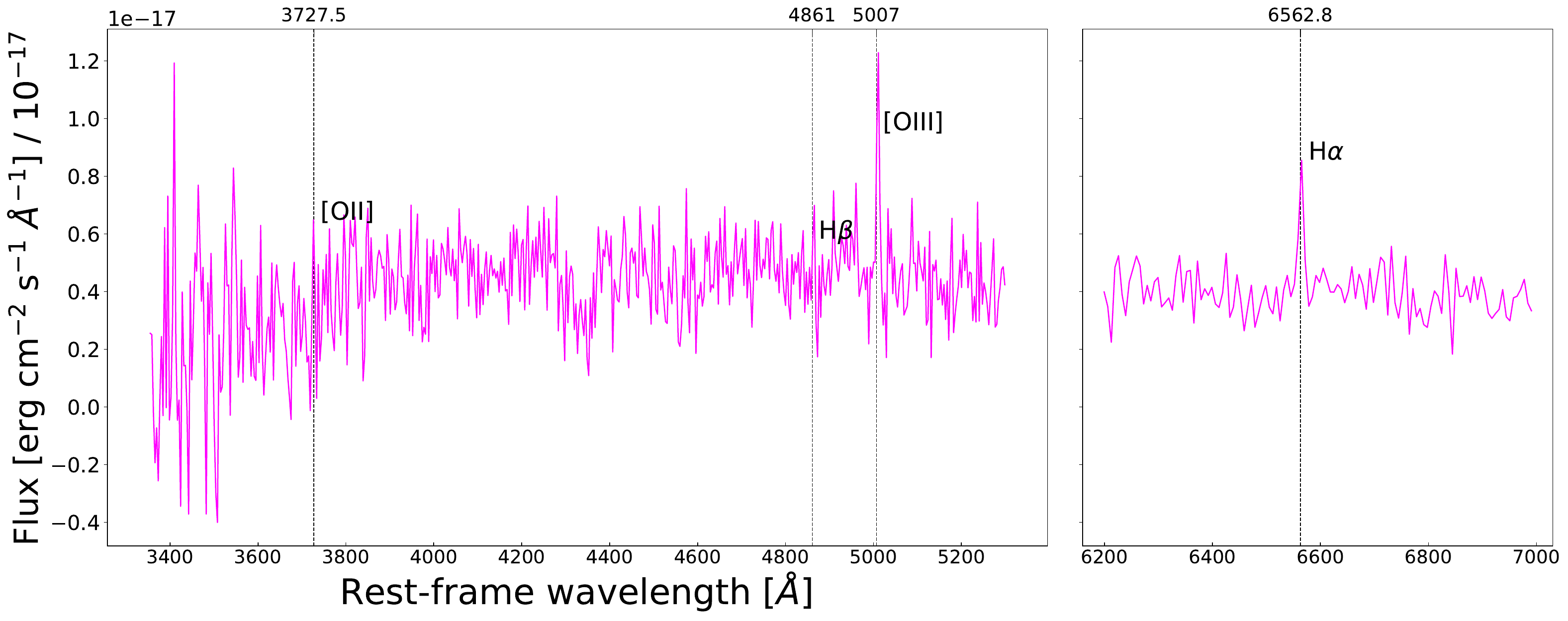}\vspace{-2.5em}
\label{435 spectrum}
\end{minipage}
\vspace{2em}
\caption*{\raggedright \textbf{Fig.~B.1.} continued.}
\label{example}
\end{figure*}

\begin{figure*}[ht]
\centering
\begin{minipage}[t]{0.27\textwidth}
\centering
\subcaption*{\textbf{LBT-12}}
\includegraphics[width=\linewidth]{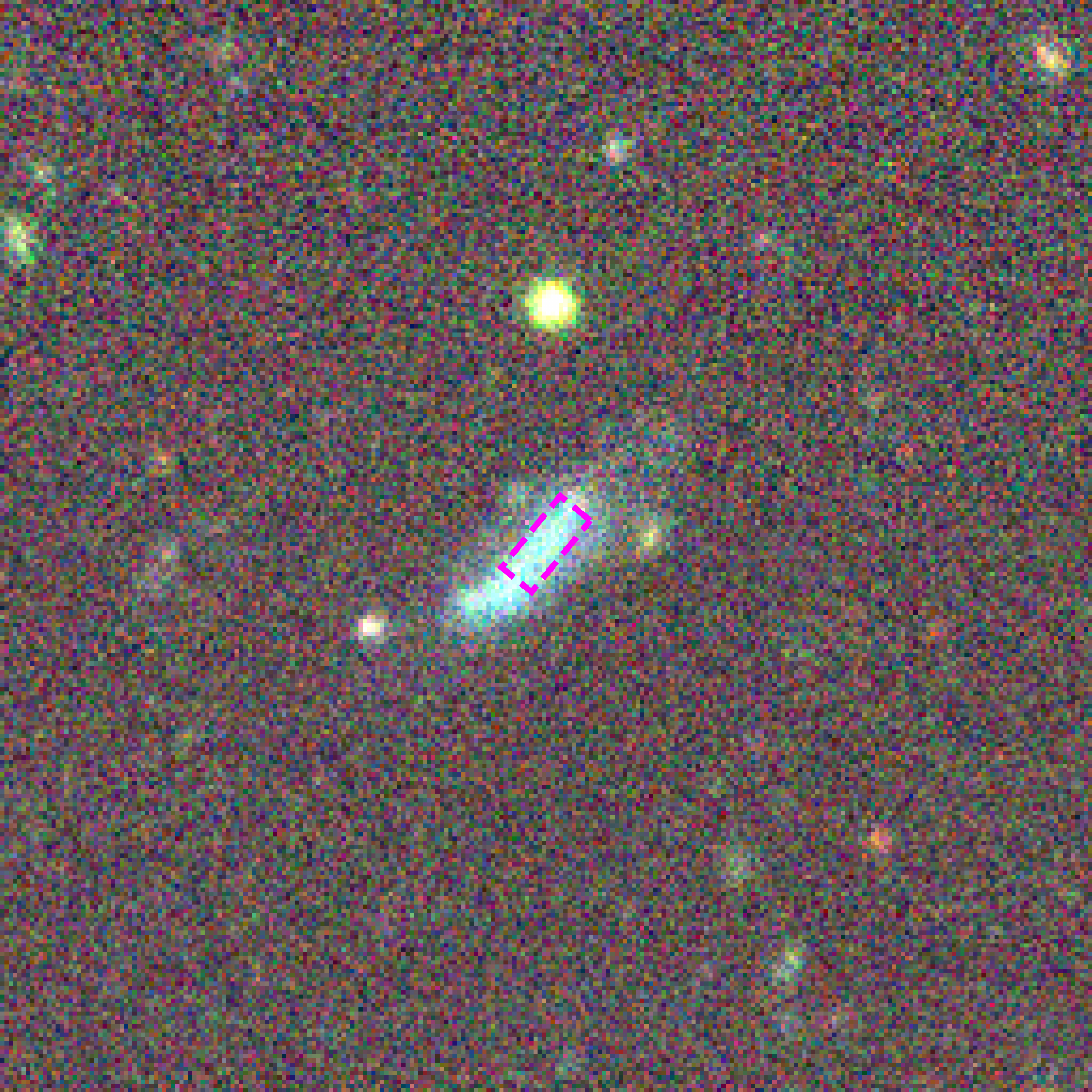}\vspace{-2.5em}
\label{435 cutout}
\end{minipage}
\hspace{0.001\textwidth}
\begin{minipage}[t]{0.71\textwidth}
\centering
\subcaption*{\textbf{LBT-12}}
\includegraphics[width=\linewidth]{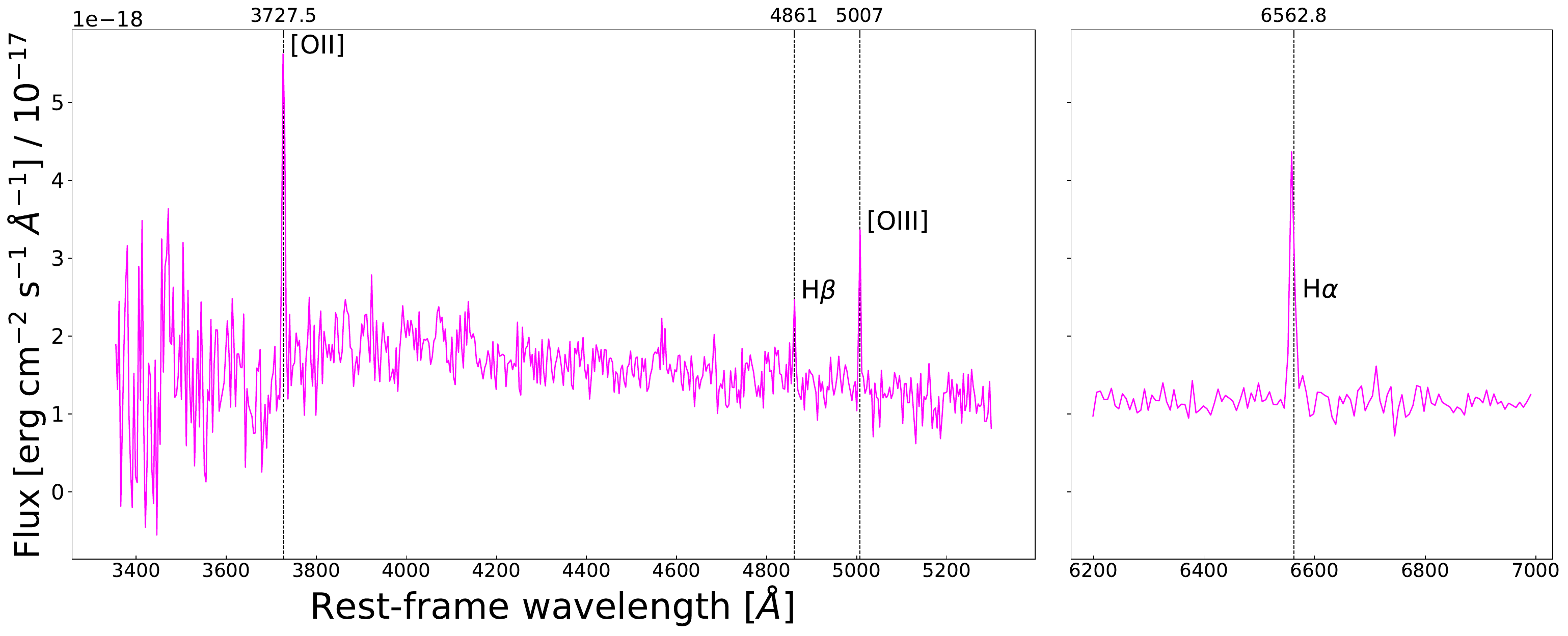}
\label{435 spectrum}\vspace{-2.5em}
\end{minipage}
\label{example}
\end{figure*}

\begin{figure*}[ht]
\centering
\begin{minipage}[t]{0.27\textwidth}
\centering
\subcaption*{LBT-13}
\includegraphics[width=\linewidth]{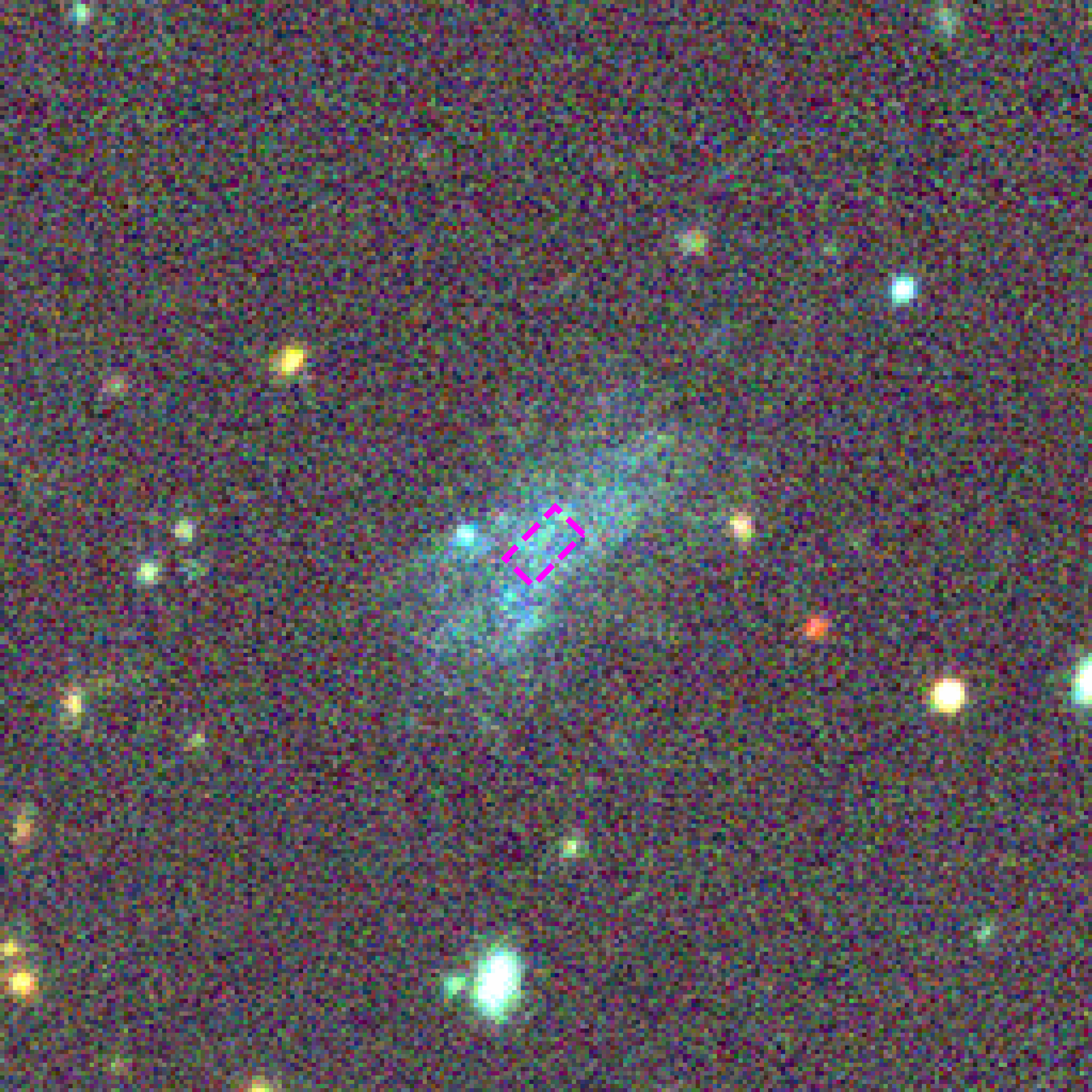}\vspace{-2.5em}
\label{435 cutout}
\end{minipage}
\hspace{0.001\textwidth}
\begin{minipage}[t]{0.71\textwidth}
\centering
\subcaption*{LBT-13}
\includegraphics[width=\linewidth]{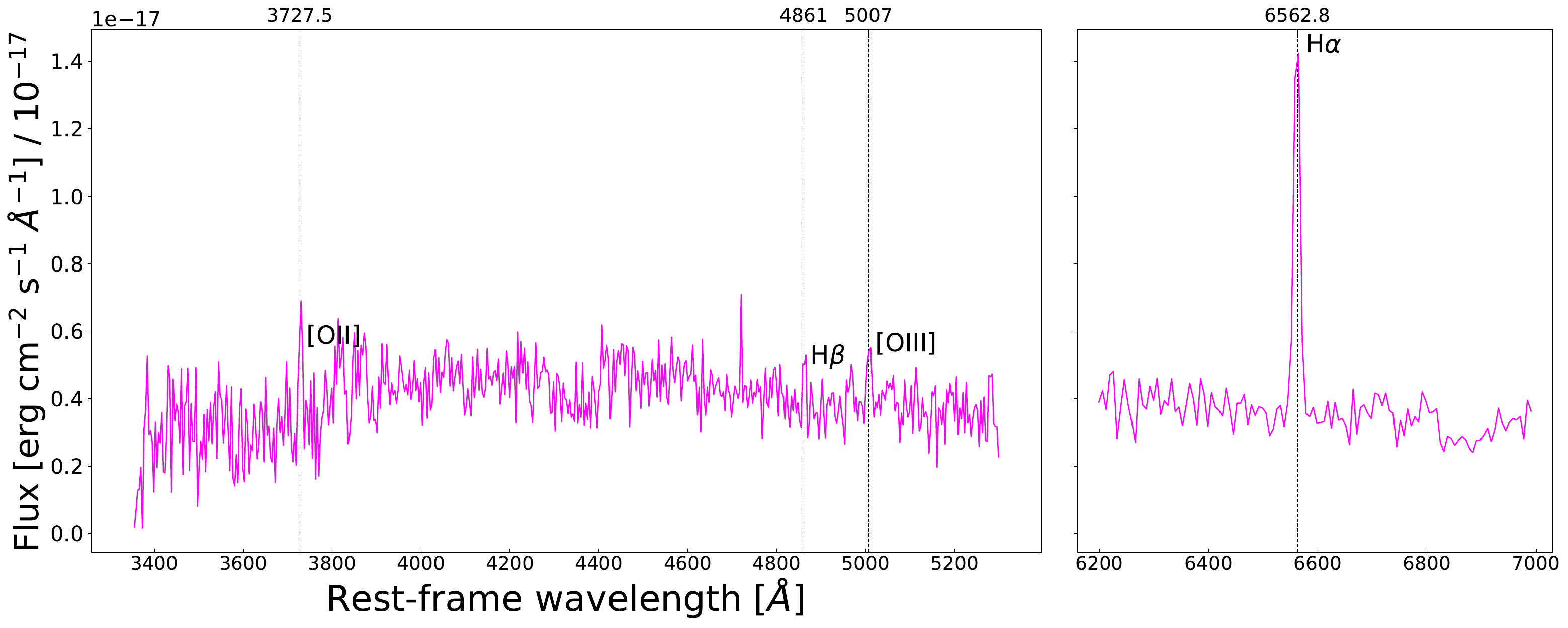}\vspace{-2.5em}
\label{435 spectrum}
\end{minipage}
\label{example}
\end{figure*}

\begin{figure*}[ht]
\centering
\begin{minipage}[t]{0.27\textwidth}
\centering
\subcaption*{OHP-1}
\includegraphics[width=\linewidth]{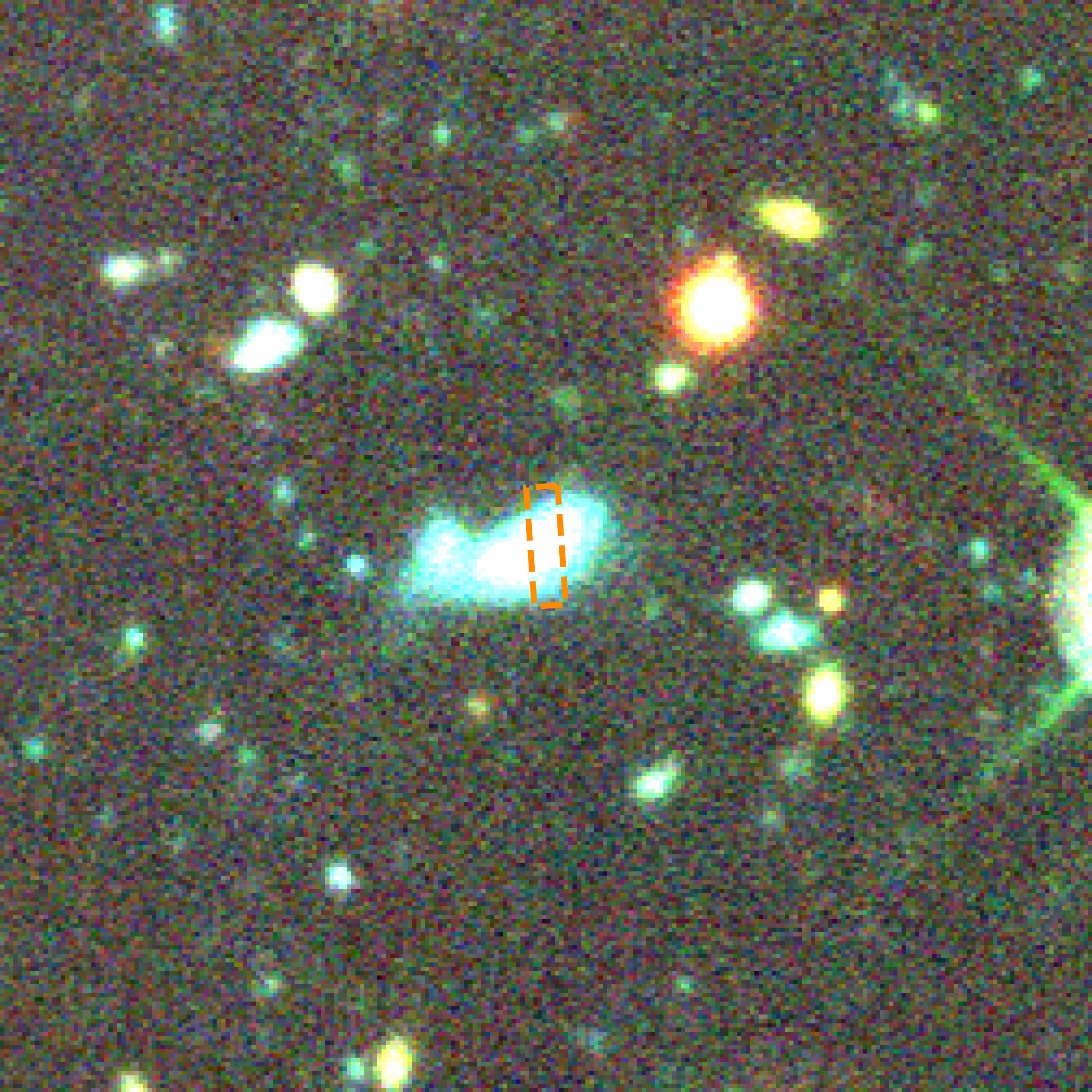}\vspace{-2.5em}
\label{435 cutout}
\end{minipage}
\hspace{0.001\textwidth}
\begin{minipage}[t]{0.71\textwidth}
\centering
\subcaption*{OHP-1}
\includegraphics[width=\linewidth]{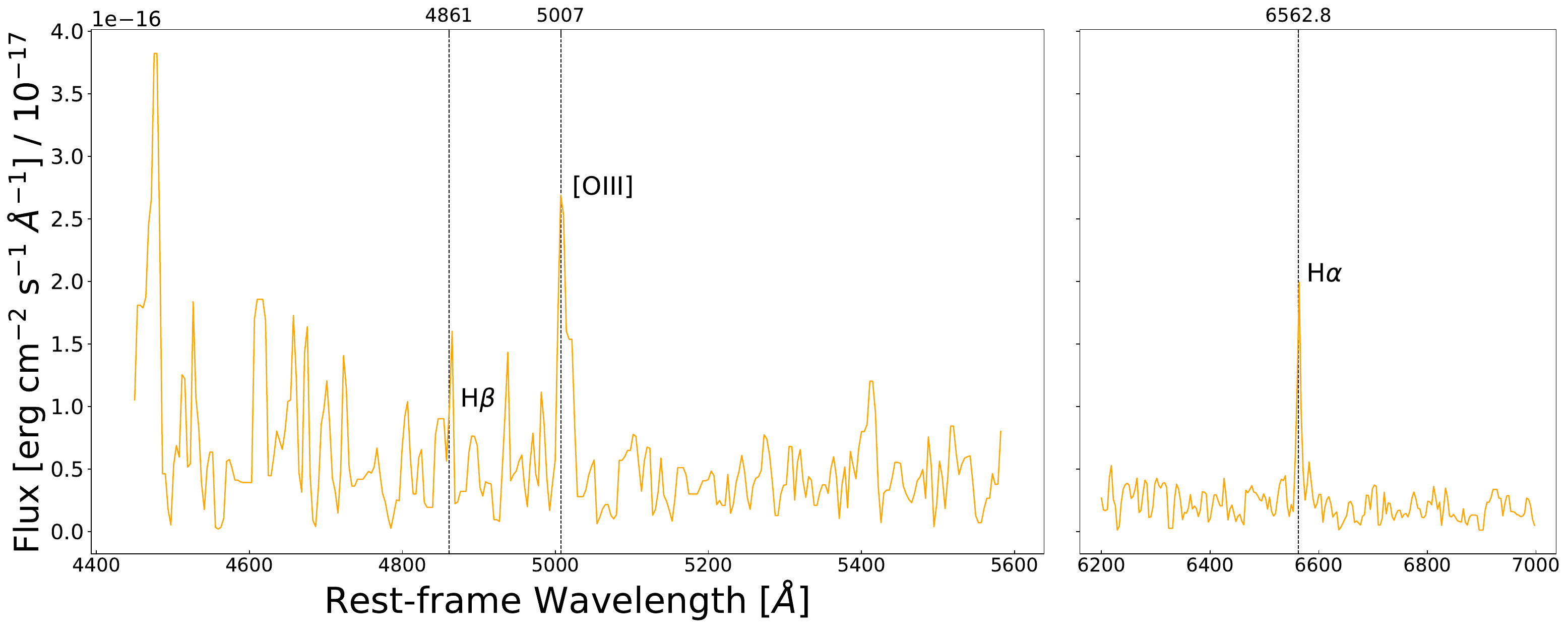}\vspace{-2.5em}
\label{435 spectrum}
\end{minipage}
\label{example}
\end{figure*}

\begin{figure*}[ht]
\centering
\begin{minipage}[t]{0.27\textwidth}
\centering
\subcaption*{OHP-2}
\includegraphics[width=\linewidth]{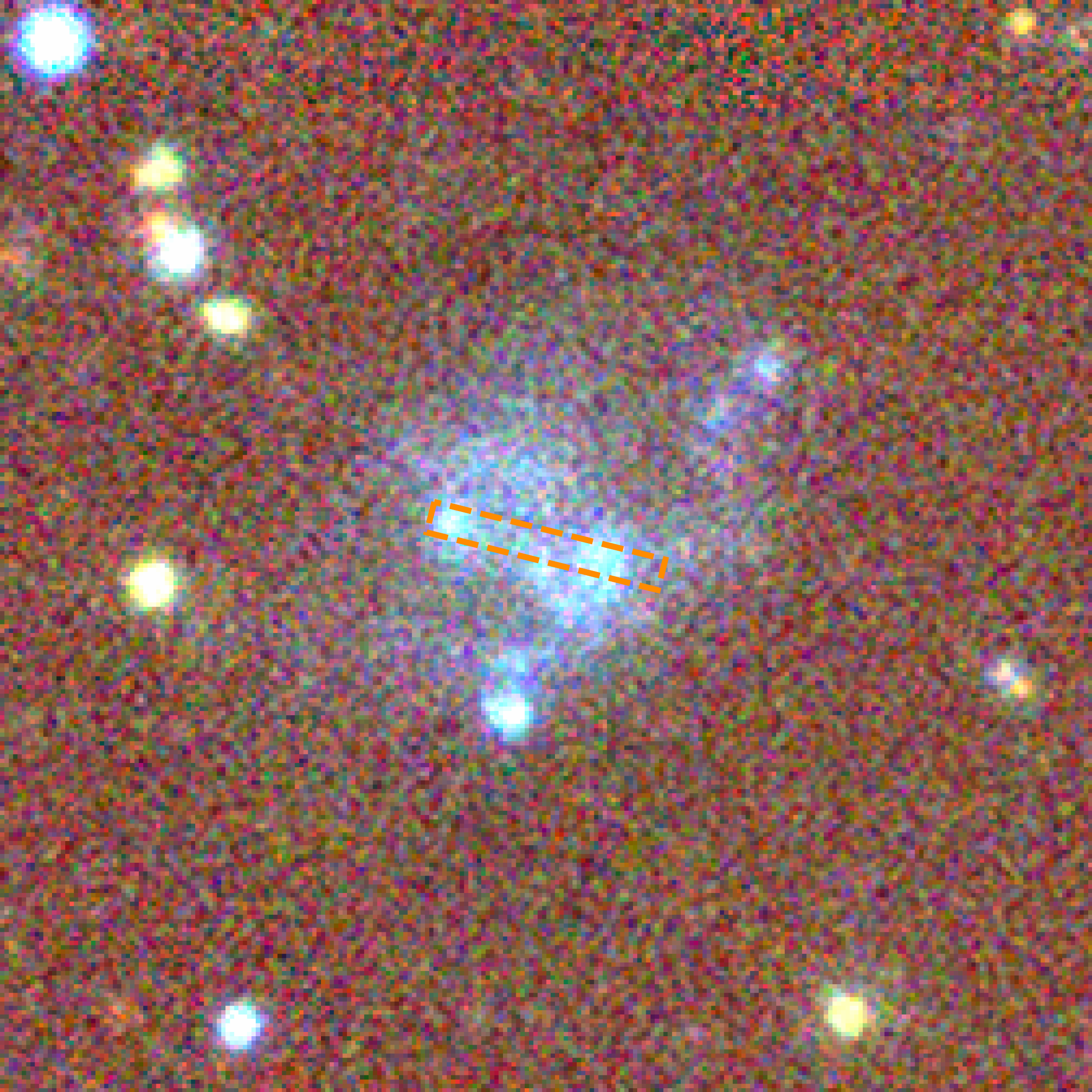}\vspace{-2.5em}
\label{435 cutout}
\end{minipage}
\hspace{0.001\textwidth}
\begin{minipage}[t]{0.71\textwidth}
\centering
\subcaption*{OHP-2}
\includegraphics[width=\linewidth]{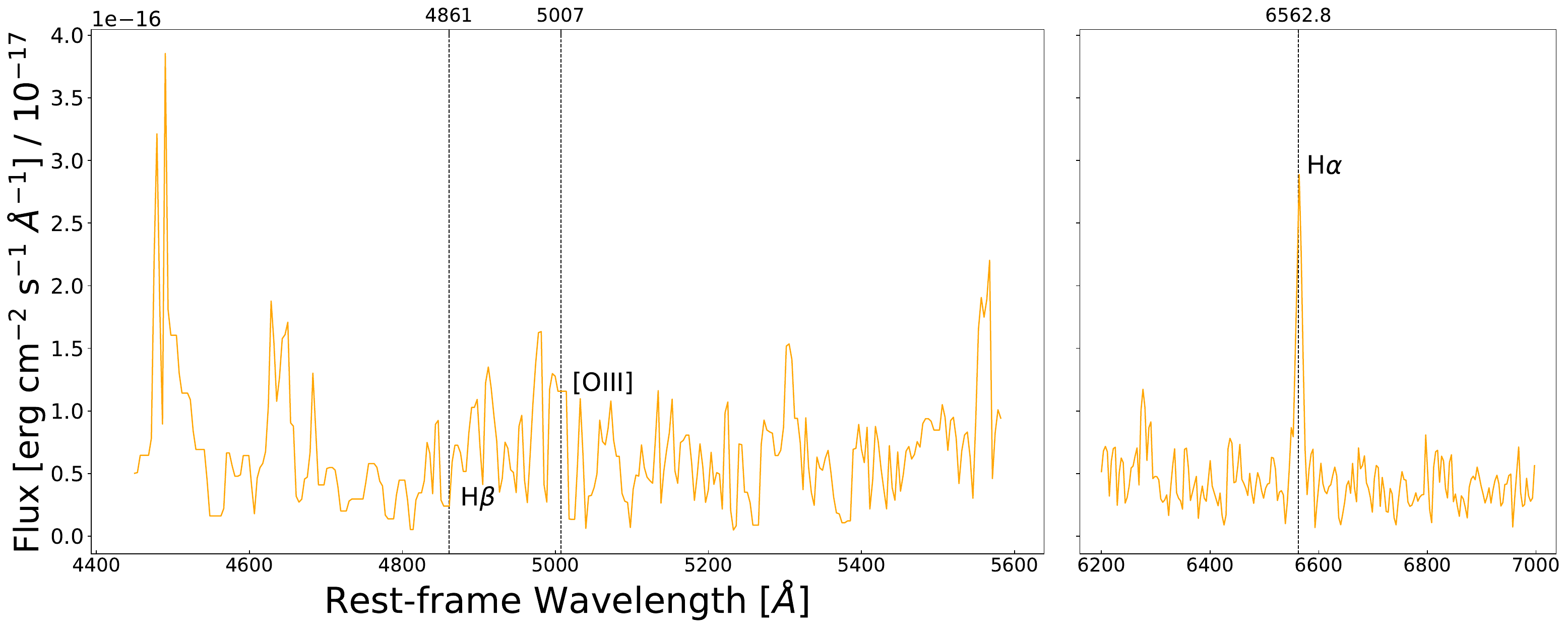}\vspace{-2.5em}
\label{435 spectrum}
\end{minipage}
\vspace{2em}
\caption*{\raggedright \textbf{Fig.~B.1.} continued.}
\label{example}
\end{figure*}

\begin{figure*}[ht]
\centering
\begin{minipage}[t]{0.27\textwidth}
\centering
\subcaption*{\textbf{OHP-3}}
\includegraphics[width=\linewidth]{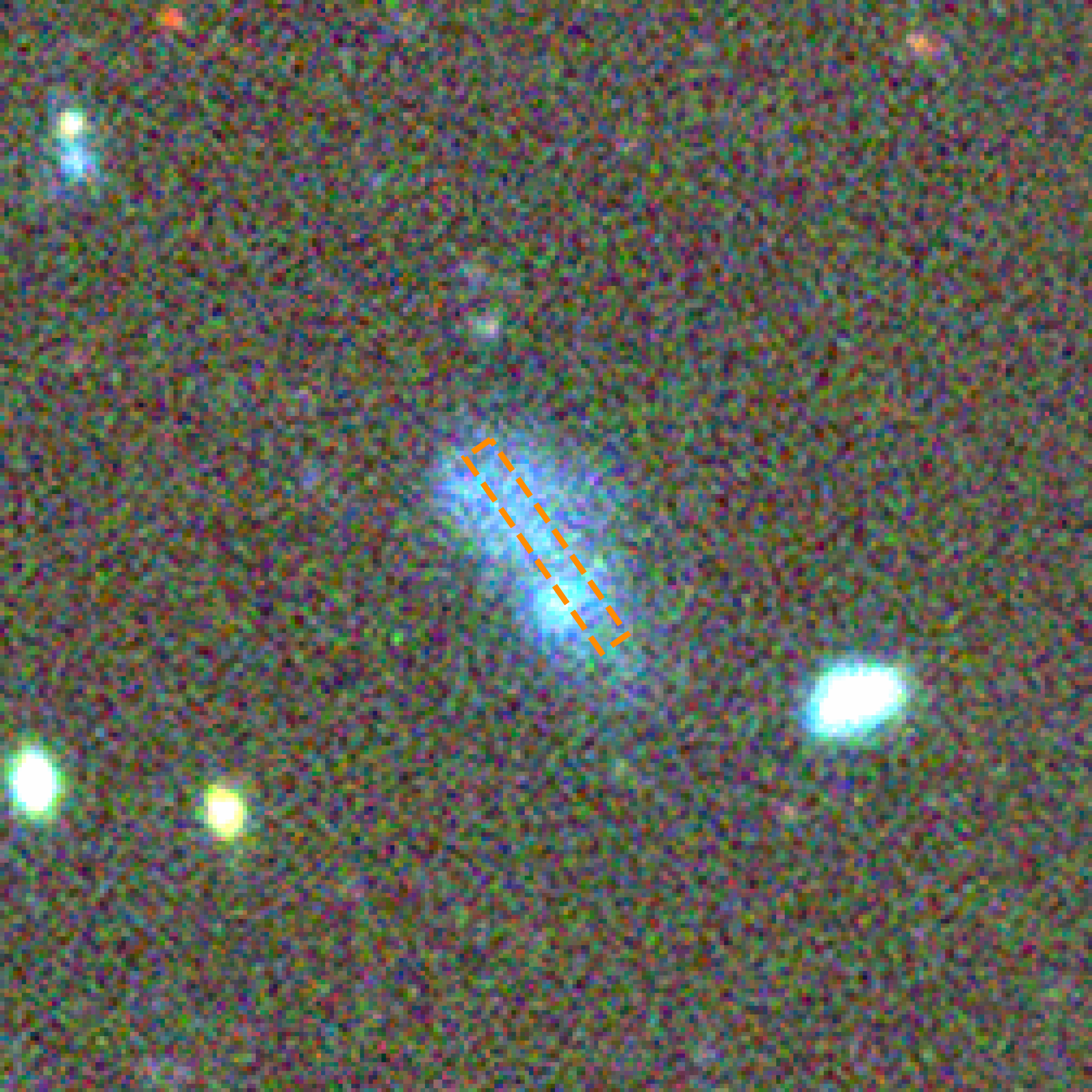}\vspace{-2.5em}
\label{435 cutout}
\end{minipage}
\hspace{0.001\textwidth}
\begin{minipage}[t]{0.71\textwidth}
\centering
\subcaption*{\textbf{OHP-3}}
\includegraphics[width=\linewidth]{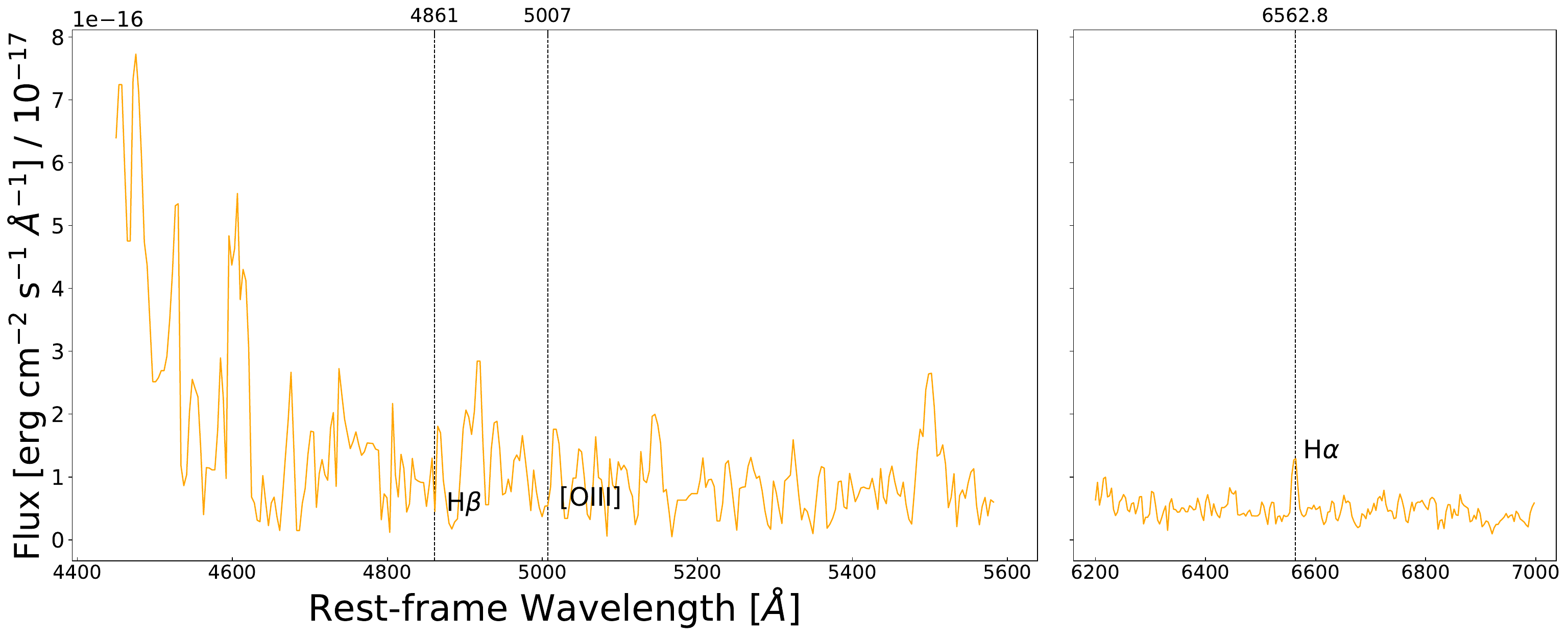}\vspace{-2.5em}
\label{435 spectrum}
\end{minipage}
\label{example}
\end{figure*}

\begin{figure*}[ht]
\centering
\begin{minipage}[t]{0.27\textwidth}
\centering
\subcaption*{OHP-4}
\includegraphics[width=\linewidth]{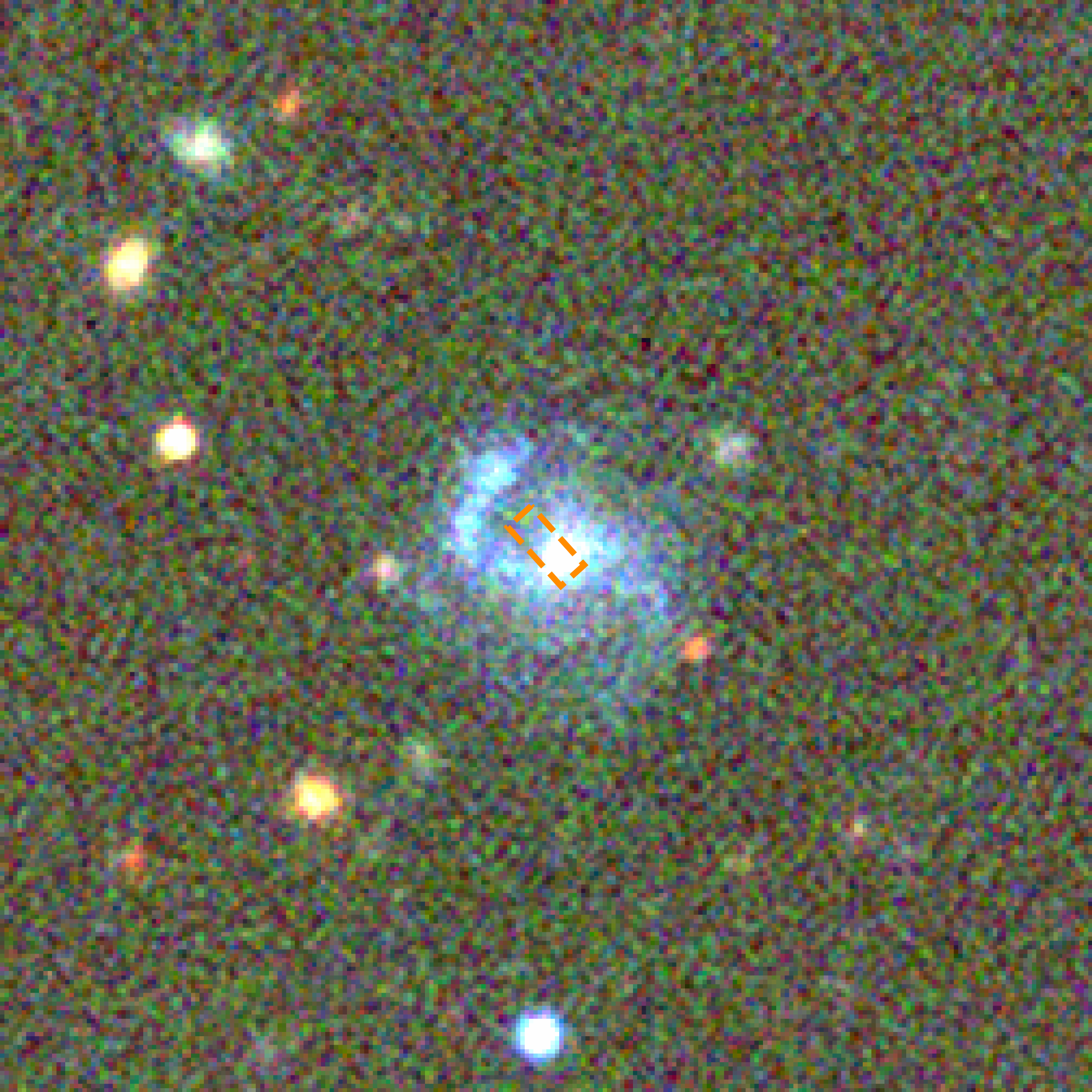}
\label{435 cutout}
\end{minipage}
\hspace{0.001\textwidth}
\begin{minipage}[t]{0.71\textwidth}
\centering
\subcaption*{OHP-4}
\includegraphics[width=\linewidth]{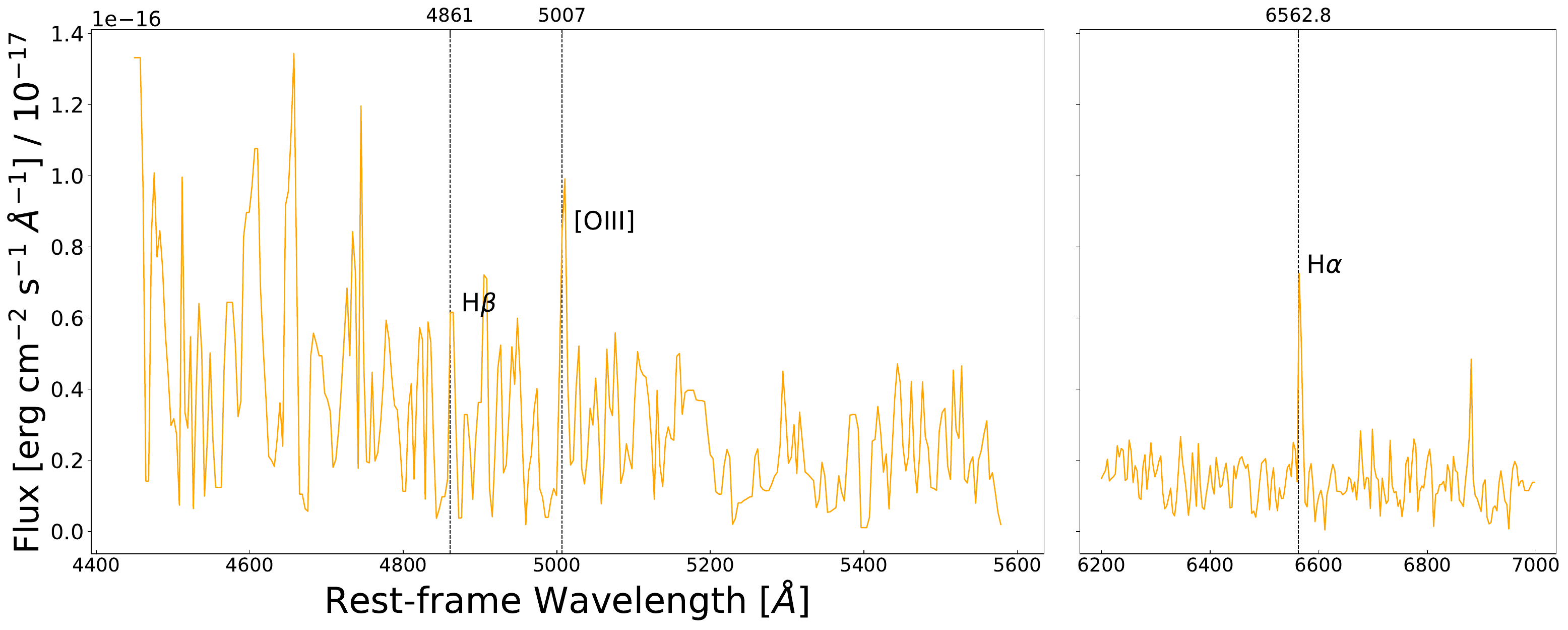}
\label{435 spectrum}
\end{minipage}
\vspace{1em}
\caption*{\raggedright \textbf{Fig.~B.1.} continued.}
\end{figure*}

\twocolumn

\section{Local peculiar velocities}
\label{Appendix_D}
The sample of galaxies studied in this work consists of nearby galaxies for which peculiar motions can potentially affect the inferred distances and consequently the effective radius estimations, fundamental for classifying a galaxy as a UDG.
To check the impact of the peculiar velocities on our analysis, we used the reconstructed velocity field created by \cite{Carrick2015}.
In their paper, \cite{Carrick2015} uses the $2M++$ redshift compilation \citep{Lavaux2011}, which maps the local galaxy distribution out to a distance of approximately $200h^{-1}\mathrm{Mpc}$, to build a model and estimate peculiar velocities of galaxies.
Assuming linear perturbation theory, the observed galaxy density contrast is converted into a prediction of the peculiar velocity field through \citep[Eq.~10 in ][]{Carrick2015}:

\begin{equation}
\label{eq:peculiar_velocity}
\mathbf{v}(\mathbf{r}) =
\frac{\beta}{4\pi}
\int
\delta_g(\mathbf{r}')
\frac{\mathbf{r}'-\mathbf{r}}
{\left|\mathbf{r}'-\mathbf{r}\right|^{3}}
\,\mathrm{d}^{3}\mathbf{r}' + \mathbf{V_{ext}}.
\end{equation}
where $\mathbf{V_{ext}}$ encapsulates contributions from beyond the reconstruction volume, and to first order, can be approximated as a dipole or residual bulk flow.
$\delta_g$ is the galaxy density contrast and $\beta=f/b$ is the ratio between $f$, the linear growth rate of cosmic structure, with $f=\Omega^{\gamma}_m$ where $\gamma = 0.55$, according to the $\Lambda$CDM model and $b$, the linear galaxy bias.
The value of $\beta$ was calibrated by comparing the reconstructed velocity field with independent peculiar velocity measurements obtained from the Tully-Fisher relation and Type Ia supernovae.
We interpolate this 3D reconstructed velocity field at the position of each galaxy in our sample and compute the corresponding line-of-sight peculiar velocity.
After computing the peculiar velocities for our sample, except for OHP-4, which lies outside the reconstructed volume ($200 h^{-1} \mathrm{Mpc}$), we found that LBT-1, LBT-3, LBT-4, LBT-5, LBT-6, LBT-7, LBT-8, LBT-9, LBT-10, LBT-11, LBT-12, OHP-1, and OHP-3 (which correspond to $76$\% of the total sample) have reconstructed peculiar velocities amounting to only $1-7\%$ of their observed recessional velocities.  
At the same time, LBT-2, LBT-13, and OHP-2 (representing the remaining $18$\% of the sample) exhibit larger corrections, corresponding to $21$\%, $33$\%, and $23$\% of their recessional velocities, respectively. 
These findings show that peculiar velocities indeed affect all our galaxies' distance estimates. 
However, for 13 galaxies, the impact is expected to be modest. 
For the remaining three galaxies, the peculiar velocities could significantly affect the distance estimation, but none of them has been considered a UDG in this work.

\section{Spectral energy distribution fitting}
\label{Appendix_C}
In this Appendix, we compare the SFR obtained from $\rm{H{\alpha}}$ measurements as well as \Mstar~from \textit{g}-\textit{r} colour with those estimated from broadband SED fitting.
We perform SED fitting to assess how accurately the SFR can be estimated without using the $\rm{H{\alpha}}$ line, instead relying on deep optical measurements. 
We took advantage of the photometry obtained in Sect.~\ref {forced_photometry} and the spectroscopic redshift measurements (Sect.~\ref{Redshift Measurement}). 
We used all this information to estimate SFRs, dust attenuations, and \Mstar. 
Given that LBT-UDG candidates have limited photometric coverage (five optical bands from DES), with no information from the ultraviolet or infrared, we expect that SFR and attenuation estimates from broadband photometry are uncertain. 
To obtain the main physical properties of the sample of LBT-UDG candidates, we used the Code Investigating GALaxy Emission  \citep[\cigale,][]{CIGALE_Burgarella2005, CIGALE_Noll2009, CIGALE_Boquien2019}. 
The \cigale~tool is designed to estimate the physical parameters (i.e. SFR, \Mstar, dust luminosity, dust attenuation, AGN fraction) by comparing modelled galaxy SEDs with observed ones. 
Many authors \citep[e.g. ][]{CIGALE_Buat2014, CIGALE_Ciesla2016, CIGALE_Malek2018} have already presented the methodology and the strategy of the code. 
A more detailed description of the code is provided in \cite{CIGALE_Boquien2019}. 
\cigale~is widely used to study low- and high-redshift galaxies, normal star-forming, starburst, and quiescent galaxies, but it has not been used in the literature for LSBG/UDGs. 

Here we discuss the choice of modelling parameters for \cigale.
In this work, we used a delayed star formation history with instantaneous recent variation of the SFR, upwards (burst) or downwards (quench). 
This module, called \texttt{SFHDelayedBQ}, introduced by \cite{ciesla2017}, is parametrised as
\begin{equation}	
    \mathrm{SFR}(t) \propto
    	\begin{cases}
    	  t  e^{-t/\tau_\mathrm{main}}, & \text{when}\ t \leq t_0 \\
    	  \mathrm{r_{SFR}} \times \mathrm{SFR}(t=t_0), & \text{when}\ t>t_0,\\
    	\end{cases}
\end{equation}	

where $\tau_\mathrm{main}$ represents the e-folding time of the main stellar population,  $t_0$ is the time when a rapid burst or quench is allowed in the SFH, and $\rm{r_{SFR}}$ is the ratio between the SFR after $t_0$ and before $t_0$. 

To create a grid of templates that fits the age of the main stellar population in the galaxy, we used 26 evenly spaced values ranging from 1000 to 13,000 Myr. 
For simplicity, only five values in the range between 1\,000 and 20\,000 Myrs of the $\tau_\mathrm{main}$ parameter were used. 
For the time of the rapid burst/quench ($t_0$), we used values between 10 and 700 Myrs, and the ratio of the SFR after/before the $t_0$ event; nine parameters in the range 0-0.3 (we included only delayed SFH and delayed SFH with possible quench).
For the stellar population, we used the library of \cite{BC2003}, with subsolar metallicity (Z=0.008), and an IMF of \cite{Chabrier2003}  to combine the values obtained with the SFR measured using the line $\rm{H{\alpha}}$ and the relation \cite{Samuel}.  
We applied the \cite{CF2000} dust attenuation law, which takes into account a differential attenuation between young stars (age $<10^7$ years) and old stars (age $>10^7$ years).
Both young and old stellar populations are attenuated in the interstellar medium (ISM); however, young stars are additionally attenuated in the birth clouds (both attenuations are
modelled by a power law). 
In our analysis, we used fixed exponents of the power laws as initially defined in the work of \cite{CF2000}, equal to $-0.7$. 
We also used the original ratio of attenuation in the V band experienced by old and young stars, which is equal to $0.3$. 
For V-band attenuation in the ISM, we used a grid of 50 values in a linear distribution between $0$ and $0.2$.
With such grid parameters, we created 4\,563\,000 models  (351\,000 models per redshift bin).

As a quality check, we inspected the reduced $\chi^2$ distribution, as well as the probability distribution functions for the parameters used and obtained the main physical properties.   

\begin{figure}[ht]
\centering
\includegraphics[width=0.45\textwidth]{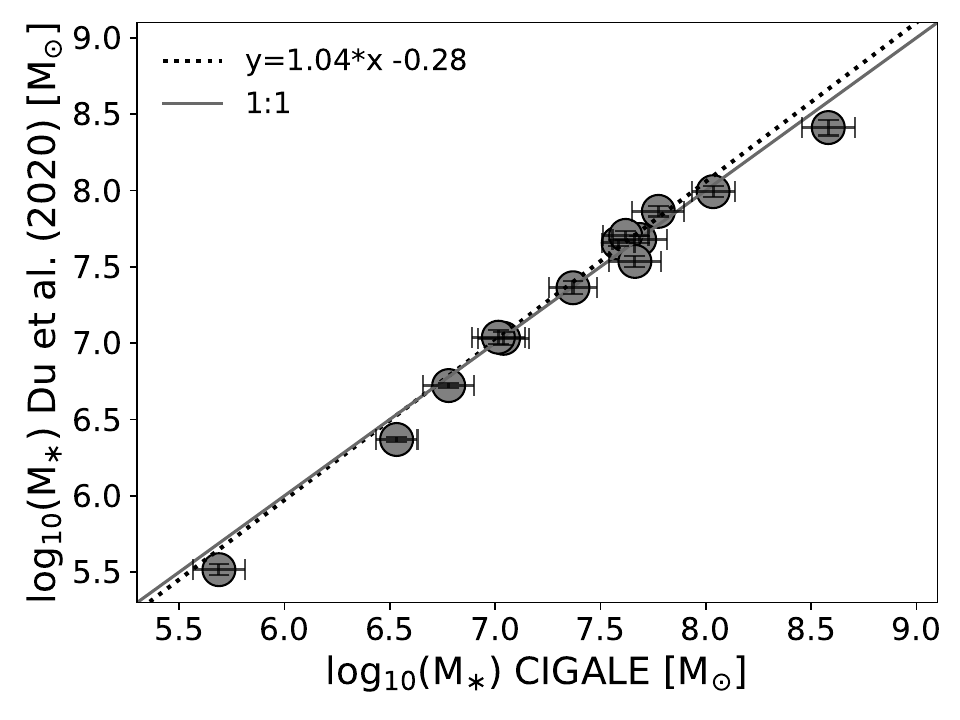}
  \caption{ Comparison of the \Mstar{ } values obtained from the \cigale~SED fitting code using five optical bands (x-axis) and those calculated using the (\textit{g}-\textit{r}) colour following the \cite{Du_2020} relation (full grey circles).
The solid line represents a 1:1 relation, while the dotted line represents a linear fit to the data.}

     \label{Mstar_cigale_Halpha}    
\end{figure}
We find agreement (with the offset lower than 0.1 dex) between \Mstar{ }estimated using the \cite{Du_2020} relation and those from the SED fitting using broadband optical data (see Fig.~\ref{Mstar_cigale_Halpha}). 
The offset between stellar masses estimated using different methods (from mass-to-light ratios to broadband SED fitting with various stellar population synthesis techniques) has already been described in the literature \citep{Conroy2009} and is estimated to reach up to 0.30 dex at $z\sim0$. 
In our case, the method of using only five optical bands for the SED fitting is very similar to a simple (\textit{g}-\textit{r}) colour method used by \cite{Du_2020}, and the final agreement was to be expected.

\begin{figure}[ht]
\centering
\includegraphics[width=0.45\textwidth]{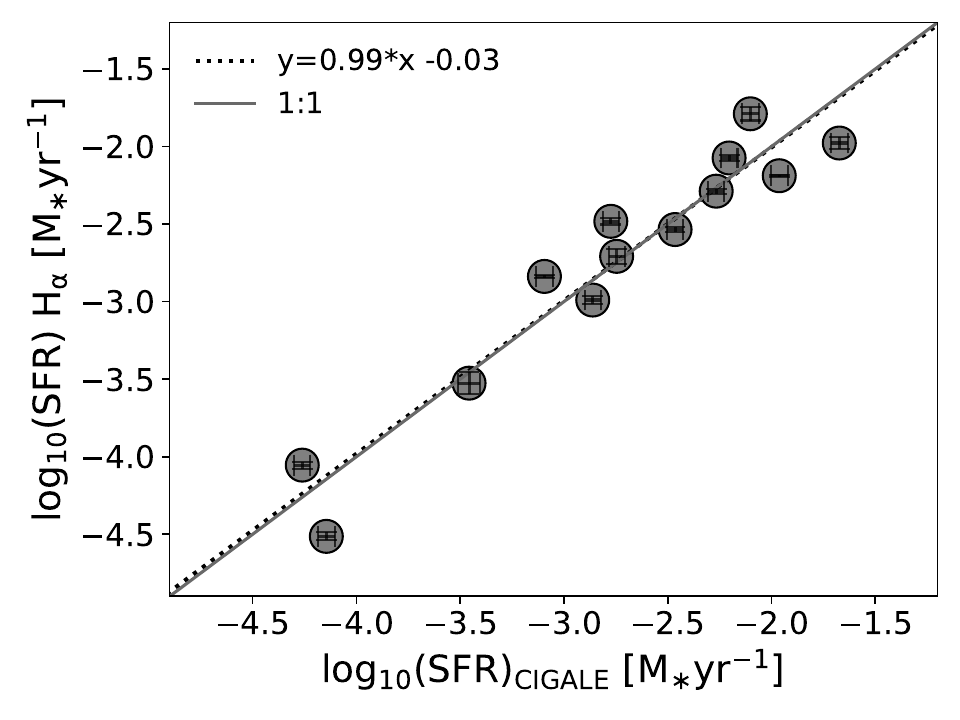}
  \caption{Comparison between the SFRs obtained from the \cigale~SED fitting code using five optical bands (x-axis), and those using $\rm{H{\alpha}}$ fluxes after dust and aperture correction (full grey circles). The solid line represents a 1:1 relation, while the dotted line represents a linear fit to the data.
  }
     \label{SFR_cigale_Halpha}    
\end{figure}
Figure~\ref{SFR_cigale_Halpha} shows a very good agreement between the SFR measured directly from $\rm{H{\alpha}}$ (corrected for dust extinction, as well as for the aperture correction) and the SFR estimated from the broadband SED fitting (black full circles). 
The SFR estimated using optical broadband photometry shows a small scatter, also lower than 0.10 dex. 
This result is more surprising, as for the \cigale{ } SFR estimation, we were able to use only five optical bands.  
A very good agreement with SFR directly measured from the H$\alpha$ line was only able to be obtained using a delayed SFH with additional quenching and using fluxes from dedicated forced photometry. 
Here, we report that using the simplest SFH, a delayed SFH without possible quenching, can result in a large overestimation (more than 0.6 dex) for analysed LSBGs. 
\end{appendix}
\end{document}